\documentclass[a4paper,11pt]{article}
\usepackage{amsfonts,amsmath,amssymb,mathrsfs}
\usepackage{float,graphicx}
\usepackage{hyperref}
\usepackage{enumitem}
\usepackage[vcentermath]{youngtab}
\usepackage{tocloft,ulem}
\usepackage{listings}
\usepackage{physics}
\usepackage[T1]{fontenc}
\usepackage{accents}

\usepackage{pgf}
\usepackage{ytableau,varwidth}

\numberwithin{equation}{section}

\newcommand{\Rep}[1]{\mathrm{Rep}(#1)}

\newcommand{\bea}{\begin{eqnarray}}
\newcommand{\eea}{\end{eqnarray}}
\newcommand{\acting}[1]{\mathfrak{#1}}
\newcommand{\cycn}[1]{\mathcal{C}_{#1}}
\newcommand{\invs}[1]{\mathcal{S} (#1)}

\newcommand{\ssp}[0]{\hspace{0.3mm}}
\newcommand{\Clebsch}[7]{S^{#1 \ssp}{}^{#2}_{#5}{}^{#3}_{#6}{}^{#4}_{#7}}

\newcommand{\CGs}[7]{S^{#4 \ssp}{}^{#3}_{#7}{}^{#1}_{#5}{}^{#2}_{#6}}

\newcommand{\mC}{ \mathbb{C}}
\newcommand{\cA}{\mathcal{A}}
\newcommand{\cN}{\mathcal{N}}
\newcommand{\cO}{\mathcal{O}}
\newcommand{\cZ}{\mathcal{Z}}
\newcommand{\cQ}{\mathcal{Q}}
\newcommand{\cB}{\mathcal{B}}
\newcommand{\cJ}{\mathcal{J}}
\newcommand{\cV}{\mathcal{V}}
\newcommand{\cM}{\mathcal{M}}
\newcommand{\cK}{\mathcal{K}}
\newcommand{\sfQ}[0]{\mathsf{Q}}
\newcommand{\sfN}[0]{\mathsf{N}}
\newcommand{\sfM}[0]{\mathsf{M}}
\newcommand{\scrE}[0]{\mathscr{E}}
\newcommand{\scrF}[0]{\mathscr{F}}

\newcommand{\bb}[1]{\mathbb{#1}}
\newcommand{\alg}[1]{\mathfrak{#1}}
\newcommand{\ol}[1]{\overline{#1}}
\newcommand{\hook}{\mathrm{hook}}
\newcommand{\Dim}{\mathrm{Dim}}
\newcommand{\wt}{\mathrm{Wt}}
\newcommand{\Vev}[1]{{\left\langle #1\right\rangle}}
\renewcommand{\atop}[2]{\begingroup \setlength{\arraycolsep}{1.1pt} \renewcommand{\arraystretch}{.9}
\begin{matrix} #1 \\ #2 \end{matrix} \endgroup}
\newcommand{\matop}[3]{{\begingroup \setlength{\arraycolsep}{1.1pt} \renewcommand{\arraystretch}{.9}\begin{matrix} #1 \\ #2 \end{matrix} \ #3 \endgroup}}
\renewcommand{\( }[0]{\left(}
\renewcommand{\) }[0]{\right)}

\begin{document} 
\thispagestyle{empty}

\ \vskip 30mm

\begin{center} 
\hspace{5mm}{\LARGE \bf 
Eigenvalue systems for integer orthogonal bases of}

\vspace{5mm}

\hspace{5mm}{\LARGE \bf 
multi-matrix invariants  at finite $N$}
\end{center} 

\vspace{3mm}

\vskip 16mm

\centerline{
{\large \bf Adrian Padellaro,${}^a$ 
Sanjaye Ramgoolam,${}^{b,c}$
Ryo Suzuki\,${}^d$
}}

\vskip 8mm

\hspace{5mm}{\it ${}^a$ 
Faculty of Physics, Bielefeld University, PO-Box 100131, D-33501 Bielefeld, Germany}

\hspace{5mm}{\it ${}^b$ 
Centre for Theoretical Physics, Department of Physics and Astronomy, }

\hspace{5mm}{\it \ \ 
Queen Mary University of London, London E1 4NS, United Kingdom}

\hspace{5mm}{\it ${}^c$ 
School of Theoretical Physics,  Dublin Institute for Advanced Studies, }

\hspace{5mm}{\it \ \ 
10 Burlington Road, Dublin 4, Ireland}

\hspace{5mm}{\it ${}^d$ Shing-Tung Yau Centre of Southeast University, No.2 Sipailou, Xuanwu district, Nanjing,}

\hspace{5mm}{\it \ \ 
Jiangsu, 210096, China}

{\let\thefootnote\relax\footnotetext{{\tt 
\href{apadellaro@physik.uni-bielefeld.de}{apadellaro(at)physik.uni-bielefeld.de}, 
\href{s.ramgoolam@qmul.ac.uk}{s.ramgoolam(at)qmul.ac.uk}, 
\href{rsuzuki.mp@gmail.com}{rsuzuki.mp(at)gmail.com}
}}}

\vskip 6mm

\centerline{\bf Abstract}

\vskip 4mm 

\hspace{1mm}\parbox{.9\linewidth}{%
Multi-matrix invariants, and in particular  the scalar multi-trace operators of $\cN=4$ SYM with $U(N)$ gauge symmetry, can be described using permutation centraliser algebras (PCA), which are generalisations of the symmetric group algebras and independent of $N$.
Free-field two-point functions define an $N$-dependent inner product on the PCA, and bases of operators have been constructed which are orthogonal at finite $N$.
Two such bases are well-known, the restricted Schur and covariant bases, and both definitions involve representation-theoretic quantities such as Young diagram labels, multiplicity labels, branching and Clebsch-Gordan coefficients for symmetric groups.    
The explicit computation of these coefficients grows rapidly in complexity as  the operator length increases.
We develop a new method for explicitly constructing all the operators with specified Young diagram labels, based on an $N$-independent integer eigensystem formulated in the PCA. 
The eigensystem construction naturally leads to orthogonal basis elements which are integer linear combinations of the multi-trace operators, and the $N$-dependence of their norms are simple known dimension factors. 
We provide examples and give computer codes in SageMath which efficiently implement the construction for operators of classical dimension up to 14. 
While the restricted Schur basis relies on the Artin-Wedderburn decomposition of  symmetric group algebras, the covariant basis relies on a variant which we refer to as the Kronecker decomposition. Analogous decompositions exist for any finite group algebra and the eigenvalue construction of integer orthogonal bases extends to the group algebra of any finite group with rational characters.
 }

\newpage
\setcounter{tocdepth}{2}

\tableofcontents

\newpage
\section{Introduction}\label{sec:intro}

An important problem in the AdS/CFT correspondence  \cite{Maldacena:1997re,Gubser:1998bc,Witten:1998qj} is to understand the map between quantum states in the CFT and in the string theory side. 
The operator-state map in CFTs \cite{Polchinski_1998} means that the quantum states in CFT correspond to local operators.
In any CFT with gauge symmetry, the local operators are constructed as gauge-invariant composites of the elementary fields. 
The study of the construction and correlators  of general half-BPS operators in four-dimensional $\cN=4$ super Yang-Mills (SYM) with $U(N)$ gauge group   $\cN=4$ SYM  \cite{Corley:2001zk} has allowed a detailed study of the physics of giant gravitons \cite{McGreevy:2000cw} and half-BPS geometries  \cite{Lin:2004nb} in ten-dimensional space-time from the dual CFT point of view. 

Half-BPS local operators  in the $U(N) $ theory are gauge-invariant holomorphic functions of a complex matrix $Z$ of size $N \times N$. 
Since the half-BPS sector is protected, the inner product on the space of half-BPS operators defined by the free-field two-point functions is valid for any coupling. 
The operators of dimension $L$, which are polynomial gauge invariants of degree $L$, have an orthogonal basis labelled by Young diagrams with $L$ boxes and no more than $N$ rows. The elements of the basis are expressible as linear combinations of multi-trace operators with coefficients given by characters of the symmetric group $S_L$ for the irreducible representation specified by the Young diagram, evaluated on a conjugacy class determined by the trace structure of the multi-trace operator. 
The norm of the operator in the free-field inner product is a polynomial in $N$ which is related to dimensions of $U(N)$ and $S_L$ representations associated with the Young diagram \cite{Corley:2001zk}.

The quarter-BPS and 1/8-BPS sectors contain gauge-invariant operators of more than one type of complex matrices, which we call multi-matrix operators or multi-matrix invariants. Multi-matrix operators are relevant to strings attached to giant gravitons and to brane-anti-brane systems. Finite $N$ bases for multi-matrix operators were constructed in \cite{Kimura:2007wy,Brown:2007xh,Bhattacharyya:2008rb,Bhattacharyya:2008xy,Brown:2008ij}, which are orthogonal under the free-field inner product, and the norms of the basis elements were calculated. 
The number of elements in these finite $N$ bases were shown (\cite{Collins:2008gc} and section 6 of \cite{Brown:2008ij}) to agree with finite $N$ counting formulae for gauge invariant operators \cite{Dolan:2007rq,Aharony:2003sx,Kinney:2005ej,Dutta:2007ws}.  The basis in \cite{Bhattacharyya:2008rb,Collins:2008gc} is known as the restricted Schur basis, and has its origins in studies of  general (BPS or non-BPS) open string excitations of giant gravitons \cite{Balasubramanian2005,deMelloKoch:2007rqf,deMelloKoch:2007nbd}. The simplest instance involves holomorphic gauge-invariant functions of two matrices. 
The construction of \cite{Kimura:2007wy} was motivated by the brane-anti-brane interpretation of operators constructed from $ Z , Z^{\dagger}$ and employed Walled Brauer algebras.  The basis in \cite{Brown:2007xh} was studied with motivations coming from the quarter-BPS sector and involved holomorphic functions of two complex matrices, which are quarter-BPS at zero Yang-Mills coupling. 
This basis is known as the covariant basis for the 
two-matrix system, since the global  $U(2)$ rotating the two matrices into each other is manifest. 
The generalisation to multi-matrix systems for general global symmetry group was described in \cite{Brown:2008ij}. The existence of different orthogonal bases was explained in terms of  Casimirs for  enhanced symmetries in the zero coupling limit in \cite{Kimura:2008ac}. The generalisation of the restricted Schur and covariant matrix bases to quiver gauge theories was given in \cite{Pasukonis:2013ts}. Reviews of work on applications of these results on field theoretic multi-matrix combinatorics are available in \cite{Ramgoolam:2008yr,Berenstein:2013md,Ramgoolam:2016ciq,deMelloKoch:2024sdf}.

It has been explained  that the space of 
two-matrix gauge-invariant operators  is closely related to the structure of an algebra $ \cA ( \mu_1,\mu_2) $ which is an instance of a permutation centraliser algebra (PCA) \cite{Mattioli:2016eyp}.
$\cA (\mu_1,\mu_2)$ is a sub-algebra of $ \mC [ S_{ L } ]$, where $ \mC [ S_{ L } ]$ is the group algebra of the symmetric group $S_L$ of all permutations of $\{ 1, 2, \cdots , L \}$. It is the sub-algebra 
which commutes with the sub-group $S_{ \mu_1} \times S_{\mu_2}$ of $S_L$, where  $ \mu_1 + \mu_2 = L $ and $ S_{ \mu_1} $ permutes $\{ 1, \cdots , \mu_1 \}$ and $ S_{\mu_2} $ permutes $ \{ \mu_1+1 , \cdots , \mu_1 + \mu_{ 2 } = L \}$:
\begin{equation}
\cA (\mu_1,\mu_2) = \Bigl\{ \sigma \in \bb{C} [S_{\mu_1+\mu_2}] \, \Big| \, \gamma \sigma \gamma^{-1} = \sigma , \ \ 
\forall \gamma \in S_{\mu_1} \times S_{\mu_2} \Bigr\} .
\end{equation}
More generally, the gauge-invariant operators of $M$ types of complex matrices are related to
\begin{equation}
\cA(\mu) = \cA(\mu_1, \mu_2, \dots, \mu_M) 
= \Big\{ \sigma \in \mathbb{C} [S_L] \, \Big| \, \gamma \sigma \gamma^{-1} = \sigma , \ \ 
\forall \gamma \in S_{\mu_1} \times S_{\mu_2} \times \dots \times S_{\mu_M} \subseteq S_L\Big\} , 
\end{equation}
$ \mu_1 + \mu_2 + \cdots + \mu_M = L$. The structure constants of PCAs have been shown to arise in the correlators of Gaussian matrix models with background fields \cite{Ramgoolam:2023vyq} and these algebras have been found to be useful in proving identities for Littlewood-Richardson coefficients which arise in quantum information theory \cite{Ramgoolam:2018ceu}.

In this paper we develop efficient algorithms to compute orthogonal bases for the permutation centraliser algebras $\cA ( \mu_1,\mu_2 ) $. 
These lead to orthogonal bases for the two-matrix system, which are closely related to the restricted Schur and covariant bases.

The restricted Schur basis for gauge-invariant composites of matrices $ Z , W $ of size $N \times N$, of degree $\mu_1$ in $Z$ and degree $\mu_2$ in $W$ takes the form 
\bea\label{reschu1} 
\cO^{ R, (r_1, r_2) }_{ \nu_-, \nu_+ }  [ Z , W ] 
\eea
where $r_1, r_2, R$ are Young diagrams with $\mu_1, \mu_2 , \mu_1+\mu_2\equiv L $ boxes respectively, with their number of rows should be less or equal to $N$.
These Young diagrams label irreducible representations of symmetric groups, denoted by $V_{r_1}^{S_{\mu_1}}  , V_{r_2}^{S_{\mu_2}}  , V_{R}^{S_{\mu_1+\mu_2}}$ respectively. 
The index $\nu_\mp$ is an integer ranging over the Littlewood-Richardson coefficient $g(r_1, r_2; R)$, which has an interpretation as the multiplicity in the reduction of the irreducible representation $V_{R}^{S_{\mu_1+\mu_2}}$ to the representation $V_{r_1}^{S_{\mu_1}}  \otimes  V_{r_2}^{S_{\mu_2}}$\,.
The restricted Schur basis is closely related to the Artin-Wedderburn decomposition of $\cA(\mu)$.
The explicit formula for the operators \eqref{reschu1} involves branching coefficients for this reduction (for the details see \cite{Bhattacharyya:2008rb,Bhattacharyya:2008xy,Kimura:2008ac}).
The explicit construction of these branching coefficients  has been discussed in the mathematical physics literature \cite{EHJ53,CCG83,PC93,McAven1998s,McAven1999s,McAven2002c,Chilla05,Chilla06} and become very complex when $\mu_1,\mu_2$ increase up to say $4,5$. Nevertheless the Young diagram structure of the restricted Schur basis has been useful in identifying new large $N$ integrable sectors in the two-matrix system (see \cite{deMelloKoch:2011wah,deMelloKoch:2012ck,Suzuki:2021sma} and references therein).

The covariant basis elements \cite{Brown:2007xh,Brown:2008ij} for gauge-invariant composites of $ Z , W $ of size $N \times N$ of total degree $L$ in the two matrices take the form 
\bea\label{covbas1}  
\cO^{ R , \Lambda , M_{ \Lambda } , \tau }  [ Z , W ] 
\eea
where $R,  \Lambda $ are Young diagrams with $L$ boxes  whose length (number of non-vanishing rows)  obey $\ell(R) \le N, \ell(\Lambda) \le 2$. 
$ \Lambda $ is associated with a representation $V_{ \Lambda }^{ S_L} $  of $S_L$ and also by Schur-Weyl duality with a representation $V_{ \Lambda }^{ U(2) } $ of $U(2)$. 
$M_{ \Lambda } $ denotes a state in the  representation $V_{ \Lambda }^{ U(2) } $ of $ U(2)$. The index $ \tau $ runs over the multiplicity of $V_{ \Lambda }^{ S_{ L} } $ in the tensor product decomposition of $ V_{ R }^{ S_L }  \otimes V_R^{ S_L }  $. 
This multiplicity is equal to the Kronecker coefficient $C(R, R, \Lambda)$ for the triple of representations $  ( R , R , \Lambda )$ of $S_L$ and is also equal to the number of times the trivial representation of $S_{L} $ appears in this tensor product $ V_{ R }^{ S_L }  \otimes V_R^{ S_L } \otimes V_{ \Lambda }^{ S_L }$. 
In this two-matrix case, the state label $M_{ \Lambda }$ can further be decomposed into a pair of non-negative integers $(\mu_1,\mu_2)$ with $ \mu_1+\mu_2 = L $, along with a branching multiplicity $\beta$. The integers $\mu_1$ and $\mu_2$ are the numbers of $Z$ and $W$ respectively in the composite operator, and the index $\beta$ ranges over the Kotska number $K_{\Lambda (\mu_1,\mu_2)}$, which represents how many times the trivial representation of $S_{\mu_1} \times S_{\mu_2}$ appears when the representation $V_{\Lambda}^{S_L}$ is restricted from $S_L$ to the subgroup $S_{\mu_1} \times S_{\mu_2}$\,.
The explicit formula for the basis elements involves Clebsch-Gordan coefficients for the overlaps of the states in $ V_{ R }^{ S_L }  \otimes V_R^{ S_L }$ with the states in the subspace $V_{ \Lambda }^{ S_L } $, as well as the branching coefficients for the trivial representation of $S_{\mu_1} \times S_{\mu_2}$ in  $V_{ \Lambda }^{ S_L  } $. 
Explicit examples at small $\mu_1,\mu_2$ in \cite{Brown:2007xh,Brown:2008ij} were given using standard group theory constructions of symmetric group Clebsch-Gordan coefficients \cite{hamermesh2012group}, but these constructions become rapidly very complex as $L$ increases.

We will explain in section \ref{sec:decomp CSL} that the covariant construction is closely related to a variant of Artin-Wedderburn decomposition of $\bb{C}[S_L]$ which we call Kronecker decomposition.

We will describe an eigenvalue system for matrices derived from a selection of central elements in $ \cA( \mu_1,\mu_2) $, which allows the construction of  multi-matrix invariants of restricted Schur type in the vector  space of dimension $g( r_1, r_2; R)^2$
\bea 
\bigoplus_{ \nu_- , \nu_+ =1 }^{ g(r_1, r_2; R) }  \hbox{ Span }  \left (  \cO^{ R, (r_1, r_2) }_{ \nu_-, \nu_+ }  [ Z, W ]   \right ) 
\eea 
The matrices in this eigenvalue system have non-negative integer entries.
The eigenvalues are also integers, as they are expressible in terms of characters of symmetric group elements in the irreducible representations $r_1 , r_2 , R$. 
This integrality of the matrices and their eigenvalues means that the eigenvectors can be constructed as integer linear combinations of multi-trace functions of $ Z , W$ by standard algorithms for null vectors of integer matrices based on Hermite normal forms. The output of the null vector algorithms can be efficiently orthogonalised, in the planar limit of the free-field inner product  by a Gram-Schmidt procedure. This  naturally produces rational linear combinations of multi-traces but can easily be adapted to produce integer combinations. 
Interestingly, the orthogonal basis for the planar inner product thus produced is automatically orthogonal under the finite $N$ inner product.

Extending these considerations to the covariant basis \eqref{covbas1} for each $ R , \Lambda , M_{ \Lambda }$, we will describe an eigenvalue system  allowing the construction of multi-matrix invariants in the  vector space of dimension $ C ( R , R , \Lambda ) $ 
\bea\label{covsubspace}  
\bigoplus_{ \tau =1}^{ C( R , R , \Lambda ) }  \hbox{ Span }  \Bigl( \cO^{ R , \Lambda , M_{ \Lambda } , \tau }  [ Z , W ]  \Bigr) .
\eea
In this case the eigenvalue system is derived by considering the matrix elements of the appropriate actions of central elements of $ \mC [ S_L]$. 
The left or right action selects $R$ and the adjoint action selects $ \Lambda $. The integrality of the eigenvalue system leads to a basis for \eqref{covsubspace} which consists of integer linear combinations of multi-traces. 
The states of different $ M_{ \Lambda } $ are related by the generators of $ U(2)$, expressed as operators on 
\bea 
\bigoplus_{ \substack{\mu_1,\mu_2 \\ \mu_1 + \mu_2 = L} }  \cA ( \mu_1,\mu_2 ) .
\eea

Similar eigenvalue methods have been developed in the context  of representation-theoretic orthogonal bases arising in tensor models with $U(N)$ symmetries \cite{BenGeloun:2020yau} as well as matrix \cite{padellaro2023} and tensor \cite{barnes2024permutation} models with $S_N$ symmetries . In a close similarity to  \eqref{reschu1} and \eqref{covbas1} the basis labels in these cases include representation labels as well as appropriate multiplicity labels.  These applications, as well as the present one, exploit properties of centres of symmetric group algebras whereby the characters of a small set conjugacy classes of $S_L$, which multiplicatively generate the centres, suffice to identify an irreducible representation  of $S_L$ \cite{Kemp:2019log,Ramgoolam:2022xfk}.

The paper is organised as follows.
In section \ref{sec:background} we review the connection between multi-matrix invariants, symmetric group algebras, permutation centraliser algebras and the previously mentioned representation bases.
An important result in section \ref{sec:EV Amn} is that the representation bases of $\cA(\mu)$ is characterised as eigenstates of an integer eigensystem constructed from 
certain linear combinations in $\mathbb{C}[S_L]$ acting on $\cA(\mu)$.
We explain how to construct integer eigenvector solutions to the eigensystem by using the Hermite normal form from the theory of integer matrices.
A similar construction is applied to the Artin-Wedderburn and the Kronecker decompositions of $\mathbb{C}[S_L]$ in section \ref{sec:EV CSL}.
Section \ref{sec:general fg} extends this dicussion to general finite group algebras, specialises to rational groups where analogous integrality properties hold.
Appendix \ref{app:notation} details the notation used for group and representation theoretic quantities.
Details on the algorithm used to compute integer bases for $\cA(\mu_1,\mu_2)$ are given in \ref{apx: algo}.
This paper is accompanied by a computer code which easily produces the results up to $L=10$ by laptop, and in principle works up to $L=14$.
We give a complete set of the restricted and covariant basis elements for $\cA(\mu_1,\mu_2)$ up to $L=5$, and give an interesting example at $L=6$ in appendix \ref{app:data Amu}.
Following section \ref{sec:EV CSL}, we give a complete set of bases elements for $\bb{C}[S_L]$ up to $L=4$ in appendix \ref{sec:data}.

\section{Symmetric group algebras and multi-matrix invariants}\label{sec:background}

Physically, we are interested in constructing an orthogonal basis of all multi-trace operators of $\cN=4$ SYM.
The $\cN=4$ SYM contains six real scalar fields in the adjoint representation of the gauge group $U(N)$, denoted by $\Phi^I_{ij}$ with $I=1,2,\dots, 6$ and $i,j=1,2,\dots, N$.
The $\alg{su}(2)$ sector of $\cN=4$ SYM is made of two complex scalars, $Z=\Phi^5 + \sqrt{-1} \, \Phi^6$ and $W = \Phi^3 + \sqrt{-1} \, \Phi^4$.
The general multi-trace operator in the $\alg{su}(2)$ sector of length $L=\mu_1+\mu_2$ can be written as
\begin{equation}
\cO_g [Z,W] = 
\sum_{i_1, i_2, \dots, i_{\mu_1+\mu_2}=1}^{N} \ 
\prod_{a = 1}^{\mu_1} Z_{i_{a} i_{g(a)}}
\prod_{b = \mu_1+1}^{\mu_1 + \mu_2} W_{i_{b} i_{g(b)}} 
\label{def:su(2) perm basis}
\end{equation}
with $g$ in the symmetric group $S_{\mu_1+\mu_2}$. 
It is straightforward to generalize \eqref{def:su(2) perm basis} to all multi-trace operators made of $M$ complex scalars as 
\begin{equation}
\cO_g (\vec a) = \sum_{i_1, i_2 , \dots , i_L=1}^N 
X^{a_1}_{i_1 i_g(1)} \, X^{a_2}_{i_2 i_g(2)} \, \dots \, X^{a_L}_{i_L i_g(L)} 
\label{def:cov perm basis}
\end{equation}
where $a_ i \in \{1, 2, \dots, M\}$. 
We call $\cO_g (\vec a)$ multi-matrix invariants, or gauge-invariant operators.
As will be shown below, the multi-matrix invariants possess a conjugation symmetry
\begin{equation}
\cO_g ( a_1, a_2 \dots a_L) = \cO_{\gamma g \gamma^{-1}} ( a_{\gamma(1)} \, a_{\gamma(2)} \, \dots a_{\gamma(L)} ) 
, \qquad \forall \gamma \in S_L \,.
\label{symm cov perm}
\end{equation}
An extra sign is needed in this equation when $X^a$ are fermionic fields. One example of such situations is the $\alg{su}(2|3)$ sector of $\cN=4$ SYM consisting of three complex scalars and two fermions \cite{Beisert:2003ys,deMelloKoch:2012sie}.

\subsection{Notation}\label{sec:notation}

The symmetric group algebra $\bb{C}[S_L]$ helps organise contractions of tensor indices of multi-matrix $U(N)$ invariants, and Young diagrams are useful for describing the irreducible representations of $\bb{C}[S_L]$.
We begin by fixing our notation about $\bb{C}[S_L]$ and Young diagrams.

\subsubsection{Group algebra actions}\label{sec:notation CSL}

Let $g_1 \,, g_2$ be elements in $S_L$\,. 
The composition of two permutations acts on $i \in \{ 1,2, \dots, L \}$ as
\begin{equation}
    (g_1 \, g_2 ) \, (i) = g_2 (g_1(i)) .
\label{def:composition g1g2}
\end{equation}
Let $\sigma \in \mathbb{C} [S_L]$ be given by
\begin{equation}
    \sigma = \sum_{g \in S_L} c_g (\sigma) \, g, \qquad c_g (\sigma) \in \mathbb{C}.
\label{def:cg of sigma}
\end{equation}
We define the $\delta$-function by
\begin{equation}
\delta (g) = 
\begin{cases}
1 &\qquad g = {\bf 1} \\
0 &\qquad {\rm otherwise}
\end{cases}
\label{def:delta fn}
\end{equation}
and extend it linearly to the group algebra $\mathbb{C}[S_L]$ by
\begin{equation}
\delta (\sigma) = \sum_{g \in S_L} c_g (\sigma) \, \delta( g ) = c_{\bf 1} (\sigma) .
\end{equation}

The left and right actions of $\bb{C}[S_L]$ are denoted by the following linear operators
\begin{equation}
    m^{\acting{L}} [\sigma] ( \tau ) = \sigma \tau, \qquad
    m^{\acting{R}} [\sigma] ( \tau ) = \tau \sigma, \qquad 
    \sigma, \tau \in \bb{C}[S_L].
\label{def:LR gen sig}
\end{equation}
The left and right actions always commute,
\begin{equation}
    m^{\acting{L}} [\sigma_1] \Big( m^{\acting{R}} [\sigma_2] ( h ) \Big)
    = m^{\acting{R}} [\sigma_2] \Big( m^{\acting{L}} [\sigma_1] ( h ) \Big)
    = \sigma_1 h \, \sigma_2  .
\end{equation}
Expanding the left or right action of $\bb{C}[S_L]$ in a basis of $S_L$\,, we obtain their matrix elements as
\begin{equation}
    \sigma h = \sum_{g \in S_L} m^{\acting{L}}_{gh} [\sigma] \, g, \qquad
    h \sigma = \sum_{g \in S_L} m^{\acting{R}}_{gh} [\sigma] \, g .
\label{def:LR matrix elements sig}
\end{equation}
which can be written as 
\begin{equation}
m^{\acting{L}}_{gh} [\sigma] = \delta \Big( g^{-1} \sigma h \Big), \qquad 
m^{\acting{R}}_{gh} [\sigma] = \delta \Big( g^{-1} h \sigma \Big) .
\label{extract mji from delta}
\end{equation}
We also introduce the involution
\begin{equation}
    \invs{\sigma} = \sum_{g \in S_L} c_g (\sigma) \, g^{-1}
\label{def:inverse of sigma}
\end{equation}
which is known as the antipode in the context of group Hopf algebras.
The adjoint action of $\bb{C}[S_L]$, often called conjugation in the literature, is denoted by
\begin{equation}
    m^{\acting{ad}}[\sigma ](h) = \sum_{g \in S_L} c_g (\sigma) \, g h g^{-1} , \qquad {\rm for} \quad
    \sigma  = \sum_{g \in S_L} c_g (\sigma) \, g, \quad
    \sigma, h \in \bb{C}[S_L].
\label{def:adj gen sig}
\end{equation}
The matrix elements of the adjoint action is denoted by
\begin{equation}
    m^{\acting{ad}}[\sigma ](h) = \sum_{g \in S_L} m^{\acting{ad}}_{gh} [\sigma] \, g .
\label{def:adj matrix sig}
\end{equation}

\subsubsection{Centre of $\mathbb{C}[S_L]$}\label{sec:centre}
The centre of $\mathbb{C}[S_L]$, denoted $\mathcal{Z}(\mathbb{C}[S_L])$, is the subalgebra defined by
\begin{equation}
    \mathcal{Z}(\mathbb{C}[S_L]) = \Big\{ z \in \mathbb{C} [S_L]  \, \Big| \, zg = gz \ \text{for all} \ g \in S_L \Big\}. 
\label{def:centre ZCSL}
\end{equation}
The centre has two important bases. The first basis is labelled by conjugacy classes of $S_L$.
Let $\rho \vdash L$ (here $\vdash L$ means an integer partition of $L$, see equation \eqref{def:R partition of L}) and $C_\rho$ the corresponding conjugacy class of $S_L$, we define
\begin{equation}
    z_\rho \equiv \frac{1}{\abs{{\rm Stab} (g_\rho)}} \sum_{\gamma \in S_L} \gamma \, g_\rho \, \gamma^{-1}
\label{def:SL equiv cycle p}
\end{equation}
for any $g_\rho \in C_\rho$, where ${\rm Stab}(g_\rho)$ is the stabiliser subgroup of $S_L$, which leaves $g_p$ invariant.
That $z_\rho$ is central is proven as follows
\begin{equation}
	h z_\rho  
	= \frac{1}{\abs{{\rm Stab} (g_\rho)}} \sum_{\gamma \in S_L} h\gamma \, g_\rho \, \gamma^{-1} 
	= \frac{1}{\abs{{\rm Stab} (g_\rho)}} \sum_{\gamma' \in S_L} \gamma' \, g_\rho \, \gamma'{}^{-1} h = z_\rho h
\label{centre proof}
\end{equation}
where we defined $\gamma' = h\gamma$.
The second basis is a set of the projection operators to the irreducible representation $R \vdash L$,
\begin{equation}
    P^R = \frac{d_R}{L!} \sum_{g \in S_L} \chi^R (g) \, g^{-1} 
\label{def:SL proj irrep R}
\end{equation}
where $\chi^R (g)$ is a character of $S_L$.
That $P^R$ is central follows from the fact that $\chi^R(g)$ is a class function.
The two bases are related by an eigensystem
\begin{equation}
	m^{\acting{L}}[z_\rho](P^R) = z_\rho P^R = \frac{\chi^R(z_\rho)}{d_R} P^R 
\end{equation}
and the eigenvalues are called normalised characters.
The last equality follows from Schur's lemma. As we will see in subsequent sections, generalisations of this eigenvalue equation exist for permutation centraliser algebras and play an important role in the eigenvalue method used in this paper.

There exists a subset of the above equations that uniquely distinguish all $P^R$.
It was proven in \cite[Section 3.4]{Kemp:2019log} (see \cite[Section 3.1]{Ramgoolam:2022xfk} for the general group algebra case) that the following two statements are equivalent
\begin{itemize}
    \item A set of central elements $z_1, z_2, \dots, z_k \in \mathcal{Z}(\mathbb{C}[S_L])$ multiplicatively generate the centre.
    \item The ordered list of normalised characters $(\frac{\chi^{R}(z_1)}{d_R}, \dots, \frac{\chi^{R}(z_k)}{d_R})$ uniquely determine $R \vdash L$.
\end{itemize}
 The following facts, which were observed in \cite{Kemp:2019log}, contribute to the efficiency of our algorithms. Let $p \in \{1, \dots, L\}$ and define
\begin{equation}
	T_p = z_{(p, 1^{L-p})},
\end{equation}
which is a sum of all the elements of $S_L$ with a single cycle of length $p$ and $L{-}p$ cycles of length $1$.
It was then observed that $T_2$ generates $\mathcal{Z}(\mathbb{C}[S_L])$ for $L=2,3,4,5,7$, and $T_2$ together with $T_3$ generates the centre for $L=6,8,9,\dots,14$.
Therefore, for $L \leq 14$ it is sufficient to solve the eigensystem
\begin{equation}
	m^{\acting{L}}[T_2](P^R) = \frac{\chi^R(T_2)}{d_R} P^R, \quad m^{\acting{L}}[T_3]( P^R) =  \frac{\chi^R(T_3)}{d_R} P^R,
\end{equation}
to find the projectors $P^R$.
More generally, there exists a $k_* < L$ such that $T_2, \dots, T_{k^*}$ generate the centre but $T_2, \dots, T_{k^*-1}$ do not.
An important observation is that $k_*$ is typically much smaller than $L$.

\subsubsection{Basis for multi-matrix $U(N)$ invariants}\label{sec:basis MMI}

Denote an $N$-dimensional vector space by $V_N = \mathrm{Span}_{\bb{C}} \, (e_i \, | \, i \in \{ 1,2, \dots, N \} )$, and its dual vector space by $\ol{V}_N = \mathrm{Span}_{\bb{C}} \, (\check{e}_j  \, | \, j \in \{ 1,2, \dots, N \} ) $.
Consider a set of $M$ linear operators 
\begin{equation}
    X^a: V_N \rightarrow V_N, \qquad a=1,2,\dots,M
\end{equation}
which defines a set of $N \times N$ matrices as $X^a e_i = \sum_{j=1}^N X^a_{ji} \, e_j$\,.
For a given tuple $\vec{a} = (a_1, a_2 , \dots, a_L) \in \{1, 2, \dots, M\}^{\times L}$, we define the tensor product of linear operators
\begin{equation}
   \mathcal{O}(\vec{a}) =  X^{a_1} \otimes X^{a_2} \otimes \dots \otimes X^{a_L}.
\end{equation}
The operator $\mathcal{O}(\vec{a})$ acts on 
\begin{equation}
    V_N^{\otimes L} = \mathrm{Span}_{\bb{C}} \, (e_{i_1} \otimes e_{i_2} \otimes \dots \otimes e_{i_L} \, | \, i_1, \dots, i_L = 1,\dots,N)
\end{equation}
as
\begin{equation}
    \mathcal{O}(\vec{a}) \, e_{i_1} \otimes \dots \otimes e_{i_L} 
    = X^{a_1}e_{i_1} \otimes \dots \otimes X^{a_L}e_{i_L}
    =\prod_{k=1}^L X^{a_k}_{j_k i_k} \ e_{j_1} \otimes \dots \otimes e_{j_L}.
\label{matrix elements of O vec}
\end{equation}
We define an operator $\mathcal{L}: S_L \rightarrow \mathrm{End}(V_N^{\otimes L})$ as the permutation of factors of the tensor product
\begin{align}
\mathcal{L} (g) \ket{e_{i_{1}} \otimes e_{i_{2}} \otimes \dots \otimes e_{i_{L}}} 
&= \big| e_{i_{g(1)}} \otimes e_{i_{g(2)}} \otimes \dots \otimes e_{i_{g(L)}} \big\rangle 
\label{permutation on tensor-1} \\[1mm]
\bra{ \check{e}_{j_{1}} \otimes \check{e}_{j_{2}} \otimes \dots \otimes \check{e}_{j_{L}} } \mathcal{L} (g^{-1}) 
&= \big\langle \check{e}_{j_{g(1)}} \otimes \check{e}_{j_{g(2)}} \otimes \dots \otimes \check{e}_{j_{g(L)}} \big|  \,.
\label{permutation on tensor-2}
\end{align}
This permutation operator obeys the composition rule $\mathcal{L} (g_1) \mathcal{L} (g_2) = \mathcal{L} (g_1 g_2)$ under the convention \eqref{def:composition g1g2}. 
It acts on the operator $\mathcal{O}(\vec{a})$ as
\begin{equation}
\mathcal{L} (g) \, \mathcal{O}(\vec{a}) \, \mathcal{L} (g^{-1})
= \mathcal{O}(g \cdot \vec{a})
\equiv X^{a_{g(1)}} \otimes X^{a_{g(2)}} \otimes \dots \otimes X^{a_{g(L)}} .
\label{permutation on tensor-3}
\end{equation}
where
\begin{equation}
    g \cdot \vec{a} = (a_{g(1)}, a_{g(2)}, \dots, a_{g(L)}).
\label{def:SL action on vec a}
\end{equation}
By combining \eqref{permutation on tensor-1}-\eqref{permutation on tensor-3}, we find
\begin{equation}
    \prod_{\ell=1}^L X^{a_{g(\ell)}}_{j_{g(\ell)} i_{g(\ell)}}
    = \prod_{k=1}^L X^{a_k}_{j_k i_k} 
\label{tensor relabelling identity}
\end{equation}
which means that the product of matrix elements are invariant under the relabelling.

By taking the trace over the tensor product space $V_N^{\otimes L}$, we obtain the multi-matrix invariant \eqref{def:cov perm basis}
\begin{equation}
    \mathcal{O}_g (\vec{a}) 
    \equiv \tr_{V_N^{\otimes L}} ( \mathcal{L} (g) \mathcal{O}(\vec{a})) 
    = \sum_{j_1, j_2 ,\dots, j_L=1}^{N} X^{a_1}_{j_1 j_{g(1)}} X^{a_2}_{j_2 j_{g(2)}} \dots X^{a_L}_{j_L j_{g(L)}} \,.
\label{def:cO g-a}
\end{equation}
It defines a map between the pairs $(g, \vec{a})$ and the space of multi-traces of multiple matrices.
For a fixed $\vec{a}$ this map can be extended linearly to elements of the group algebra $\mathbb{C} [S_L]$.
Let $\sigma \in \mathbb{C} [S_L]$ be given by \eqref{def:cg of sigma}.
The corresponding sum of multi-matrix invariants is given by
\begin{equation}
    \mathcal{O}_\sigma (\vec{a}) = \sum_{g \in S_L} c_g (\sigma) \, \mathcal{O}_g(\vec{a}).
\label{def:cO sig-a}
\end{equation}
From the cyclic property of the trace, we have
\begin{equation}
\sum_{i_1 , i_2 , \dots , i_L} \bra{ \check{e}_{i_1} \otimes \dots } \mathcal{L} (\sigma) \cO(\vec a) \ket{e_{i_1} \otimes \dots }
=
\sum_{i_1 , i_2 , \dots , i_L} \bra{ \check{e}_{i_1} \otimes \dots } \mathcal{L} (\gamma) \mathcal{L} (\sigma) \cO(\vec a) \mathcal{L} (\gamma^{-1}) \ket{e_{i_1} \otimes \dots } 
\end{equation}
for any $\gamma \in S_L$\,.
By evaluating this equation with the help of \eqref{permutation on tensor-1}-\eqref{permutation on tensor-3}, we can derive the identity for the multi-matrix invariants \eqref{symm cov perm},
\begin{equation}
\mathcal{O}_{\sigma} ( \vec a ) = \mathcal{O}_{\gamma \sigma \gamma^{-1}} (\gamma \cdot \vec a)
, \qquad \forall \gamma \in S_L \,.
\label{def:ad-gamma O}
\end{equation}

\subsubsection{Representations and Young diagrams}\label{sec:basis YD}

The irreducible representations of $S_L$ are labelled by integer partitions of $L$, or equivalently Young diagrams with $L$ boxes.
Young diagrams with $L$ boxes are also denoted by
\begin{equation}
R = [ R_1, R_2, \dots, R_\ell] \vdash L , \qquad
R_1 \ge R_2 \ge \dots \ge R_\ell , \qquad
\sum_{i=1}^{\ell} R_i = L , \qquad
\ell \equiv \ell(R).
\label{def:R partition of L}
\end{equation}

The dimension of the irreducible representation $V^{S_L}_R$ for $R\vdash L$ can be computed combinatorially using Young diagrams.
Given a Young diagram $R$, we label the boxes using coordinates $(i,j)$ where for example $(1,1)$ labels the top-left box.
Then the dimension $d_R = \dim V^{S_L}_R$ is
\begin{equation}
d_R = \frac{L!}{\hook_R} \,, \qquad
\hook_R = \prod_{(i,j) \in R} \Big( \text{hook length at }(i,j) \Big).
\label{def:dR hookR} 
\end{equation}
From Schur-Weyl duality, $R$ also labels an irreducible representation of $U(N)$.
The dimension of the irreducible representation $V^{U(N)}_R$ labelled by $R$ is 
\begin{equation}
\Dim_{N} (R) = \frac{d_R}{L!} \, \wt_{N} (R) \,, \qquad
\wt_{N} (R) = \prod_{(i,j) \in R} (N+i-j) .
\label{def:DimNR wtNR} 
\end{equation}

We use the symbol $e^R_I$ as the $I$-th basis element of $V^{S_L}_R$ with $I=1,2,\dots, d_R$\,.
We introduce the inner product of $S_L$ by $(e^R_I, e^S_J)$.
The representation $R$ is called unitary if the group action respects the inner product
\begin{equation}
( g \, e^R_I, g \, e^R_J) = (e^R_I, e^R_J) , \qquad  \forall g \in S_L \,.
\label{def:unitary inner}
\end{equation}
When the inner product is diagonal, $(e^R_I, e^S_J) = \delta^{RS} \, \delta_{IJ}$, we use Dirac's bracket notation
\begin{equation}
e^S_J = \ket{ \atop{S}{J} } , \qquad
(e^R_I, \cdot ) = \bra{ \atop{R}{I} }, \qquad
\Big\langle \, \atop{R}{I} \, \Big| \, \atop{S}{J} \, \Big\rangle 
= \delta^{RS} \, \delta_{IJ} \,.
\label{def:SL irrep inner prod}
\end{equation}
The group element $g \in S_L$ acts on this basis as
\begin{equation}
g \, \Big| \atop{R}{I} \, \Big\rangle = \sum_{J=1}^{d_R} \ 
D^R_{JI} (g) \, \Big| \atop{R}{J} \, \Big\rangle 
\label{def:g action DRIJ}
\end{equation}
where $D^R_{JI}(g)$ is the matrix element of the irreducible representation $R$\,. The matrix element for $g_1 g_2$ is
\begin{equation}
D^R_{JI}(g_1 g_2) = \sum_{K=1}^{d_R} D^R_{JK}(g_1) D^R_{KI}(g_2) .
\end{equation}
It follows that
\begin{equation}
\Vev{ \atop{R}{I} \, \Big| \, \sigma \, \Big| \, \atop{S}{J} } 
= \delta^{RS} \, D^R_{IJ} (\sigma) .
\label{def:DRIJ g}
\end{equation}
The unitary representation $R$ can also be defined by using the matrix elements as
\begin{equation}
D^R_{IJ} (g^{-1}) = D^{\ol{R}}_{JI} (g) 
 \label{def:unitary inverse}
\end{equation}
where $\ol{R}$ is the complex conjugate representation of $R$.
The character of $V^{S_L}_R$ is defined by
\begin{equation}
    \chi^R (\sigma) = \sum_{I=1}^{d_R} D^R_{II} (\sigma) .
\label{def:SL char}
\end{equation}
Since \eqref{def:SL irrep inner prod} is a complete basis, the projection operator to $V^{S_L}_R$ can also be written as
\begin{equation}
  P^R \ = \ \sum_{I=1}^{d_R}  \Big| \, \atop{R}{I} \, \Big\rangle \Big\langle \, \atop{R}{I} \, \Big| .
\label{app:projector PR}
\end{equation}
The dimension $d_R$ is equal to the number of standard Young tableaux of shape $R$. For example, when $R = [3,1] = \raisebox{-1.4mm}{\tiny \yng(3,1)}$\,, we find $d_R = 3$ and
\ytableausetup{boxsize=1em,centertableaux}
\begin{equation}
\Biggl\{
\ket{ \atop{R}{1} } , \ \ket{ \atop{R}{2} } , \ \ket{ \atop{R}{3} }
\Biggr\}
\ = \ 
\Biggl\{ \, \begin{ytableau}
1 & 2 & 3 \\ 4
\end{ytableau} \ , \ \begin{ytableau}
1 & 2 & 4 \\ 3
\end{ytableau} \ , \ \begin{ytableau}
1 & 3 & 4 \\ 2
\end{ytableau} \, 
\Biggr\}  .
\end{equation}

\subsection{Two decompositions of $\mathbb{C}[S_L]$}\label{sec:decomp CSL}

We will discuss two interesting decompositions of $\bb{C}[S_L]$. In the first decomposition, $\bb{C}[S_L]$ decomposes into a direct sum of matrix sub-algebras, which transform as $  V^{S_L}_R \otimes  V^{ S_L}_{  R }  $   under the simultaneous left and right action of $\bb{C}[S_L]$ on itself. We refer to this as the Artin-Wedderburn decomposition since it is an example of the Artin-Wedderburn decomposition of general semi-simple associative algebras into simple algebras. It is  also known, in the context of finite groups, as the Maschke's decomposition, since Maschke's decomposition establishes the semi-simplicity of the group algebras of finite groups \cite{Wiki-Wedderburn-Artin,Wiki-Maschke}. 
In the second decomposition, the simultaneous left and right actions of $\bb{C}[S_L]$ on $\bb{C}[S_L]$ are recognised  as an action of $ \bb{C}[S_L] \otimes \bb{C}[S_L] $ and we decompose the matrix blocks $V^{S_L}_R \otimes  V^{S_L}_{R }$ under the diagonal embedding $ \bb{C}[S_L] \hookrightarrow \bb{C}[S_L] \otimes \bb{C}[S_L] $. 
We call it Kronecker decomposition, as it involves a multiplicity space whose dimension is equal to the Kronecker coefficient $C(R, R, \Lambda)$.
These  Kronecker coefficient multiplicities have appeared in the construction of covariant bases of matrix operators \cite{Brown:2007xh,Brown:2008ij}.

\subsubsection{Artin-Wedderburn decomposition}

Any irreducible representations of $S_L$ satisfy the orthogonality relation
\begin{equation}
    \sum_{g \in S_L} D^R_{IJ}(g^{-1}) \, D^S_{KL}(g) 
    = \frac{|S_L|}{d_R} \, \delta^{RS} \, \delta_{IL} \, \delta_{JK} \,.
\label{grand orthogonality SL}
\end{equation}
All irreducible representations of $S_L$ can be chosen real and unitary, so that $D^R_{IJ}(g^{-1}) = D^R_{JI}(g)$.
With this assumption, we define the following elements of $\bb{C}[S_L]$,
\begin{equation}
    Q_{IJ}^R = \frac{d_R}{|S_L|} \sum_{g \in S_L} D^R_{JI}(g) \, g^{-1} .
\label{def:AW basis}
\end{equation}
Using \eqref{grand orthogonality SL}, one can prove that they satisfy the relation \cite[Proposition 11, page 49]{Serre1977},
\begin{equation}
    Q_{IJ}^R \, Q_{KL}^S = \delta^{RS} \, \delta_{JK} \, Q^R_{IL} \,.
\label{def:matrix units SL}
\end{equation}
These elements are called matrix units (borrowing the language in \cite[Theorem 3.7]{Ram1991}).
The matrix units form a complete basis of $\bb{C}[S_L]$ making the following group theoretical identity manifest
\begin{equation}
|S_L| = L! = \sum_{R \, \vdash L}  d_R^2 .
\end{equation}

The existence of a complete basis of matrix units is guaranteed by the Artin-Wedderburn theorem.
The Artin-Wedderburn theorem, applied to $\mathbb{C}[S_L]$, states that $\mathbb{C}[S_L]$ decomposes into the direct sum of simple subalgebras, and each simple subalgebra is a matrix algebra over an irreducible representation of $S_L$ \cite[Section 6.2]{Serre1977}.
The Artin-Wedderburn decomposition of $\mathbb{C}[S_L]$ is written as
\begin{equation}
\mathbb{C}[S_L] \cong \bigoplus_{R \, \vdash L} {\rm Mat} \, (V_R^{S_L}) 
\label{eq: AW CSL}
\end{equation}
where $\mathrm{Mat}(V_R^{S_L})$ is the matrix algebra over $V_R^{S_L}$\,, consisting of $d_R \times d_R$ matrices.
The matrix units $Q_{IJ}^R$ in \eqref{def:matrix units SL} can be identified as a complete basis of ${\rm Mat} \, (V_R^{S_L})$.

The matrix algebra over $V_R$ can be written as a tensor product
\begin{equation}
    {\rm Mat}(V^{S_L}_R) \cong V^{S_L}_R \otimes V^{S_L}_{R} \,.
\label{mat to VRxVR}
\end{equation}
To see this, consider group actions on the matrix units. From the definition in \eqref{def:AW basis}, we have
\begin{equation}
\begin{aligned}
    h \, Q^R_{IJ} 
    &= \frac{d_R}{|S_L|} \sum_{g \in S_L} D^R_{JI}(g) \, hg^{-1} \\
    &= \frac{d_R}{|S_L|} \sum_{k \in S_L} \sum_{K=1}^{d_R} D^R_{JK}(k) D^R_{KI}(h) \, k^{-1} 
    = \sum_{K=1}^{d_R} D^R_{KI}(h) \, Q^R_{KJ}.
\end{aligned}
\label{left action on QRIJ}
\end{equation}
where in the second equality we defined $k^{-1}=hg^{-1}$.
Following similar steps we also have
\begin{equation}
    Q^R_{IJ} \, h^{-1} 
    =\sum_{K=1}^{d_R} D^R_{JK} (h^{-1}) \, Q^R_{IK}
    =\sum_{K=1}^{d_R} D^R_{KJ} (h) \, Q^R_{IK} \,.
\label{right inverse action on QRIJ}
\end{equation}
where we used the fact that the representation matrices $D^R_{IJ}(h)$ are real and unitary.
This argument shows that under the following group action
\begin{equation}
    (h,h') \cdot Q^R_{IJ} \equiv h \, Q^R_{IJ} \, (h')^{-1} \,, \qquad (h, h') \in S_L \times S_L 
\end{equation}
the subscripts of $Q^R_{IJ}$ separately form a basis for irreducible representations $V^{S_L}_R \otimes V^{S_L}_{R}$ of $S_L \times S_L$\,, which is \eqref{mat to VRxVR}.

The centre $\mathcal{Z}(\mathbb{C}[S_L])$ acts on the matrix units as the multiplication of a constant known as a normalised character,
\begin{equation}
    m^{\acting{L}}[z](Q^R_{IJ}) = m^{\acting{R}}[z](Q^{R}_{IJ}) = \frac{\chi^R (z)}{d_R} \, Q^R_{IJ} 
\label{centre acting on MU}
\end{equation}
where we used \eqref{left action on QRIJ}, \eqref{right inverse action on QRIJ} and Schur's lemma which implies that $D^R_{IJ} (z) = \chi^R(z) \, \delta_{IJ}/d_R$ for $z \in \cZ(\bb{C}[S_L])$.

\subsubsection{Kronecker decomposition}
\label{subsubsec: Kronecker decomp}
Recall that the matrix algebra over the irreducible representation $V_R^{S_L}$ can be regarded as a tensor product of a pair of $V_R^{S_L}$'s \eqref{mat to VRxVR}. We decompose this tensor product into a sum of irreducible representations,
\begin{equation}
    \mathbb{C}[S_L] \ \cong \ \bigoplus_{R \, \vdash L} \ V_R^{S_L} \otimes V_R^{S_L}
    \ \cong \ 
    \bigoplus_{R \, \vdash L} \bigoplus_{\Lambda \, \vdash L} \ 
    V_\Lambda^{S_L} \otimes V_{R,R,\Lambda} 
\label{eq: RRLambda decomp}
\end{equation}
where the multiplicity space $V_{R,R,\Lambda}$ has the dimension equal to the Kronecker coefficient,
\begin{equation}
    \dim V_{R,R,\Lambda} = C(R,R,\Lambda)
    = \frac{1}{\abs{S_L}} \sum_{g \in S_L} \chi^R (g) \chi^R(g) \chi^\Lambda (g) .
\label{def:Kronecker coef}
\end{equation}
We take an explicit basis of $V_R^{S_L}$ as in Section \ref{sec:basis YD} and rewrite the irreducible decomposition \eqref{eq: RRLambda decomp} as
\begin{equation}
\ket{ \atop{R}{I} } \otimes \ket{ \atop{R}{J} }
 = \sum_{\Lambda \, \vdash L} \sum_{\tau=1}^{C(R,R,\Lambda)} \sum_{K=1}^{d_\Lambda}  
\CGs{R}{R}{\Lambda}{\tau}{I}{J}{K} \, 
\ket{ \matop{\Lambda}{K}{\tau} } 
\label{def:CG coeff}
\end{equation}
where $\CGs{R}{R}{\Lambda}{\tau}{I}{J}{K}$ is called the Clebsch-Gordan coefficient and $\tau$ is a label for an orthogonal basis in the multiplicity space. You can read more about these in appendix \ref{app:CG of SL}.

Now consider the following transformation of the matrix units,
\begin{equation}
\cQ^{R,\Lambda, \tau}_{\ \ K} = \sum_{I,J=1}^{d_R} \CGs{R}{R}{\Lambda}{\tau}{I}{J}{K} \, Q^R_{IJ} \,.
\label{def:kr basis SL}
\end{equation}
Since this is a unitary transformation, the elements $\{ \cQ^{R,\Lambda, \tau}_{\ \ K} \}$ also form a complete basis of $\bb{C}[S_L]$,
\begin{equation}
\bb{C} [S_L] \cong \mathrm{Span}_{\bb{C}} \, \Big( \cQ^{R,\Lambda, \tau}_{\ \ K} \, \Big| \, R, \Lambda \vdash L , 
K\in \{1, \dots , d_\Lambda\}, \ \tau \in  \{1, \dots, C(R,R,\Lambda) \}\Big) .
\label{def:Kronecker decomp}
\end{equation}
We call \eqref{def:kr basis SL} Kronecker basis and \eqref{def:Kronecker decomp} Kronecker decomposition.
This equation is consistent with the identity
\begin{equation}
|S_L| = L! = \sum_{R , \Lambda \, \vdash L} C(R,R,\Lambda) \, d_\Lambda \,.
\end{equation}
It turns out that the subscript $K$ of the Kronecker basis forms an irreducible representation $V^{S_L}_\Lambda$. In particular, if we consider the adjoint action of $\gamma \in S_L$ defined in \eqref{def:adj gen sig}, we will find
\begin{equation}
    m^{\acting{ad}}[\gamma ](\cQ^{R,\Lambda, \tau}_{\ \ K}) 
    = \sum_{K'} \cQ^{R,\Lambda, \tau}_{\ \ K'} \, D^\Lambda_{K' K} (\gamma) .
\label{adjoint action on Kr basis-1}
\end{equation}
To see this, we combine \eqref{def:AW basis} and \eqref{def:kr basis SL} as
\begin{equation}
\begin{aligned}
    m^{\acting{ad}}[\gamma ](\cQ^{R,\Lambda, \tau}_{\ \ K}) 
    &= \frac{d_R}{|S_L|} \sum_{g \in S_L} \sum_{I,J} 
    \CGs{R}{R}{\Lambda}{\tau}{I}{J}{K} \, D^R_{JI}(g) \, \gamma g^{-1} \gamma^{-1}
    \\
    &= \frac{d_R}{|S_L|} \sum_{g \in S_L} \sum_{I,J,I',J'} 
    \CGs{R}{R}{\Lambda}{\tau}{I}{J}{K} \, D^R_{JJ'} (\gamma^{-1}) D^R_{J'I'} (g) D^R_{I'I} (\gamma) \, g^{-1} \,.
\end{aligned}
\label{adjoint action on Kr basis-2}
\end{equation}
Since $R$ is a real and unitary representation, we have $D^R_{I'I} (\gamma) = D^R_{II'} (\gamma^{-1})$.
Using the equivariance property of Clebsch-Gordan coefficients \eqref{app:SDD=SD}, the equation \eqref{adjoint action on Kr basis-2} simplifies as
\begin{equation}
\begin{aligned}
    m^{\acting{ad}}[\gamma ](\cQ^{R,\Lambda, \tau}_{\ \ K}) 
    = \frac{d_R}{|S_L|} \sum_{g \in S_L} \sum_{I',J',K'} 
    D^{\Lambda}_{K K'} (\gamma^{-1}) \, \CGs{R}{R}{\Lambda}{\tau}{I'}{J'}{K'} \, D^R_{J'I'} (g) \, g^{-1} 
    = \sum_{K'} \cQ^{R,\Lambda, \tau}_{\ \ K'} \, D^\Lambda_{K' K} (\gamma) 
\end{aligned}
\label{adjoint action on Kr basis-3}
\end{equation}
which is \eqref{adjoint action on Kr basis-1}.

Let us compute the inner product of the Kronecker basis using the $\delta$-function defined by \eqref{def:delta fn}.
We write
\begin{align}
\invs{\cQ^{R,\Lambda, \tau}_{\ \ K}}&= \frac{d_R}{|S_L|} \sum_{\sigma \in S_L} \sum_{I,J}
\CGs{R}{R}{\Lambda}{\tau}{I}{J}{K} \, D^R_{JI} (\sigma) \, \sigma
\\
\cQ^{R', \Lambda', \tau'}_{\ \ K'} &= \frac{d_{R'}}{|S_L|} \sum_{\sigma' \in S_L} \sum_{I',J'}
\CGs{R'}{R'}{\Lambda'}{\tau'}{I'}{J'}{K'} \, D^{R'}_{J'I'} (\sigma') \, \sigma'{}^{-1}
\end{align}
and use the relation
\begin{equation}
\sum_{\sigma, \sigma' \in S_L} \delta \Big( D^R_{JI} (\sigma) \, D^{R'}_{J'I'} (\sigma') \, \sigma \, \sigma'{}^{-1} \Big)
= \sum_{\sigma} D^R_{JI} (\sigma) \, D^{R'}_{J'I'}  (\sigma) 
= \frac{|S_L|}{d_R} \, \delta^{RR'} \delta_{JJ'} \, \delta_{II'} 
\label{MU inner prod}
\end{equation}
together with the orthogonality of the Clebsch-Gordan coefficients \eqref{BHR_160}.
Then we find
\begin{equation}
\delta \Big( \invs{\cQ^{R, \Lambda, \tau}_{\ \ K}} \, \cQ^{R', \Lambda', \tau'}_{\ \ K'} \Big)
= \frac{d_R}{|S_L|} \, \delta^{RR'} \delta^{\Lambda\Lambda'} \, \delta^{\tau \tau'} \, \delta_{K K'} \,.
\label{Kron basis norm}
\end{equation}
This $\delta$-function inner product is the same as the planar two-point function of $\cN=4$ SYM at zero coupling, as will be discussed in Section \ref{sec:orth finite N}.

Let us consider the adjoint action of the centre on $ \mC [ S_L ]$ and in particular on the basis
 $\cQ^{R,\Lambda, \tau}_{\ \ K}$.
Following the notation in Section \ref{sec:notation CSL}, we write
\begin{equation}
m^{\acting{ad}}[z ](\sigma) = \sum_{g \in S_L} c_g (z) \, g^{-1}\sigma g , \qquad {\rm for} \quad
z = \sum_{g \in S_L} c_g (z) \, g \ \in \ \mathcal{Z}(\mathbb{C} [S_L] ) .
\label{def:adj centre sig}
\end{equation}
The left and right action of  the center $ \mathcal{Z}(\mathbb{C} [S_L] ) $ on $ \mC [ S_L ]$ are identical. This left/right action commutes with the adjoint action of $ \mathcal{Z}(\mathbb{C} [S_L] ) $ on $ \mC [ S_L]$ . 
For $z_1 \,, z_2 \in \cZ (\bb{C} [S_L])$ and $ \sigma \in \mC [S_L]$
\begin{equation}
m^{\acting{ad}}[{z_1} ](m^{\acting{L}}[z_2](\sigma)) = m^{\acting{ad}}[{z_1} ](m^{\acting{R}}[z_2](\sigma)) 
= m^{\acting{L}}[z_2](m^{\acting{ad}}[{z_1} ](\sigma)) = m^{\acting{R}}[z_2](m^{\acting{ad}}[{z_1} ](\sigma)) .
\end{equation}
Equivalently as linear operators on $ \mC [S_L]$, 
\bea 
m^{\acting{ad}}[{z_1} ] m^{\acting{L}}[z_2] = m^{\acting{L}}[z_2]m^{\acting{ad}}[{z_1} ]
\eea 
It follows from \eqref{adjoint action on Kr basis-1} and \eqref{centre acting on MU} that the centre acts on the Kronecker basis as
\begin{equation}
m^{\acting{L}}[z_1](\cQ^{R,\Lambda, \tau}_{\ \ K}) = m^{\acting{R}}[z_1](\cQ^{R,\Lambda, \tau}_{\ \ K}) = \frac{\chi^R (z_1)}{d_R} \, \cQ^{R,\Lambda, \tau}_{\ \ K} \,, \qquad
m^{\acting{ad}}[{z_2} ]( \cQ^{R,\Lambda, \tau}_{\ \ K}  ) = \frac{\chi^\Lambda(z_2)}{d_\Lambda}  \cQ^{R,\Lambda, \tau}_{\ \ K}  
\label{adj centre acting on MU}
\end{equation}
In other words, the Kronecker basis simultaneously diagonalises the left/right action and the adjoint action of the centre $\mathcal{Z}(\mathbb{C} [S_L] )$.

The state label $I,J$ in the matrix units $Q^R_{IJ}$ and $K$ in the Kronecker basis $\cQ^{R,\Lambda, \tau}_{\ \ K}$ can be determined by the action of non-central elements such as the Young-Jucys-Murphy elements. We will elaborate on this in Section \ref{sec:EV CSL}.

\subsection{Permutation centraliser algebra $\cA(\mu)$}\label{sec:PCA}

Consider the multi-matrix invariants $\cO_g (\vec a)$ in \eqref{def:cO g-a} which consist of $M$ types of $N \times N$ matrices. We fix the field content of $\cO_g (\vec a)$ as having $\mu_1$ $X^1$'s, $\mu_2$ $X^2$'s and so on up to $\mu_M$ $X^M$'s. 
Each multi-matrix invariant corresponds to a choice of the field contents $\mu = (\mu_1, \dots, \mu_M)$ within the range
\begin{equation}
\mu_\ell \in \{0,1, \dots, L\} \quad \text{such that} \quad \sum_{\ell=1}^M \mu_\ell = L .
\end{equation}
For simplicity we assume that the matrix size is larger than or equal to the operator length, $N \ge L$.

Using the conjugation symmetry \eqref{symm cov perm}, we may rearrange $\vec a$ as
\begin{equation}
    \vec a_\mu = ( \underbrace{1,1, \dots, 1}_{\mu_1} , \underbrace{2,2, \dots, 2}_{\mu_2} , \dots, 
    \underbrace{M,M,\dots, M}_{\mu_M} ) .
\label{def:vec amu}
\end{equation}
We denote the multi-matrix invariant for the field content $\mu = (\mu_1, \dots, \mu_M)$ by
\begin{equation}
    \mathcal{O}_g (\vec a_\mu) 
    = \tr_{V_N^{\otimes L}} \Big( \mathcal{L} (g) \, 
    (X^1)^{\otimes \mu_1} \otimes (X^2)^{\otimes \mu_2} \otimes \dots \otimes (X^M)^{\otimes \mu_M} \Big).
\label{def:cOg mu}
\end{equation}
This $\mathcal{O}_g (\vec a_\mu)$ is invariant under conjugation by the subgroup,
\begin{equation}
    \mathcal{O}_g ( \vec a_\mu ) = \mathcal{O}_{hgh^{-1}} ( \vec a_\mu ), \qquad \forall (g, h) \in S_L \times S_\mu \,. 
\label{inv adj Smu}
\end{equation}
where $S_\mu$ is called the Young subgroup indexed by $\mu$,
\begin{equation}
    S_\mu = S_{\mu_1} \times S_{\mu_2} \times \dots \times S_{\mu_M} \subseteq S_L \,.
\label{def:Young subgroup}
\end{equation}
The sum over the orbits of the equivalence class of $S_\mu$ in \eqref{inv adj Smu} defines an interesting subalgebra of $\bb{C}[S_L]$ called permutation centraliser algebra \cite{Mattioli:2016eyp}, 
\begin{equation}
    \cA(\mu) = \cA(\mu_1, \mu_2, \dots, \mu_M) 
    = \Big\{ \sigma \in \mathbb{C} [S_L] \, \Big| \, h \sigma h^{-1}= \sigma \ \ \text{for all} \ h \in S_{\mu} \Big\} 
\label{def:cAmu}
\end{equation}
One may also write $\cA(\mu) = \bb{C} [S_L]^{S_\mu}$, where $G^H$ is the $H$-invariant subspace of $G$.

For any $g \in S_L$ we define the $S_\mu$-orbit
\begin{equation}
    \{g\}_\mu = \{ hgh^{-1} \, | \, \forall h \in S_\mu \}.
\end{equation}
Let $l = \dim \cA(\mu)$ and $g_1, g_2, \dots, g_l$ be a complete set of representatives.
That is, a set of elements satisfying 
\begin{equation}
    \{g_1\}_\mu \cup \{g_2\}_\mu \dots \cup \dots \{g_l\}_\mu = S_L 
\label{eq: Amu representatives}
\end{equation}
and $\{g_i\}_\mu \cap \{g_j\}_\mu$ is empty if $i\neq j$.
Let us define a linear operator $P_\mu: S_L \rightarrow A(\mu)$ by
\begin{equation}
    P_\mu (g) = \frac{1}{|\mathrm{Stab} (g)|} \sum_{\gamma \in S_\mu} \gamma g \gamma^{-1}
\label{def:Pmu sig}
\end{equation}
where $\mathrm{Stab} (g)$ is the subgroup of $S_\mu$ that leaves $g$ invariant under conjugation.\footnote{The stabilizer subgroup $\mathrm{Stab} (\sigma)$ is also called Automorphism group, denoted by $\mathrm{Aut} (\sigma)$ in \cite{Ramgoolam:2023vyq}.}
The algebra $A(\mu)$ has a linear basis labelled by the $S_\mu$-orbits,
\begin{equation}
    \{ P_\mu(g_1), P_\mu(g_2), \dots, P_\mu(g_l) \}
    \label{eq: Amu orbit basis}
\end{equation}
that we call the orbit basis.
It follows that
\begin{equation}
    \mathcal{O}_{P_\mu (\sigma)} ( \vec a_\mu ) 
    = \abs{ \mathrm{Orb} (\sigma) } \mathcal{O}_\sigma ( \vec a_\mu ), \qquad 
    \sigma \in \bb{C}[S_L]
\label{cO on orbit basis P}
\end{equation}
where $| \mathrm{Orb} (\sigma)|$ is the number of different elements in $\bb{C}[S_L]$ which belong to the same equivalence class of $\cA(\mu)$. We have $\abs{\mathrm{Orb} (\sigma)} \abs{\mathrm{Stab} (\sigma)} = \abs{ S_\mu}$ according to the orbit-stabilizer theorem.

There are two interesting cases where $\cA(\mu)$ reduces to well-known algebras.
The first case is
\begin{equation}
    \cA(L) = \Big\{ \sigma \in \mathbb{C} [S_L] \, \Big| \, h \sigma h^{-1}= \sigma \ \ \text{for all} \ h \in S_L \Big\} 
    = \mathcal{Z}(\mathbb{C}[S_L])
\label{case of cAL}
\end{equation}
which corresponds to the centre of the symmetric group algebra and describes the half-BPS operators of $\cN=4$ SYM. The second case is
\begin{equation}
    \cA( \underbrace{1,1,\dots,1}_L ) = \mathbb{C} [S_L] 
\label{case of cA11...}
\end{equation}
which is the symmetric group algebra discussed in Section \ref{sec:decomp CSL}.

 The dimension of $ \cA ( \mu ) $ is equal to the number of multi-matrix invariants at large $N$, i.e. for operators with $L \le N$. 
\begin{equation}
    \sigma \in \cA(\mu) \quad \stackrel{\text{large $N$}}{\Longleftrightarrow} \quad
    \mathcal{O}_{\sigma} ( \vec a_\mu ) \in 
    \Bigl\{ \text{$M$-matrix invariants of degree $L \leq N$ and field content $\mu$} \Bigr\} .
\label{eq:Omu correspondence} 
\end{equation}
The restricted Schur basis of multi-matrix operators \cite{Bhattacharyya:2008rb,Bhattacharyya:2008xy} corresponds to a restricted basis in
 $ \cA ( \mu ) $ with labels $R \vdash L, r_i \vdash \mu_i$, in addition to multiplicity labels $1 \le  \nu_\mp \le g ( r_1, r_2 ; R ) $.  The finite $N$ constraint is  the condition $ \ell (R) \le N$ on this basis. Equivalently the space of finite $N$ multi-matrix invariants is in 1-1 correspondence with a quotient of $ \cA ( \mu ) $ defined by setting to zero all the elements of the restricted basis with $ l( R ) > N$. In the next two sub-sections we will arrive at the restricted basis in $ \cA ( \mu ) $ by starting with the Artin-Wedderburn basis in $ \mC [ S_L ] $ and projecting with branching coefficients for the subgroup $ S_{ \mu } \rightarrow S_{L}$.

 The covariant basis \cite{Brown:2007xh,Brown:2008ij} corresponds to a basis in $ \cA ( \mu ) $  with labels $ R , \Lambda \vdash L $, along with a branching multiplicity coefficient for branching of the irreducible representation of $S_L$ labelled by $ \Lambda $ into the trivial representation of $ S_{ \mu}$, and a Clebsch-Gordan (or Kronecker) multiplicity for $ V^{ S_L}_{ \Lambda} $ in the tensor product $ V_{R}^{ S_L} \otimes V_{ R}^{ S_L} $. The finite $N$ constraint is again the condition $ \ell (R) \le N$ on this basis. Equivalently the space of finite $N$ multi-matrix invariants is in 1-1 correspondence with a quotient of $ \cA ( \mu ) $ defined by setting to zero all the elements of the covariant basis with $ \ell ( R ) > N$.

\subsection{Restricted Schur basis}\label{sec: res basis}

We review the restricted Schur basis \cite{deMelloKoch:2007rqf} emphasising the connection with the Artin-Wedderburn basis for $\mathbb{C}[S_L]$ in Section \ref{sec:decomp CSL}.

Given an irreducible representation of $S_L$, we can restrict $S_L \rightarrow S_\mu$ and look at the corresponding decomposition into irreducible representations of $S_\mu$
\begin{equation}
    V_R^{S_L} \cong \bigoplus_{r_1 \vdash \mu_1} \dots \bigoplus_{r_M \vdash \mu_M} \Bigl(
    V^{S_{\mu_1}}_{r_1} \otimes \dots \otimes V^{S_{\mu_M}}_{r_M} \otimes V_{R \rightarrow (r_1 \dots r_M)} \Bigr) .
    \label{eq: smu decomp}
\end{equation}
Here $V_{R \rightarrow (r_1 \dots r_M)}$ is the multiplicity space of this decomposition, and its dimension is equal to the Littlewood-Richardson coefficient
\begin{equation}
    \dim V_{R \rightarrow (r_1 \dots r_M)} = g(r_1 \dots r_m; R ) .
\end{equation}
Concretely, we can write an orthonormal basis of states for the irreducible representation of $S_\mu$ as
\begin{equation}
\ket{ \matop{r_1 & \dots & r_M}{i_1 & \dots &i_M}{\nu} } 
= \sum_{I=1}^{d_R} B^{R \to (r_1 \dots  r_M)}_{I \to (i_1  \dots i_M), \nu} \, \ket{\atop{R}{I}} 
\label{def:restricted irrep inv}
\end{equation}
where $i_k =1,\dots, d_{r_k}$ and $\nu=1,\dots, g(r_1 \dots r_M; R)$. The transformation matrix $B^{R \to (r_1 \dots  r_M)}_{I \to (i_1  \dots i_M), \nu}$ is called the branching coefficients, whose properties are discussed in detail in Appendix \ref{sec:branching coeff}.
The restricted Schur basis is built from this decomposition as 
\begin{equation}
\begin{gathered}
    \cA (\mu) \ \cong \ \mathrm{Span}_{\bb{C}} \, \Big( 
    Q^{R , (r_1 \dots r_M)}_{\nu_+ \nu_-} \, \Big| \, R \vdash L, r_k \vdash \mu_k \,, \nu_\pm \in \{ 1, \dots, g(R; r_1, \dots, r_M) \} \Big) 
\\[1mm]
    Q^{R , (r_1 \dots r_M)}_{\nu_+ \nu_-} = \sum_{I,J} \sum_{i_1, \dots, i_M} 
        B^{R \rightarrow (r_1 \dots r_M)}_{I \rightarrow (i_1 \dots i_M), \nu_-} Q^{R}_{IJ} \,
        B^{R \rightarrow (r_1 \dots r_M)}_{J \rightarrow (i_1 \dots i_M), \nu_+}
\end{gathered}
\label{eq: AW Amu}
\end{equation}
Tracing over the representations $(r_1 \otimes \dots \otimes r_M)$ ensures that the restricted Schur basis satisfies
\begin{equation}
    h \, Q^{R , (r_1 \dots r_M)}_{\nu_+,\nu_-} \, h^{-1} = Q^{R , (r_1 \dots r_M)}_{\nu_+ \nu_-} , \qquad \forall h \in S_\mu.
\end{equation}
The numerical coefficient of $Q^{R , (r_1 \dots r_M)}_{\nu_+ \nu_-}$ in front of $g^{-1} \in S_L$ is called the restricted Schur character \cite{deMelloKoch:2007rqf},
\begin{equation}
    \chi^{R, (r_1 \dots r_M)}_{\nu_+ \nu_-} (g) = \frac{d_R}{\abs{S_L}} 
    \sum_{I,J} \sum_{i_1 \dots i_M}
    B^{R \rightarrow (r_1 \dots r_M)}_{I \rightarrow (i_1 \dots i_M);\nu_-} D^{R}_{JI} (g) \,
    B^{R \rightarrow (r_1 \dots r_M)}_{J \rightarrow (i_1 \dots i_M); \nu_+} \,.
\label{def:restricted Schur}
\end{equation}
Note that the restricted Schur character is invariant under the permutation of $(r_1 \,, \dots, r_M)$.
The restricted Schur basis is a matrix unit of $\cA(\mu)$ as they satisfy the relation \cite{Mattioli:2016eyp,Mattioli:2016gyl}
\begin{equation}
    Q^{R , (r_1 \dots r_M)}_{\nu_+ \nu_-} \, Q^{S , (s_1 \dots s_M)}_{\xi_+ \xi_-} 
    = \delta^{RS} \, \Big( \prod_{k=1}^M \delta^{r_k s_k} \Big) \, \delta_{\nu_- \xi_+} \,
    Q^{R , (r_1 \dots r_M)}_{\nu_+ \xi_-} \,. 
\label{def:matrix units Amu}
\end{equation}
The $\delta$-function inner product of the restricted Schur basis can be computed as
\begin{equation}
\delta \Big( \invs{Q^{R , (r_1 \dots r_M)}_{\nu_+ \nu_-}} \, Q^{S , (s_1 \dots s_M)}_{\xi_+ \xi_-} \Big) 
= \delta^{RS} \, \Big( \prod_{k=1}^M \delta^{r_k s_k} \Big) \, \delta_{\nu_+ \xi_+} \,
\delta ( Q^{R , (r_1 \dots r_M)}_{\nu_- \xi_-} )
= \delta^{RS} \, \Big( \prod_{k=1}^M \delta^{r_k s_k} \, d_{r_k} \Big) \, \delta_{\nu_+ \xi_+} \, \delta_{\nu_- \xi_-} 
\label{res Schur norm}
\end{equation}
where we used the real unitarity of $R$,
\begin{equation}
\invs{Q^{R , (r_1 \dots r_M)}_{\nu_+ \nu_-}} = \sum_{I,J} \sum_{i_1, \dots, i_M} 
        B^{R \rightarrow (r_1 \dots r_M)}_{I \rightarrow (i_1 \dots i_M), \nu_-} D^{R}_{IJ} (\sigma) \,
        B^{R \rightarrow (r_1 \dots r_M)}_{J \rightarrow (i_1 \dots i_M), \nu_+} \, \sigma
= Q^{R , (r_1 \dots r_M)}_{\nu_- \nu_+} \,.
\end{equation}

In fact, this basis is a Artin-Wedderburn decomposition of $\cA(\mu)$ \cite{Kimura:2008ac}
\begin{equation}
    \cA(\mu) \cong \bigoplus_{\substack{R \, \vdash L \\ r_i \vdash \mu_i}} 
    {\rm Mat} ( V_{R \rightarrow (r_1 \dots r_M)} )
\label{WA decomp Amu}
\end{equation}
and it gives the formula
\begin{equation}
    \dim \cA(\mu) = \sum_{\substack{R \, \vdash L \\ r_i \vdash \mu_i}}   g(r_1, \dots, r_M ; R)^2.
\end{equation}
With the help of \eqref{eq:Omu correspondence}, we write the corresponding multi-matrix invariants as
\begin{equation}
    \mathcal{O}^{R, (r_1 \dots r_m)}_{\nu_+ \nu_-}(\vec a_\mu) 
    = \tr_{V_N^{\otimes L}} \Bigl( \mathcal{L} ( Q^{R (r_1 \dots r_M)}_{\nu_+ \nu_-} ) \,
    (X^1)^{\otimes \mu_1} \otimes (X^2)^{\otimes \mu_2} \otimes \dots \otimes (X^M)^{\otimes \mu_M}
    \Bigr) 
\label{def:res Schur op mu}
\end{equation}
which is called the restricted Schur operators.
The operators $\mathcal{O}^{R, (r_1 \dots r_m)}_{\nu_+ \nu_-}(\vec a_\mu)$ form a basis of multi-matrix invariants for $L \le N$. At finite $N$, only the subset of operators with $\ell (R) \leq N$ remain linearly independent \cite{Corley:2001zk,deMelloKoch:2007rqf}.
The restricted Schur operators form an orthonormal basis with respect to the two-point functions of $\cN=4$ SYM at zero coupling. The operators above the cutoff $\ell (R) > N$ have zero norm.
Further discussion will be given in Section \ref{sec:orth finite N} about the two-point functions at finite $N$.

Let us investigate the properties of the restricted Schur basis as in Section \ref{sec:decomp CSL}.
The following basic observations were made in \cite{Kimura:2008ac}.
First, it follows from \eqref{eq: AW Amu} that for $z \in \mathcal{Z}(\mathbb{C}[S_L])$
\begin{equation}
    z \, Q^{R , (r_1 \dots r_M)}_{\nu_+ \nu_-} = \frac{\chi^R(z)}{d_R} \, Q^{R , (r_1 \dots r_M)}_{\nu_+ \nu_-}.
    \label{eq: L-centre on res schur}
\end{equation}
Secondly, the irreducible representations $r_k$ can be determined in a similar manner.
For this, the relevant subalgebras are the centres of the group algebras $\mathbb{C}[S_{\mu_k}] \subseteq \mathbb{C}[S_L]$.
As we will now show, for any $z_k \in \mathcal{Z}(\mathbb{C}[S_{\mu_k}])$
\begin{equation}
    z_k \, Q^{R , (r_1 \dots r_M)}_{\nu_+ \nu_-}  
    = \frac{\chi^{r_k}(z_k)}{d_{r_k}} \, Q^{R , (r_1 \dots r_M)}_{\nu_+ \nu_-} \, .
    \label{eq: m-centre on res schur}
\end{equation}
To prove this, consider
\begin{equation}
z_k = \sum_{\gamma \in S_{\mu_k}} c_\gamma (z_k) \, \gamma \ \in \ \mathcal{Z}(\mathbb{C}[S_{\mu_k}])
\end{equation}
and compute its left action on the restricted Schur basis
\begin{equation}
\begin{aligned}
&z_k \, Q^{R , (r_1 \dots r_M)}_{\nu_+ \nu_-}  
= \sum_{\gamma \in S_{\mu_k} } \sum_{\sigma \in S_L} \sum_{I,J,K} \sum_{i_1 \dots i_M}
c_\gamma (z_k) \,
B^{R \rightarrow (r_1 \dots r_M)}_{I \rightarrow (i_1 \dots i_M);\nu_-} D^{R}_{JK} (\sigma) \, D^{R}_{KI} (\gamma) \,
B^{R \rightarrow (r_1 \dots r_M)}_{J \rightarrow (i_1 \dots i_M); \nu_+} \sigma^{-1}
\\
&= \sum_{\gamma \in S_{\mu_k} } \sum_{\sigma \in S_L} \sum_{J,K} \sum_{\substack{i_1 \dots i_M \\ j_1 \dots j_M}}
c_\gamma (z_k) \,
B^{R \rightarrow (r_1 \dots r_M)}_{K \rightarrow (j_1 \dots j_M);\nu_-} 
D^{r_k}_{j_k i_k} (\gamma) \, \Big( \prod_{\ell \neq k}^M \delta_{j_\ell \, i_\ell} \Big) \,
D^{R}_{JK} (\sigma) \, 
B^{R \rightarrow (r_1 \dots r_M)}_{J \rightarrow (i_1 \dots i_M); \nu_+} \sigma^{-1}
\end{aligned}
\label{app: m-centre on res schur-1}
\end{equation}
where we used the equivariance property of branching coefficients \eqref{branching equivariance-1}. 
Since the central element $z_k \in \mathcal{Z}(\mathbb{C}[S_{\mu_k}])$ satisfies $D^{r_k}_{j_k i_k} (z_k) = \chi^{r_k} (z_k) \, \delta_{j_k i_k}/d_{r_k}$, we find
\begin{equation}
\begin{aligned}
z_k \, Q^{R , (r_1 \dots r_M)}_{\nu_+ \nu_-}  
&= \frac{\chi^{r_k}(z_k)}{d_{r_k}} \, \sum_{J,K} \sum_{i_1 \dots i_M}
B^{R \rightarrow (r_1 \dots r_M)}_{K \rightarrow (i_1 \dots i_M);\nu_-} Q^R_{KJ}
B^{R \rightarrow (r_1 \dots r_M)}_{J \rightarrow (i_1 \dots i_M); \nu_+}  
\\
&= \frac{\chi^{r_k}(z_k)}{d_{r_k}} \, Q^{R , (r_1 \dots r_M)}_{\nu_+ \nu_-}
\end{aligned}
\label{app: m-centre on res schur-2}
\end{equation}
which is \eqref{eq: m-centre on res schur}.
The actions of $\mathcal{Z}(\mathbb{C}[S_L]),\mathcal{Z}(\mathbb{C}[S_{\mu_1}]), \dots, \mathcal{Z}(\mathbb{C}[S_{\mu_M}])$ all commute and can be simultaneously diagonalised.
The corresponding eigenspaces, which we will construct in section \ref{sec:EV Amn} are
\begin{equation}
    \cA^{R, r_1, \dots, r_M}(\mu) = \mathrm{Span} \Big( Q^{R, (r_1 \dots r_M)}_{\nu_+ \nu_-} \, | \, 
    \nu_+,\nu_- = 1, \dots g(r_1, \dots, r_M ; R ) \Big). 
    \label{eq: R r1 r2}
\end{equation}

\subsection{Covariant basis} \label{sec: cov basis}

In a similar manner we will review the covariant basis \cite{Brown:2007xh,Brown:2008ij}, which can be viewed as generalisation of the Kronecker basis of $\mathbb{C}[S_L]$.

Recall that the Kronecker basis was organised in terms of irreducible representations $V^{S_L}_\Lambda$ of the adjoint action of $S_L$.
Restricting to the adjoint action of $S_\mu$ gives the following decomposition
\begin{equation}
    V_\Lambda^{S_L} 
    \cong \bigoplus_{r_1 \vdash \mu_1} \dots \bigoplus_{r_M \vdash \mu_M} \Bigl(
    V^{S_{\mu_1}}_{r_1} \otimes \dots \otimes V^{S_{\mu_M}}_{r_M} \otimes 
    V_{\Lambda \rightarrow (r_1 \dots r_M)}
    \Bigr) 
\label{def:part cov VSmu}
\end{equation}
where $V_{\Lambda \rightarrow (r_1 \dots r_M)}$ is the multiplicity space.
Since we are interested in the states invariant under the adjoint action of $S_\mu$\,, we pick out the trivial representation as
\begin{equation}
    V_\Lambda^{S_L} \Big|_{S_\mu\text{-inv}} \ \cong \ 
    V^{S_\mu}_{\text{triv} (\mu)} \otimes V_{\Lambda \rightarrow ([\mu_1] \dots [\mu_M])} \,, \qquad
    V^{S_\mu}_{\text{triv} (\mu)} \equiv V^{S_{\mu_1}}_{[\mu_1]} \otimes \dots \otimes V^{S_{\mu_M}}_{[\mu_M]}
\label{def:VLam Smu-inv}
\end{equation}
where ${\rm triv} (\mu) = ([\mu_1] \dots [\mu_M])$ is the trivial representation of $S_{\mu}$. 
The multiplicity space has the dimension
\begin{equation}
    \dim V_{\Lambda \rightarrow ([\mu_1] \dots [\mu_M])} = g( [\mu_1], \dots, [\mu_M] ; \Lambda ) \equiv  K_{\Lambda \mu},
\label{def:Kotska number}
\end{equation}
which is also known as the Kostka number. 
In concrete terms, we take orthonormal bases of states and express \eqref{def:part cov VSmu} as
\begin{equation}
\ket{ \matop{r_1 & \dots & r_M}{k_1 & \dots &k_M}{\beta} } 
= \sum_{K=1}^{d_{\Lambda}} B^{\Lambda \to (r_1 \dots  r_M)}_{K \to (k_1  \dots k_M), \beta} \, \ket{\atop{\Lambda}{K}} 
\label{def:restricted irrep Lambda}
\end{equation}
with $\beta = 1, \dots, K_{\Lambda \mu}$ as in \eqref{def:restricted irrep inv}. 
We are interested in the $S_{\mu}$-invariant subspace as in \eqref{def:VLam Smu-inv}. 
Let us write the branching coefficients \eqref{def:restricted irrep inv} to the trivial representation as
\begin{equation}
\ket{ \matop{[\mu_1] & \dots & [\mu_M]}{1 & \dots &1}{\beta} } 
= \sum_{K=1}^{d_{\Lambda}} 
B^{\Lambda \rightarrow {\rm triv}(\mu)}_{K, \beta}  \, \ket{\atop{\Lambda}{K}} ,\qquad
B^{\Lambda \rightarrow {\rm triv}(\mu)}_{K, \beta} \equiv 
B^{\Lambda \rightarrow ([\mu_1] \dots [\mu_M])}_{K \rightarrow (1 \dots 1), \beta} \,.
\label{def:branching Lambda Mu}
\end{equation}

From \eqref{def:branching Lambda Mu} we obtain the covariant basis for $\cA(\mu)$,
\begin{equation}
\begin{gathered}
    \cA (\mu) \ \cong \ \mathrm{Span}_{\bb{C}} \, \Big( 
    \cQ^{R, \Lambda, \mu, \tau}_{\beta} \, \Big| \, R \vdash L, \Lambda \vdash L,
    \tau \in \{ 1,\dots,C(R,R,\Lambda) \}, \beta \in \{ 1,\dots,K_{\Lambda \mu} \} \Big) 
\\[1mm]
    \cQ^{R, \Lambda, \mu, \tau}_{\beta} = \sum_{K}
    B^{\Lambda \rightarrow {\rm triv}(\mu)}_{K, \beta} \, \cQ^{R,\Lambda, \tau}_{\ \ K} 
\end{gathered}
\label{eq: cov Amu}
\end{equation}
which implies 
\begin{equation}
    \dim \cA(\mu) = \sum_{R, \Lambda \vdash L} C(R,R,\Lambda) K_{\Lambda \mu} \,.
\end{equation}
The corresponding matrix invariants
\begin{equation}
    \mathcal{O}^{R, \Lambda, \mu, \tau}_{\beta}(\vec a_\mu) 
    = \tr_{V_N^{\otimes L}} \Bigl( \mathcal{L} ( \cQ^{R, \Lambda, \mu ,\tau}_{\beta} ) \,
    (X^1)^{\otimes \mu_1} \otimes (X^2)^{\otimes \mu_2} \otimes \dots \otimes (X^M)^{\otimes \mu_M}
    \Bigr) 
\label{def:cov op mu}
\end{equation}
is called the covariant basis of operators.
Again, the covariant basis of operators diagonalises the two-point functions of $\cN=4$ SYM at zero coupling. 
When $L > N$ we need a cut-off on $R$ given by $\ell(R) \leq N$.
We may impose an extra condition $\ell(\Lambda) \le M$ to the equation \eqref{eq: cov Amu}, because the Kotska number $K_{\Lambda\mu}$ vanishes if $\ell(\Lambda) > M$.\footnote{The symbol $K_{\Lambda\mu}$ counts the number of semi-standard Young tableau of shape $\lambda$ and contents $\mu$. In other words, one must fill the numbers $1 \dots M$ in the Young diagram $\Lambda$. If $\ell(\Lambda) > M$ the Young tableau cannot be semi-standard.}

The real unitarity of $R$ gives
\begin{equation}
\invs{\cQ^{R, \Lambda, \mu, \tau}_{\beta}} = \cQ^{R, \Lambda, \mu, \tau}_{\beta} 
\end{equation}
and the $\delta$-function inner product of the covariant basis is
\begin{equation}
\delta \Big( \invs{\cQ^{R, \Lambda, \mu, \tau}_{\beta}} \, \cQ^{R', \Lambda', \mu', \tau'}_{\beta'} \Big)
= \frac{d_R}{|S_L|} \, \delta^{RR'} \delta^{\Lambda\Lambda'} \, \delta^{\tau \tau'} \, \delta^{\mu \mu'} \,
\delta_{\beta \beta'} \,.
\label{cov basis norm}
\end{equation}
This result follows from \eqref{Kron basis norm} and the orthogonality relation of the branching coefficients \eqref{product branching}.

It immediately follows from the fact that we started with the Kronecker basis that, 
\begin{alignat}{9}
        z \, \cQ^{R, \Lambda, \mu, \tau}_{\beta} &= \frac{\chi^{R}(z)}{d_R} \, \cQ^{R, \Lambda, \mu, \tau}_{\beta}, 
        &\quad z &\in \mathcal{Z}(\mathbb{C}[S_L])
        \label{eq: L-centre on kronecker}\\
        m^{\acting{ad}}[z ]( \cQ^{R, \Lambda, \mu, \tau}_{\beta} ) &= \frac{\chi^{\Lambda}(z)}{d_\Lambda} \, \cQ^{R, \Lambda, \mu, \tau}_{\beta}, 
        &\quad z &\in \mathcal{Z}(\mathbb{C}[S_L]).
        \label{eq: adjoint on kronecker}
\end{alignat}
We give a name to the corresponding eigenspaces
\begin{equation}
    \cA^{R, \Lambda}(\mu) 
    = \mathrm{Span} \Big( \cQ^{R, \Lambda, \mu, \tau}_{\beta} \, | \, \tau \in \{ 1,\dots,C(R,R,\Lambda) \}, \beta \in \{ 1,\dots,K_{\Lambda \mu} \} \Big).
\label{def:cA RLam mu}
\end{equation}
They will be constructed in section \ref{sec:EV Amn}.

\subsection{General covariant basis}
\label{subsec: gen cov basis}

Recall that the multi-matrix invariants in \eqref{def:cO g-a} respect the global $U(M)$ symmetry.
The tuple $\vec a = (a_1, \dots, a_L)$ belongs to the tensor product $V_M^{\otimes L}$, which decomposes into
\begin{equation}
V_M^{\otimes L} = \bigoplus \limits_{\substack{\Lambda \vdash L \\ \ell(\Lambda) \le M}} \Big( V^{S_L}_\Lambda \otimes V^{U(M)}_{\Lambda} \Big)
\end{equation}
under the Schur-Weyl duality of $S_L \times U(M)$ symmetry.
By taking an explicit basis of states, we can rewrite it as
\begin{equation}
\ket{a_1 \,, \dots \,, a_L}_{U(M)} = \sum_{\substack{\Lambda \vdash L \\[.5mm] \ell(\Lambda) \le M}} \sum_{M_\Lambda=1}^{\Dim_M (\Lambda)}
\sum_{K=1}^{d_\Lambda}
C^{\Lambda}_K {}^{\Lambda}_{M_\Lambda}  (\vec a) \, 
\ket{ \atop{\Lambda}{K} } \ket{ \atop{\Lambda}{M_\Lambda} }_{U(M)}  
\label{def:CG coeff SLxU(M)}
\end{equation}
where $C^{\Lambda}_K {}^{\Lambda}_{M_\Lambda} (\vec a)$ is the Clebsch-Gordan coefficient of $S_L \times U(M)$, whose properties will be explained in detail in Appendix \ref{app:CG of UM}.
Following \cite{Brown:2008ij}, we define the general covariant basis of operators as
\begin{equation}
    \cO^{R, \Lambda, M_\Lambda, \tau} = \sum_{a_1, \dots, a_L=1}^M
    \sum_{K} 
    C^{\Lambda}_K {}^{\Lambda}_{M_\Lambda} (\vec a) \ 
    \tr_{V_N^{\otimes L}} \Bigl( \mathcal{L} ( \cQ^{R,\Lambda, \tau}_{\ \ K} ) \,
    \cO (\vec a) 
    \Bigr) 
\label{def:general cov basis}
\end{equation}
where $\Lambda$ is the irreducible representation of $SU(M)$ with its dimension equal to $\Dim_M (\Lambda)$.
The general covariant basis respects the global $U(M)$ symmetry, and hence the name `covariant'.

The parameters $\mu=(\mu_1 \,, \dots \,, \mu_M)$ in \eqref{def:cov op mu} represent the $U(1)$ charges of the Cartan generators in $U(M)$.\footnote{Our $U(1)$ charges take non-negative integer values.}
If the states with different $M_\Lambda$ carry the same $U(1)^M$ charges, the number of such states is counted by the Kotska number. 
Therefore, we can interchangeably specify a state in $\Lambda$ by $M_\Lambda$ or $(\mu, \beta)$.
Concretely we can write
\begin{equation}
    \ket{ \atop{\Lambda}{M_\Lambda} }_{U(M)}
    = \sum_{\substack{\mu_1 , \dots , \mu_M \ge 0 \\ \mu_1 + \dots + \mu_M = L} } 
    \sum_{\beta=1}^{K_{\Lambda\mu}}
    \cB^{\Lambda \to (\mu_1 \dots \mu_M)}_{M_\Lambda \to \beta}
    \ket{ \atop{\Lambda}{\mu,\beta} }_{U(M)} 
\label{branching Lambda Mu SUM}
\end{equation}
where $\cB^{\Lambda \to (\mu_1 \dots \mu_M)}_{M_\Lambda \to \beta}$ is the coefficients of some unitary transformation.
This equation implies
\begin{equation}
    \Dim_M (\Lambda) = \sum_{\substack{\mu_1 , \dots , \mu_M \ge 0 \\ \mu_1 + \dots + \mu_M = L} }
    K_{\Lambda\mu} \,.
\label{Dim as Kotska sum}
\end{equation}
We stress that the multiplicity space of $U(M) \to U(1)^M$ has the same dimension as the multiplicity space of the restriction $S_L \to S_\mu$ in \eqref{def:Kotska number},
\begin{equation}
\dim V^{U(M) \rightarrow U(1)^M}_{\Lambda \to \mu }
= \dim V_{\Lambda \rightarrow ([\mu_1] \dots [\mu_M])} 
= K_{\Lambda \mu} 
\label{Kotska SW dual}
\end{equation}
which can be explained by Schur-Weyl duality \cite{deMelloKoch:2012ck}.

By using \eqref{branching Lambda Mu SUM}, we can relate the original covariant basis and the general covariant basis as
\begin{equation}
    {\rm Span} \, \Big( \mathcal{O}^{R, \Lambda, \mu, \tau}_{\beta}(\vec a_\mu) \, \Big| \, \beta \in \{ 1, ..., K_{\Lambda\mu} \} \Big)
    = {\rm Span} \, \Big( \sum_{M_\Lambda} \cB^{\Lambda \to (\mu_1 \dots \mu_M)}_{M_\Lambda \to \beta} \cO^{R, \Lambda,M_\Lambda, \tau} 
    \, \Big| \, \beta \in \{ 1, ..., K_{\Lambda\mu} \} \Big) .
\label{identify cov gen cov-1}
\end{equation}
In other words, the original covariant basis is a fixed-charge projection of the general covariant basis.

\section{Eigenvalue method for $\cA(\mu_1,\mu_2)$}\label{sec:EV Amn}

Having described the general representation theoretic structure of the restricted Schur and covariant bases of the PCA $ \cA ( \mu_1 , \cdots , \mu_M )$  we now turn our attention to concrete construction algorithms.
These involve integer matrix algorithms, in particular Hermite normal forms. 
We will see that these constructions give finite $N$ integer orthogonal bases (with respect to the free two-point function in $\mathcal{N}=4$ SYM).
Similar integer matrix methods were used in \cite{BenGeloun:2020yau} to describe and construct representation theoretic subspace of algebras related to tensor invariants and Kronecker coefficients.

For concreteness, we will consider the case of $M=2, L=\mu_1+\mu_2 \leq 14$ and $\mu = (\mu_1,\mu_2)$, but the generalisation is straightforward. In this section, we describe the mathematical aspects of the construction, while a detailed description of the code is given in Appendix \ref{apx: algo}. For specified $\mu_1 , \mu_2 $, the code produces as output : 
\begin{itemize} 
\item A restricted basis $\cA ( \mu_1 , \mu_2 )$, labelled by Young diagrams $ R \vdash \mu_1 + \mu_2 = L , r_1 \vdash \mu_1 , r_2 \vdash \mu_2 $, presented as  $ g ( r_1 , r_2 ; R )^2 $ linear combinations of multi-traces, where $ g ( r_1, r_2 ; R)$ is the Littlewood-Richardson coefficient for the triple. 
\item A covariant basis of $\cA ( \mu_1 , \mu_2 )$, labelled by Young diagrams $ R \vdash L $ and $ \Lambda \vdash L$, with $ \ell ( \Lambda ) \le 2$, presented as $ C  ( R , R , \Lambda ) K_{\Lambda \mu}$ with $\mu = (\mu_1,\mu_2)$ linear combinations of multi-traces, where   $ C  ( R , R , \Lambda ) $ is the Kronecker coefficient (or Clebsch-Gordan multiplicity)  for the triple of Young diagrams. 
\end{itemize} 
For $ N > L$, these span the space of 2-matrix invariants and diagonalise the free field inner product. For $ N < L$, the basis of 2-matrix invariants is obtained by keeping only basis states of $  \cA ( \mu_1 , \mu_2 ) $ with $ \ell ( R ) \le N$. The basis states of $ \cA ( \mu_1  , \mu_2 ) $ labelled by Young diagrams with $ \ell (R) >  N $ span the vector space of multi-traces with $( \mu_1  , \mu_2) $ copies of the two matrices which vanish by finite $N$ trace relations (which are also known as Mandelstam identities in the physics literature \cite{Mandelstam:1978ed,Giles:1981ej,Berenstein:1993gb}).

The finite $N$ relations for the 2-matrix invariants are also encoded in 2-variable partition functions, or Hilbert series, for this problem. The form of these partition functions rapidly increase in complexity as a function of $N$, and are explicitly available for $ N $ up to $7$ in the mathematics and physics literature 
\cite{PROCESI1976306,drensky2012polynomial,djokovic2006poincare,Kristensson:2020nly}. These partition functions are of interest in the thermodyamics of the 2-matrix system which captures the deconfinement property of $ \cN =4$ SYM 
with $U(N) $ gauge group and is important in the AdS/CFT correspondence. Discussion of the thermodynamic implications of finite $N$ relations in the context of these partition functions are available in \cite{Kristensson:2020nly,Berenstein:2018hpl,OConnor:2024udv}. The description of the basis of null states given by the code for a fixed value of $L$, contains information related to the finite $N$ relations  for all $ N < L$. For $L=14$, this includes $N$ up to 13. This illustrates that the algebraic method presented here is a substantial explicit information 
about finite $N$ relations in  the 2-matrix system.

\subsection{Regular representation of $\cA(\mu_1,\mu_2)$}\label{sec:regular rep Amn}

To decompose $\cA(\mu_1,\mu_2)$ we will study its regular representation and a set of relevant eigenvalue systems.
The regular representation of $\cA(\mu_1,\mu_2)$ is defined as follows.

Let $g_1, \dots, g_l \in S_L$ be a complete set of representatives, which forms an orbit basis of $\cA(\mu_1,\mu_2)$, defined in \eqref{eq: Amu representatives}.
We apply $\sigma \in \cA(\mu_1,\mu_2)$ to $P_{\mu_1,\mu_2} (g_i)$ from left, and expand it in the orbit basis as
\begin{equation}
   m^{\acting{L}}[\sigma](P_{\mu_1,\mu_2}(g_i)) = \sigma P_{\mu_1,\mu_2}(g_i ) \equiv \sum_{j=1}^l m^{\acting{L}}_{ji}[\sigma] P_{\mu_1,\mu_2}(g_j) .
\label{def:m acting on Vmnreg}
\end{equation}
The matrix elements $m^{\acting{L}}_{ji}[\sigma]$ define the (left) regular representation of $\cA(\mu_1,\mu_2)$. Borrowing the notation in Section \ref{sec:notation CSL}, we define the adjoint action of $\sigma = \sum_{h} c_h(\sigma) \, h$ on the orbit basis as
\begin{equation}
    m^{\acting{ad}}[\sigma ](P_{\mu_1,\mu_2}(g_i)) = \sum_{h \in S_L} c_h(\sigma) \, h \, P_{\mu_1,\mu_2}(g_i) \, h^{-1} 
\equiv \sum_j m^{\acting{ad}}_{ji} [\sigma] P_{\mu_1,\mu_2}(g_j)
 \label{def:mAdj on Vmnreg}
\end{equation}
We can extract the matrix elements from \eqref{def:m acting on Vmnreg} or \eqref{def:mAdj on Vmnreg} using the $\delta$-function as in \eqref{extract mji from delta}, keeping in mind that we should sum over the conjugacy class of $S_\mu$ when working with $\cA(\mu_1,\mu_2)$.

When $\sigma$ is an integer linear combination of the basis elements $P_{\mu_1,\mu_2}(g_i)$, the corresponding representation matrices $m^{\acting{L}}_{ji} [\sigma], m^{\acting{ad}}_{ji} (\sigma)$ are all integer matrices. 
This follows from the fact that the structure constants $C_{ijk}$ are integers
\begin{equation}
P_{\mu_1,\mu_2}(g_i ) \, P_{\mu_1,\mu_2}(g_j) 
= \sum_{k=1}^l  C_{ijk} \, P_{\mu_1,\mu_2}(g_k) 
\label{def:Cijk PPP}
\end{equation}
To show that $C_{ijk}$ are integers, we rewrite $P_\mu (g_i )$ as
\begin{equation}
P_\mu(g_i ) = \frac{1}{|\mathrm{Stab} (g_i)|} \sum_{\gamma \in S_\mu} \gamma g_i \gamma^{-1}
= \sum_{a=1}^{\abs{ {\rm Orb} (g_i)}} g_{i,a} 
\end{equation}
with $g_i = g_{i,1}$\,.
By using the multiplication rule of $S_L$\,, we find
\begin{equation}
P_\mu(g_i ) \, P_\mu(g_j) 
= \sum_{a=1}^{\abs{ {\rm Orb} (g_i)}} \sum_{b=1}^{\abs{ {\rm Orb} (g_j)}} g_{i,a} \, g_{j,b}
= \sum_{a,b} \sum_{k=1}^l \sum_{c=1}^{\abs{ {\rm Orb} (g_k)}}  \delta \Big( g_{i,a} \, g_{j,b} \, g_{k,c}^{-1} \Big) \, g_{k,c} \,.
\label{comp PgI-PgJ}
\end{equation}
The sum over the $\delta$-functions can be written as
\begin{equation}
\begin{aligned}
\sum_{a,b} \delta \Big( g_{i,a} \, g_{j,b} \, g_{k,c}^{-1} \Big)
&= \frac{1}{|\mathrm{Stab} (g_i)| \, |\mathrm{Stab} (g_j)|} \sum_{\gamma_i , \gamma_j \in S_\mu} 
\delta \Big( \gamma_i \, g_i \, \gamma_i^{-1} \, \gamma_j \, g_j \, \gamma_j^{-1} \, g_{k,c}^{-1} \Big)
\\
&= \frac{1}{|\mathrm{Stab} (g_i)| \, |\mathrm{Stab} (g_j)|} \sum_{\gamma_i , \eta \, \in S_\mu} 
\delta \Big(  g_i \, \eta \, g_j \, \eta^{-1} \, \gamma_i^{-1} g_{k,c}^{-1} \gamma_i \Big)
\\
&= \frac{|\mathrm{Stab} (g_k)|}{|\mathrm{Stab} (g_i)| \, |\mathrm{Stab} (g_j)|} \sum_{\eta \, \in S_\mu} \sum_{c} 
\delta \Big(  g_i \, \eta \, g_j \, \eta^{-1} \, g_{k,c}^{-1} \Big)
\end{aligned}
\end{equation}
with $\eta = \gamma_i^{-1} \, \gamma_j$\,. This equation shows that the quantity
\begin{equation}
C_{ijk} \equiv \sum_{a,b} \delta \Big( g_{i,a} \, g_{j,b} \, g_{k,c}^{-1} \Big) 
\label{comp:Cijk}
\end{equation}
is independent of $c$. The equation \eqref{comp PgI-PgJ} simplifies as
\begin{equation}
P_\mu(g_i ) \, P_\mu(g_j) 
= \sum_{k=1}^l  C_{ijk} \, \sum_{c=1}^{\abs{ {\rm Orb} (g_k)}} \,g_{k,c} 
= \sum_{k=1}^l  C_{ijk} \, P_{m,n}(g_k) 
\label{comp PgI-PgJ-PgK}
\end{equation}
which is \eqref{def:Cijk PPP}. From \eqref{comp:Cijk} we find that $C_{ijk}$ is an integer.

\subsection{Generating sets of central elements and integer eigenvalues}\label{sec:EV of centres}

Having defined the representation matrices corresponding to left and adjoint actions, we will now formulate the eigenspaces $\cA^{R, r_1, r_2}(\mu_1,\mu_2), \cA^{R, \Lambda}(\mu_1,\mu_2)$ defined in \eqref{eq: R r1 r2}, \eqref{def:cA RLam mu} as the intersections of kernels of explicit integer matrices acting on $\cA(\mu_1,\mu_2)$.

Define the following elements in $\cA(\mu_1,\mu_2)$
\begin{align}
    &T^{(L)}_2 = \sum_{1 \leq i < j \leq L} (ij), \quad 
    &&T^{(\mu_1)}_2 = \sum_{1 \leq a < b \leq \mu_1} (ab), \quad 
    &&T^{({\mu_2})}_2 = \sum_{\mu_1+1 \leq p < q \leq L} (pq)
\label{eq: casimirs} \\
    &T^{(L)}_3 = \sum_{1 \leq i < j < k \leq L} (ijk) + (ikj), \quad 
    &&T^{(\mu_1)}_3 = \sum_{1 \leq a < b < c \leq \mu_1} (abc) + (acb), \quad 
    &&T^{({\mu_2})}_3 = \sum_{\mu_1+1 \leq p < q < r \leq L} (pqr) + (prq) .
\notag
\end{align}
We call $T_p^{(k)}$ Casimir operators, because they are the central elements we encountered in \eqref{eq: L-centre on res schur} and \eqref{eq: m-centre on res schur},
\begin{equation}
        T^{(L)}_p \in \mathcal{Z}(\mathbb{C}[S_L]), \qquad  
        T^{(\mu_1)}_p \in \mathcal{Z}(\mathbb{C}[S_{\mu_1}]), \qquad  
        T^{({\mu_2})}_p \in \mathcal{Z}(\mathbb{C}[S_{\mu_2}]) .
\end{equation}
As discussed in Section \ref{sec:centre}, only the generating set of Casimir operators is needed to distinguish different representations.
Furthermore, the elements in \eqref{eq: casimirs} are integer combinations of the basis elements $P_{\mu_1,\mu_2}(g_i)$ and therefore the corresponding $m^{\acting{L}}_{ji}, m^{\acting{ad}}_{ji}$ are integer matrices.

First, consider the case of $\cA^{R,r_1,r_2}(\mu_1,\mu_2)$ related to the restricted Schur basis.
We will use the following shorthand notation for these eigenvalues,
\begin{equation}
    \widehat{\chi}^{R}_{p} = \frac{\chi^R(T_p^{(L)})}{d_R}, \quad 
    \widehat{\chi}^{r_1}_{p} = \frac{\chi^{r_1}(T_p^{(\mu_1)})}{d_{r_1}},\quad
    \widehat{\chi}^{r_2}_{p} = \frac{\chi^{r_2}(T_p^{({\mu_2})})}{d_{r_2}},
    \qquad p=2,3
\end{equation}
which are normalised characters.
The normalised characters of symmetric groups are known to have integer values \cite{simon1996representations}.
Now we have a collection of integer matrices with integer eigenvalues acting on (the regular representation of) $\cA(\mu_1,\mu_2)$.
According to the discussion in Section \ref{sec:centre}, 
these matrices and corresponding eigenvalues are sufficient, for $\mu_1+{\mu_2} \leq 14$, to distinguish all the subspaces $\cA^{R,r_1,r_2} (\mu_1,{\mu_2})$ in the sense that
\begin{equation}
    \begin{aligned}
        \cA^{R,r_1,r_2}(\mu_1,{\mu_2}) \cong \,
        &\mathrm{Ker}\qty( m^{\acting{L}}[T^{(L)}_{2}]-\widehat{\chi}^{R}_{2} )\cap
        \mathrm{Ker}\qty(m^{\acting{L}} [T^{(L)}_{3}]-\widehat{\chi}^{R}_{3})\cap 
        \mathrm{Ker}\qty(m^{\acting{L}} [T^{(\mu_1)}_{2}]-\widehat{\chi}^{r_1}_{2})\cap \\
        &\mathrm{Ker}\qty(m^{\acting{L}} [T^{(\mu_1)}_{3}]-\widehat{\chi}^{r_1}_{3})\cap 
        \mathrm{Ker}\qty(m^{\acting{L}}[T^{({\mu_2})}_{2}]-\widehat{\chi}^{r_2}_{2})\cap 
        \mathrm{Ker}\qty(m^{\acting{L}} [T^{({\mu_2})}_{3}]-\widehat{\chi}^{r_2}_{3})
    \end{aligned} \label{eq: A R r1 r2 kernel}
\end{equation}
where $\mathrm{Ker}(M)$ for a linear operator $M$ is the subspace of vectors $x$ satisfying $Mx=0$.

For completeness, let us now describe the situation for $\cA^{R,\Lambda}$ related to the covariant basis, which is analogous.
From \eqref{eq: adjoint on kronecker} we know that the matrices $m^{\acting{ad}}_{ij} \, [ T_2^{(L)} ], m^{\acting{ad}}_{ij} \, [T_3^{(L)} ]$ have eigenvalues given by normalised characters 
\begin{equation}
    \widehat{\chi}^{\Lambda}_{p} = \frac{\chi^{\Lambda}(T_p^{(L)})}{d_\Lambda}, \quad p=2,3.
\end{equation}
As before, $\Lambda \vdash L \leq 14$ is uniquely determined by these eigenvalues and it follows that
\begin{equation}
    \begin{aligned}
        \cA^{R,\Lambda}(\mu_1,{\mu_2}) \cong \, 
        &\mathrm{Ker}\qty(m^{\acting{L}} [T^{(L)}_{2}]-\widehat{\chi}^{R}_{2})\cap 
        \mathrm{Ker}\qty(m^{\acting{L}} [T^{(L)}_{3}]-\widehat{\chi}^{R}_{3})\cap \\
        &\mathrm{Ker}\qty(m^{\acting{ad}} [T_2^{(L)}] -\widehat{\chi}^{\Lambda}_{2}) \cap
        \mathrm{Ker}\qty(m^{\acting{ad}} [T_3^{(L)}] -\widehat{\chi}^{\Lambda}_{3}) .
    \end{aligned} \label{eq: A R Lambda kernel}
\end{equation}

\subsection{Hermite normal forms and kernels of integer matrices}\label{sec:integrality}
Computing the intersection of kernels of square matrices corresponds to computing the kernel of a single rectangular matrix.
For example, given a pair of $l \times l$ matrices $M_1, M_2$ we find the intersection $\mathrm{Ker}(M_1) \cap \mathrm{Ker}(M_2)$ by finding the kernel of the matrix
\begin{equation}
    M = \left[
        \begin{aligned}
        &M_1 \\ &M_2
        \end{aligned}
        \right].
\end{equation}
This follows since $Mx = 0$ if and only if $M_1x = M_2 x = 0$ and this generalises to the intersection of kernels of several matrices.
Therefore, the space \eqref{eq: A R r1 r2 kernel} corresponds to solutions to
\begin{equation}
    \begin{bmatrix}
        m^{\acting{L}} [T^{(L)}_{2}]-\widehat{\chi}^{R}_{2}     \\
        m^{\acting{L}} [T^{(L)}_{3}]-\widehat{\chi}^{R}_{3}      \\
        m^{\acting{L}} [T^{(\mu_1)}_{2}]-\widehat{\chi}^{r_1}_{2}   \\
        m^{\acting{L}} [T^{(\mu_1)}_{2}]-\widehat{\chi}^{r_1}_{2}    \\
        m^{\acting{L}} [T^{({\mu_2})}_{2}]-\widehat{\chi}^{r_2}_{2}    \\
        m^{\acting{L}} [T^{({\mu_2})}_{3}]-\widehat{\chi}^{r_2}_{3}   
    \end{bmatrix}x = 0
\label{eq:matrix kernels1}
\end{equation}
We will now give a procedure for solving this equation over the integers.
In other words, for fixed $(R, r_1, r_2)$ we will give a procedure for constructing an integer basis for the space satisfying the simultaneous equations. 
We will see that this can be turned into an integer orthogonal basis of two-matrix invariants related to the restricted Schur basis.
The method works identically for the system of equations
\begin{equation}
    \begin{bmatrix}
        m^{\acting{L}} [T^{(L)}_{2}]-\widehat{\chi}^{R}_{2}\\
        m^{\acting{L}} [T^{(L)}_{3}]-\widehat{\chi}^{R}_{3}\\
        m^{\acting{ad}} [T^{(L)}_{2}]-\widehat{\chi}^{\Lambda}_{2}\\
        m^{\acting{ad}} [T^{(L)}_{3}]-\widehat{\chi}^{\Lambda}_{3}
    \end{bmatrix}x = 0
\label{eq:matrix kernels2}
\end{equation}
corresponding to \eqref{eq: A R Lambda kernel}, which gives an integer orthogonal basis related to the covariant basis for two-matrix invariants.

Computing an integer basis for $\mathrm{Ker}(M)$ uses Hermite normal forms, which are the analog of the more familiar echelon forms, adapted to integer matrix problems, as we will now describe.
We now make some general remarks about Hermite normal forms.
The fundamental result on Hermite normal forms is the following (see \cite[Theorem 2.4.3]{Cohen}).
Let $M$ be an $m \times n$ integer matrix, then there exists a unimodular $n \times n$ matrix $U$ -- an integer matrix with $\det(U)=\pm1$ -- such that
\begin{equation}
    MU= H = \mqty[0 \; h],\label{eq: HMN decomp}
\end{equation}
where $h$ is an invertible, upper triangular integer matrix and $0$ represents are rectangular matrix of zeros. By upper triangular we mean $h_{ij}=0$ for $i>j$.
Further conditions are put on $h$ to make it unique, see \cite[Definition 2.4.2]{Cohen}.
The matrix $H$ is called the Hermite normal form of $M$.
Note that, despite $H$ being unique, $U$ is not unique in general.
In particular, $H$ is invariant under permutations of the zero columns; multiplication of any zero column by $\pm 1$; addition of any scalar multiple of a zero column to any other column. 
These operations can be implemented by right multiplication with unimodular matrices $V$ and we get
\begin{equation}
    H = HV = MUV \equiv MU'
\end{equation}
where $U' = UV$ is another unimodular matrix giving an Hermite decomposition of $M = H (U')^{-1} $.
A Hermite decomposition gives a basis for the kernel of an integer matrix $M$ as follows \cite[Proposition 2.4.9]{Cohen}.
Let $H = MU$ be a Hermite decomposition such that the first $r$ columns of $H$ are equal to zero
\begin{equation}
    [\underbrace{0}_{r} \; h] = MU.
\end{equation}
Then the first $r$ columns of $U$ form a basis for the kernel of $M$.
In other words, define the vectors $v_i^{(s)} = U_{si}$, then
\begin{equation}
    \mathrm{Ker}(M) = \mathrm{Span}( v^{(1)}, v^{(2)}, \dots, v^{(r)}).
\label{sol kernel vectors}
\end{equation}
The number of independent solutions $r$ is precisely equal to the dimensions of the subalgebra, namely
\begin{equation}
\dim \cA^{R,r_1,r_2} (\mu_1,{\mu_2}) = g(r_1,r_2; R)^2
\end{equation}
for the restricted Schur basis and
\begin{equation}
\dim \cA^{R,\Lambda}(\mu_1,{\mu_2}) = C(R,R, \Lambda) K_{\Lambda (\mu_1,\mu_2)} 
\end{equation}
for the covariant basis.
In Appendix \ref{app:hermite algorithm} we will explain an algorithm to compute an Hermite normal form.

\subsection{Simple example}
\label{sec:simple example}
As a simple example to illustrate the procedure, we will consider the Schur case $\mathcal{A}(3,0) = \mathcal{Z}(\mathbb{C}[S_3])$ in detail.
We have a set of representatives given by the identity permutation $g_1 = (1)(2)(3)$ and $g_2=(12)(3), g_3=(123)$ with orbit basis
\begin{equation}
\begin{aligned}
    &P_{3,0}(g_1) = (1)(2)(3), \\
    &P_{3,0}(g_2) = (12)(3)+(13)(2)+(23)(1), \\
    &P_{3,0}(g_3) = (123)+(132).
\end{aligned}
\end{equation}
We will only need the left action of $T_2^{(3)}=P_{3,0}(g_2)$ in this basis, it is straightforward to compute
\begin{align}
    &T_2^{(3)} P_{3,0}(g_1) = P_{3,0}(g_2), \\
    &T_2^{(3)} P_{3,0}(g_2) = 3P_{3,0}(g_1)+3P_{3,0}(g_3), \\
    &T_2^{(3)} P_{3,0}(g_3) = 2P_{3,0}(g_2),
\end{align}
and equivalently, the left action matrix in this basis is
\begin{equation}
    m^{\mathfrak{L}}[T_2^{(3)}] = 
    \begin{pmatrix}
    0 & 3 & 0 \\
    1 & 0 & 2 \\
    0 & 3 & 0
    \end{pmatrix}
\end{equation}

We will only consider the eigenspace corresponding to $R=[3]$.
This has normalised eigenvalue $\widehat{\chi}^{R}(T_2^{(3)}) = 3$ and so we are looking for the kernel of the matrix
\begin{equation}
    m^{\mathfrak{L}}[T_2^{(3)}] - 3 I_3= 
    \begin{pmatrix}
    -3 & 3 & 0 \\
    1 & -3 & 2 \\
    0 & 3 & -3
    \end{pmatrix}
\end{equation}
It is easy to confirm that the vector $(1,1,1)$ lies in the kernel and forms an integer basis (this is derived in appendix \ref{apx:example kernel} using an integer matrix algorithm).
In the orbit basis, this corresponds to
\begin{equation}
    P_{3,0}(g_1) + 
    P_{3,0}(g_2) + 
    P_{3,0}(g_3) = \sum_{g \in S_3}  g
\end{equation}
or in terms of traces
\begin{equation}
    \Tr(Z)^3 + 3 \Tr(Z)\Tr(Z^2) + 2\Tr(Z^3).
\end{equation}

\subsection{Orthogonality from planar to finite $N$}\label{sec:orth finite N}

We will show that the finite $N$ two-point functions in $\mathcal{N}=4$ SYM at zero coupling are proportional to the planar two-point functions in any operator bases which are labeled by the irreducible representation $R$ under the left/right action of the centre of $\bb{C}[S_L]$.

The finite $N$ free-field two-point functions in the permutation basis is given by 
\begin{equation}
\label{bastwpt}  
    \langle \cO_{ g_1 }[Z,W] \cO_{g_2 }[Z,W]^\dagger \rangle 
    = \langle \cO_{g_1}[Z,W] \cO_{ g_2^{-1}  }[Z^{\dagger}, W^{ \dagger} ] \rangle  
    = \sum_{ \gamma \in S_{\mu_1} \times S_{\mu_2} } \sum_{ g_3 \in S_{ L } } \delta ( g_1 \gamma g_2^{-1} \gamma^{-1} g_3 ) N^{ \cycn{g_3} },
\end{equation}
where the operators have $\mu_1$ $Z$'s and $\mu_2$ $W$'s, and $\cycn{g}$ is the number of cycles in $g \in S_L$\,.
The two-point functions vanish if $\cO_{ g_1 }[Z,W]$ and $\cO_{g_2 }[Z,W]$ have different $(\mu_1,\mu_2)$'s.
This defines an inner product on $\cA(\mu_1,\mu_2)$.
Let $\sigma_1, \sigma_2 \in \cA(\mu_1,\mu_2)$, then
\begin{equation}
    (\sigma_1, \sigma_2)_{\text{finite\,$N$}} 
    = \sum_{\gamma \in S_{\mu_1} \times S_{\mu_2}} \sum_{g \in S_L} \delta(\sigma_1 \gamma \invs{\sigma_2} \gamma^{-1} g) N^{\cycn{g}} 
    = \mu_1! n! \sum_{g \in S_L} \delta(\sigma_1 \invs{\sigma_2}  g) N^{\cycn{g}},
\end{equation}
where $\invs{\sigma_2}$ is defined in \eqref{def:inverse of sigma}.
Clearly, this inner product is symmetric, $(\sigma_1, \sigma_2)_{\text{finite\,$N$}} = (\sigma_2, \sigma_1)_{\text{finite\,$N$}}$.

The planar inner product is defined by the large $N$ limit
\begin{equation}
    (\sigma_1, \sigma_2)_{\text{planar}} = \lim_{N \rightarrow \infty} \frac{1}{N^L} (\sigma_1, \sigma_2)_{\text{finite\,$N$}} = {\mu_1}! {\mu_2}! \, \delta(\sigma_1 \invs{\sigma_2}).
\label{def:planar inner product}
\end{equation}
The planar inner product is essentially the same as the $\delta$-function inner product computed in \eqref{cov basis norm}.
Conversely, the finite $N$ inner product can be obtained by deforming the planar inner product
\begin{equation}
    (\sigma_1, \sigma_2)_{\text{finite\,$N$}}  = (\Omega \, \sigma_1, \sigma_2)_{\text{planar}} \,.
\label{deform inner Omega}
\end{equation}
where we defined $\Omega \in \mathbb{C}[S_L]$ by
\begin{equation}
    \Omega = \sum_{g \in S_L} N^{\cycn{g}} g = L! \sum_{R \, \vdash L} \frac{\Dim_N (R) }{d_R }  P^R .
    \label{eq: omega}
\end{equation}
Here $P^R$ is the projector to the representation $R$ in \eqref{app:projector PR}, and $\Dim_{N} (R)$ is the $U(N)$ dimension of the representation $R$ as in \eqref{def:DimNR wtNR}. See \cite{Corley:2001zk} for the derivation of the second equality.
As we will now see, orthogonalisation with respect to the planar limit is very closely related to orthogonalisation in the finite $N$ case.

The left and adjoint actions of $T_p \in \cZ (\bb{C}[S_L])$ used in \eqref{eq:matrix kernels1}, \eqref{eq:matrix kernels2} are self-adjoint with respect to the planar as well as finite $N$ inner products. For the left action we have
\begin{equation}
    (T_p \, \sigma_1, \sigma_2)_{\text{planar}} =
     {\mu_1}! {\mu_2}! \, \delta(T_p \,\sigma_1 \invs{\sigma_2}) 
    = {\mu_1}! {\mu_2}! \, \delta( \sigma_1 \invs{T_p \sigma_2} ) 
    = (\sigma_1, T_p \, \sigma_2)_{\text{planar}},
    \label{eq: self-adjoint T2}
\end{equation}
where we have used $T_p = \invs{T_p}$ because every conjugacy class of $S_L$ contains its inverse elements and $\invs{\sigma \tau} = \invs{\tau} \invs{\sigma}$ for $\sigma, \tau \in \mathbb{C}[S_L]$.
For the adjoint action, define $C_p$ to be the conjugacy class $C_\rho$ with $\rho = (p, 1^{L-p}) \vdash L$, then we have
\begin{equation}
\begin{aligned}
    (m^{\acting{ad}}[{T_p}](\sigma_1), \sigma_2)_{\text{planar}} &=
    \sum_{g \in C_p} (g \sigma_1 g^{-1}, \sigma_2)_{\text{planar}} =
     {\mu_1}! {\mu_2}! \, \sum_{g \in C_p}  \delta( g\,\sigma_1 g^{-1} \, \invs{\sigma_2}) \\
     &={\mu_1}! {\mu_2}! \, \sum_{g \in C_p}  \delta( \sigma_1 \, \invs{g^{-1} \sigma_2 g})
     ={\mu_1}! {\mu_2}! \, \sum_{g \in C_p}  \delta( \sigma_1 \, \invs{g \sigma_2 g^{-1}}) \\
     &=( \sigma_1, m^{\acting{ad}}[{T_p}](\sigma_2))_{\text{planar}}.
\end{aligned}
    \label{eq: self-adjoint T2 2}
\end{equation}
In the penultimate step we used the fact that if $g \in C_p$ then so is $g^{-1}$ and therefore we can relabel the sum under the transformation $g \mapsto g^{-1}$.
The proofs are similar for the finite $N$ case.

\subsubsection{Restricted Schur basis}

Let the eigenvectors of the first equation in \eqref{eq:matrix kernels1} be $\sfQ^{R , r_1 r_2}_{A}$, where $A$ corresponds to the multiplicity in a fixed eigenspace of dimension $g(r_1,r_2; R)^2$. Explicitly, we write them as linear combinations of the restricted Schur basis
\begin{equation}
    \sfQ^{R , r_1 r_2}_{A} = \sum_{\nu_+ , \nu_- =1}^{g(R;r_1,r_2)}  \sfN^{R , r_1, r_2}_{A, \nu_+ \nu_-} \, Q^{R , r_1, r_2}_{\nu_+ \nu_-} 
\end{equation}
where $\sfN^{R , r_1, r_2}_{A, \nu_+ \nu_-}$ is a numerical coefficient.
With the help of \eqref{res Schur norm}, the planar inner product is given by
\begin{equation}
(\sfQ^{R, r_1,r_2}_A, \sfQ^{S, s_1, s_2}_B)_{\text{planar}} 
= \delta^{RS} \delta^{r_1 s_1} \delta^{r_2 s_2} \, d_{r_1} d_{r_2} \, g^{R,r_1,r_2}_{AB} \,, \qquad
g^{R,r_1,r_2}_{AB} \equiv \sum_{\nu_+ , \nu_- =1}^{g(R;r_1,r_2)}
\sfN^{R , r_1, r_2}_{A, \nu_+ \nu_-} \, \sfN^{R , r_1, r_2}_{B, \nu_+ \nu_-} 
\label{inner planar resSchur}
\end{equation}
where $g^{R,r_1,r_2}_{AB}$ is a Gram (real symmetric) matrix of dimensions $g(r_1, r_2; R)^2$.
Note that the inner product must vanish if $(R, r_1, r_2) \neq (S, s_1, s_2)$, because they label the eigenvectors of the self-adjoint operators $(T_p^{(L)} \,, T_p^{({\mu_1})} \,, T_p^{({\mu_2})})$.

To compute the finite $N$ inner product, consider the deformation by $\Omega$ in \eqref{deform inner Omega}.
Because $\Omega$ is central, using \eqref{eq: L-centre on res schur} we get
\begin{equation}
    \Omega \, \sfQ^{R , r_1, r_2}_{A} = \frac{\chi^{R}(\Omega)}{d_R} \, \sfQ^{R , r_1, r_2 }_{A} \,.
\end{equation}
The normalised character is readily computed using equation \eqref{eq: omega}
\begin{equation}
    \frac{\chi^R(\Omega)}{d_R} 
    = \frac{L!}{d_R} \sum_{S \vdash L} \frac{\Dim_N (S) }{d_S}  \chi^R(P^S) 
    = \frac{L!}{d_R}  \sum_{S \vdash L} \frac{\Dim_N (S) }{d_S} d_S \, \delta^{RS} 
    = L! \, \frac{\Dim_N (R)}{d_R} .
\end{equation}
Consequently,
\begin{equation}
    (\sfQ^{R, r_1,r_2}_A, \sfQ^{S, s_1, s_2}_B)_{\text{finite\,$N$}}    
    = (\Omega \, \sfQ^{R, r_1,r_2}_A, \sfQ^{S, s_1, s_2}_B)_{\text{planar}} 
    = L! \, \frac{\Dim_N (R)}{d_R} \, (\sfQ^{R, r_1,r_2}_A, \sfQ^{S, s_1, s_2}_B)_{\text{planar}} 
\label{inner fN resSchur}
\end{equation}
which shows that the finite $N$ inner product is proportional to the planar inner product in the restricted Schur basis.
Note that all the $N$-dependence is captured by the factor $\Dim_N (R)$.

\subsubsection{Covariant basis}
Analogous statements hold for the eigenvectors of \eqref{eq:matrix kernels2}, which we denote by $\sfQ^{R, \Lambda, \mu}_{A}$ where \, \, \, \, \, \, \,\, $A \in \{1,\dots, C(R,R,\Lambda) K_{\Lambda \mu}\}$.
Explicitly, they are linear combinations of covariant bases elements
\begin{equation}
    \sfQ^{R, \Lambda, \mu}_{A} = \sum_{\tau=1}^{C(R,R,\Lambda)} \sum_{\beta=1}^{K_{\Lambda \mu}}
    \sfN^{R, \Lambda, \tau, \mu, \beta}_{A} \, \cQ^{R, \Lambda, \mu, \tau}_{\beta} \,.
\end{equation}
By using \eqref{cov basis norm} we find that the planar inner product is given by
\begin{equation}
(\sfQ^{R, \Lambda, \mu}_A, \sfQ^{S, \Lambda',\mu}_B)_{\text{planar}} 
= \frac{d_R}{|S_L|} \, \delta^{RS} \delta^{\Lambda \Lambda'} \, G^{R,\Lambda,\mu}_{AB}, \qquad
G^{R,\Lambda,\mu}_{AB} \equiv \sum_{\tau=1}^{C(R,R,\Lambda)} \sum_{\beta=1}^{K_{\Lambda \mu}} 
\sfN^{R, \Lambda, \tau, \mu, \beta}_{A} \, \sfN^{R, \Lambda, \tau, \mu, \beta}_{B} 
\label{inner planar cov}
\end{equation}
where $G^{R,\Lambda,\mu}_{AB}$ is a Gram matrix of dimensions $C(R,R,\Lambda) \, K_{\Lambda \mu}$.
As before, the inner product vanishes if $(R, \Lambda) \neq (S, \Lambda')$.
From \eqref{eq: L-centre on kronecker}
\begin{equation}
    \Omega \, \sfQ^{R, \Lambda, \mu}_{A} = \frac{\chi^{R}(\Omega)}{d_R} \sfQ^{R, \Lambda}_{A}.
\end{equation}
This gives
\begin{equation}
    (\sfQ^{R,\Lambda, \mu}_A, \sfQ^{S, \Lambda',\mu}_B)_{\text{finite\,$N$}}  
    =  (\Omega \, \sfQ^{R,\Lambda, \mu}_A, \sfQ^{S, \Lambda',\mu}_B)_{\text{planar}}  
    =  L! \, \frac{\Dim_N (R)}{d_R} \, ( \sfQ^{R,\Lambda,\mu }_A, \sfQ^{S, \Lambda',\mu}_B)_{\text{planar}}    \,.
\label{inner fN cov}
\end{equation}

\subsubsection{Gram-Schmidt orthogonalisation}

The orbit basis is orthogonal with respect to the planar inner product \eqref{def:planar inner product} since
\begin{equation}
    \delta(P_{{\mu_1},{\mu_2}}(g_i) \, \invs{P_{{\mu_1},{\mu_2}}(g_j)}) = \sum_{\substack{ g \in \{g_i\}_{{\mu_1},{\mu_2}} \\ h \in \{g_j\}_{{\mu_1},{\mu_2}}} } \delta(g h^{-1}) = |\mathrm{Orb}(g_i)| \delta_{ij}.
\end{equation}
It follows that
\begin{equation}
    (P_{{\mu_1},{\mu_2}}(g_i), P_{{\mu_1},{\mu_2}}(g_j))_{\text{planar}} = {\mu_1}! {\mu_2}! \, |\mathrm{Orb}(g_i)| \, \delta_{ij} \,,
\end{equation}
which is integer valued.

As we have seen, the integer eigenvectors $\sfQ^{R,r_1,r_2}_A$ or $\sfQ^{R,\Lambda,\mu}_A$ are not automatically orthogonal.
Integer orthogonal eigenvectors are readily constructed using Gram-Schmidt orthogonalisation.
We simply define
\begin{equation}
\begin{aligned}
	\widehat \sfQ^{R,r_1,r_2}_1 &= \sfQ^{R,r_1,r_2}_1,\\
	\widehat \sfQ^{R,r_1,r_2}_A &= \sfQ^{R,r_1,r_2}_A- \sum_{B=1}^{A-1} \frac{(\sfQ^{R,r_1,r_2}_A, \widehat \sfQ^{R,r_1,r_2}_B)_{\text{planar}}}{(\widehat \sfQ^{R,r_1,r_2}_B,\widehat \sfQ^{R,r_1,r_2}_B)_{\text{planar}}} \, \widehat\sfQ^{R,r_1,r_2}_B \, , \quad A\in \{2,\dots,g(r_1,r_2;R)^2\}
\end{aligned}
\label{GS process}
\end{equation}
and it follows that
\begin{equation}
	(\widehat \sfQ^{R,r_1,r_2}_A, \widehat \sfQ^{R,r_1,r_2}_B)_{\text{planar}} \propto \delta_{AB} \quad \Rightarrow \quad  (\widehat \sfQ^{R,r_1,r_2}_A, \widehat \sfQ^{R,r_1,r_2}_B)_{\text{finite N}} \propto L! \, \frac{\Dim_N (R)}{d_R} \, \delta_{AB}.
\end{equation}
Similarly for the eigenvectors $\sfQ^{R,\Lambda,\mu}_A$.
Note that, even if $\sfQ^{R,r_1,r_2}_A$, $\sfQ^{R,\Lambda,\mu}_A$ are integer vectors, the orthogonalised vectors $\widehat \sfQ^{R,r_1,r_2}_A$, $\widehat \sfQ^{R,\Lambda,\mu}_A$ are in general rational.
This is easily remedied by multiplying each basis element by an appropriate factor.

\subsubsection{General basis}

We have discovered that the finite $N$ inner product is proportional to the planar inner product in a basis diagonalising the left action of the centre of $\mathbb{C}[S_L]$. 
The planar inner products are proportional to the Gram matrix, $g^{R,r_1,r_2}_{AB}$ in \eqref{inner planar resSchur} and $G^{R,\Lambda,\mu}_{AB}$ in \eqref{inner planar cov}.
If we orthogonalise the Gram matrices, for example using Gram-Schmidt \eqref{GS process}, we obtain orthogonal finite $N$ bases.

In both \eqref{inner fN resSchur} and \eqref{inner fN cov}, the proportionality constant only depends on the label $R \vdash L$.
Let $\sfQ^R_\alpha \in \cA(\mu_1,\mu_2)$ be solutions to the eigenvalue equations only for the left action of $T_p^{(L)}$. Written explicitly, they are
\begin{equation}
\sfQ^R_\alpha 
= \sum_{\substack{r_1 \vdash {\mu_1} \\ r_2 \vdash {\mu_2}}} \sum_{A=1}^{g(R;r_1,r_2)^2} 
\sfM^{r_1 r_2}_{\alpha A} \, \sfQ^{R , r_1 r_2}_{A} 
= \sum_{\substack{\Lambda \vdash L \\[1mm] \ell(\Lambda) \le 2}} \sum_{A=1}^{C(R,R,\Lambda) K_{\Lambda \mu}} 
\sfM^{\Lambda}_{\alpha A} \, \sfQ^{R, \Lambda, \mu}_{A} 
\end{equation}
for some numerical constants $\sfM^{r_1 r_2}_{\alpha A}$ and $\sfM^{\Lambda}_{\alpha A}$\,.
If we define the Gram matrix $\mathcal{G}^R_{\alpha \beta}$ by
\begin{equation}
	(\sfQ^R_\alpha \,, \sfQ^S_\beta )_{\text{planar}} 
    = \delta^{RS} \mathcal{G}^{R}_{\alpha \beta} \, 
\end{equation}
from arguments similar to those above we find
\begin{equation}
   (\sfQ^R_\alpha \,, \sfQ^S_\beta )_{\text{finite\,$N$}} 
   =  L! \, \frac{\Dim_N (R)}{d_R} (\sfQ^R_\alpha \,, \sfQ^S_\beta )_{\text{planar}} \, .
\end{equation}
Now the dimensions of the Gram matrix $\mathcal{G}^R_{\alpha \beta}$ is
\begin{equation}
    \sum_{\substack{\Lambda \vdash L \\[1mm] \ell(\Lambda) \leq 2}} C(R,R, \Lambda) K_{\Lambda \mu} 
    = \sum_{\substack{r_1 \vdash {\mu_1} \\ r_2 \vdash {\mu_2}}} g(r_1,r_2; R)^2.
\end{equation}
Orthogonalising $\mathcal{G}_{\alpha \beta}^R$ gives a finite $N$ orthogonal basis, which is a linear combination of the restricted Schur or covariant basis.

\section{Eigenvalue method for $\mathbb{C}[S_L]$}\label{sec:EV CSL}

In section \ref{sec:EV Amn} we developed eigenvalue systems for $\cA(\mu)$ which give bases of multi-matrix invariants related to the restricted Schur and covariant bases.
Similarly, in this section we develop two eigensystems for $\cA(1,\dots,1) = \mathbb{C}[S_L]$.
The first one gives the Artin-Wedderburn decomposition, and the second gives the Kronecker decomposition of $\mathbb{C}[S_L]$.

\subsection{Decomposition of the regular representation}\label{sec:decomp SL reg}

We will construct and diagonalise a set of mutually commuting Hermitian operators in order to decompose the regular representation of $\mathbb{C}[S_L]$ denoted by $\cV_{\rm reg}$\,.
A set of commuting operators
\begin{equation}
\bigl\{ m^X [O_1], \dots, m^X [O_\Omega] \bigr\} \ \in \ {\rm End} \, (\cV_{\rm reg}), \qquad
X \in \{ \acting{L} , \acting{R} , \acting{ad} \}
\end{equation}
 is called complete if their eigenvalues specify a unique vector in $\cV_{\rm reg}$ up to normalisation.
That is, for
\begin{equation}    
    \sum_{j=1}^{L!} m^{X}_{ij} [O_w] ( \alpha_j ) = \lambda_w^{(\alpha)}  \, \alpha_i
\label{def:eigensystem Vreg}
\end{equation}
and any pair of eigenstates $\alpha, \beta \in \cV_{\rm reg}$
\begin{equation}
    \alpha \neq \beta
    \ \ \Leftrightarrow \ \ 
    \{ \lambda_1^{(\alpha)} \,, \dots \,, \lambda_\Omega^{(\alpha)} \} \neq 
    \{ \lambda_1^{(\beta)} \,, \dots \,, \lambda_\Omega^{(\beta)} \} .
\label{def:spec decomp cH}
\end{equation}

We are interested in finding a set of commuting operators $\{ m^X [O_1], \dots, m^X [O_\Omega] \}$ whose eigenstates correspond to the two decompositions of $\mathbb{C}[S_L]$ discussed in Section \ref{sec:decomp CSL}. These decompositions allow us to construct orthogonal bases of operators of $\cN=4$ SYM for the case of $\cA(1,\dots,1) = \bb{C}[S_L]$.
This is a good starting point for considering the general case of $\cA(\mu)$.

One natural choice of the commuting operators comes from $\mathbb{C}[S_L]$ itself.
The elements
\begin{equation}
\cJ_1 = 0, \qquad
\cJ_k = \sum_{i=1}^{k-1} (i,k), \qquad (k=2,3, \dots , L)
\label{def:YJM elements}
\end{equation}
are known as Young-Jucys-Murphy elements \cite{Jucys1974,Murphy1981y}
They generate a maximal commuting subalgebra\footnote{This subalgebra is called Gelfand-Tsetlin algebra in \cite{vershik2005new}.}, denoted $\cM(\bb{C}[S_L])$, of the full group algebra and $\mathcal{Z}(\mathbb{C}[S_L])$ is a subalgebra
\begin{equation}
\cZ (\bb{C}[S_L]) \ \subset \ \cM (\bb{C}[S_L]) \ \subset \ \bb{C}[S_L] .
\label{def:ZM subalgebras-1}
\end{equation}
They satisfy $[\cJ_k \,, \cJ_\ell]=0$ for any $k, \ell$ \cite{Jucys1974,Murphy1981y} and the following relation with $s_k = (k,k+1) \in S_L$
\begin{equation}
	s_k \, \cJ_{k+1} = \cJ_{k} \, s_k + 1.
\label{rel YJM-Cox}
\end{equation}
In the regular representation, every element of $\mathbb{C}[S_L]$ corresponds to an element of $\rm End(\cV_{\rm reg})$ and the YJM elements correspond to a complete set of commuting matrices.

The YJM elements are not central in general, but power symmetric polynomials 
\begin{equation}
	\sum_{k=1}^L \cJ_k, \quad \sum_{k=1}^{L} \cJ_k^2, \dots
\end{equation}
are related to central elements by \cite{Jucys1974,Lassalle2007,Corteel2004}
\begin{alignat}{9}
T_2 
&= \sum_{i<j}^{L} \ (i,j) & 
&= \sum_{k}^L \cJ_k
\label{SL Casimir T2} \\
T_3 
&= \sum_{i<j<k}^L \Big\{ (i,j,k) + (j,i,k) \Big\} & 
&= \sum_{k}^L \cJ_k^2 - \frac{L(L-1)}{2} \,, \qquad (L \ge 3).
\label{SL Casimir T3}
\end{alignat}

Because the YJM elements generate a maximal commuting subalgebra, the eigenvalues of the corresponding matrices uniquely specify a standard Young tableau. More precisely, the left or right eigenvalue of $\cJ_k$ is equal to the content of the box $k$ inside the Young tableau $I$ of shape $R$, 
\begin{equation}
\cJ_k \ \ket{\atop{R}{I} } 
= {\tt Cont}^R_I (k) \ket{\atop{R}{I} } 
\qquad \Longleftrightarrow \qquad
 D^R_{JI} ( \cJ_k ) = \delta_{JI} \, {\tt Cont}^R_I (k) 
\label{def:JM eigenval}
\end{equation}
where the content function ${\tt Cont}$ assigns $(x-y)$ for the box $k$ found at the $x$-th column and the $y$-th row,
\ytableausetup{boxsize=1.7em,centertableaux}
\begin{equation}
{\tt Cont} \( \ 
\begin{ytableau}
\phantom{.} & & & & \\
 & & & \\
 & &
\end{ytableau} \ \)
\ = \ 
\begin{ytableau}
0 & 1 & 2 & 3 & 4 \\
-1 & 0 & 1 & 2 \\
-2 & -1 & 0 
\end{ytableau} \ .
\label{def:content}
\end{equation}
One can also show that the adjoint actions of $\cJ_k$ have the same eigenvalues as in the left or right actions by using \eqref{rel YJM-Cox}.\footnote{See the argument in \cite{MOF-RT-2011}.}

The relations \eqref{SL Casimir T2} and \eqref{SL Casimir T3} make it clear that the eigenbasis {\small $\{ \ket{\atop{R}{I}} \}$} in \eqref{def:JM eigenval} diagonalises the central elements $T_p$\,, and their eigenvalues depend only on the Young diagram $R$.
Furthermore, it was shown that this eigenbasis coincides with the Young seminormal representation by computing the action of $s_k$ \cite{Murphy1981y,vershik2005new}.
To obtain the Young-Yamanouchi orthonormal representation, which is real and unitary, we need to rescale the basis elements of {\small $\{ \ket{\atop{R}{I}} \}$} in the Young seminormal representation.

\subsection{Eigenvalue system for the Artin-Wedderburn decomposition}

Recall that the Artin-Wedderburn decomposition of $\bb{C}[S_L]$ gives the matrix unit denoted by $Q^R_{IJ}$\,.
The matrix unit satisfies the relation
\begin{equation}
    Q^{R}_{IJ} \, Q^{S}_{KL} = \delta^{R S} \, \delta_{JK} \, Q^{R}_{IL} \,.
\label{alg for YSN basis}
\end{equation}
As shown in \eqref{left action on QRIJ} and \eqref{right inverse action on QRIJ}, the left and right actions of $\bb{C}[S_L]$ on the matrix unit can be written as the multiplication of the irreducible representation matrix. 
Let us identify the subscripts of $Q^R_{IJ}$ as a basis of Young seminormal representations.
If we apply the YJM elements from left and right, we obtain
\begin{equation}
\begin{aligned}
\cJ_k \, Q^R_{IJ} &= \sum_{K} D^R_{KI} (\cJ_k) \, Q^R_{KJ} = {\tt Cont}^R_I (k) \, Q^R_{IJ} 
\\
Q^R_{IJ} \, \cJ_\ell &= \sum_{K} Q^R_{IK} \, D^R_{JK} (\cJ_\ell) = {\tt Cont}^R_J (k) \, Q^R_{IK} 
\end{aligned}
\end{equation}
where we used \eqref{def:JM eigenval}.

The matrix unit $Q^R_{IJ}$ satisfies the following set of eigenvalue system,
\begin{equation}
\begin{aligned}
T_p \, Q^R_{IJ} = Q^R_{IJ} \, T_p &= \frac{\chi^R (T_p)}{d_R} \, Q^R_{IJ} 
\\
\cJ_k \, Q^R_{IJ} &= {\tt Cont}^R_I (k) \, Q^R_{IJ} 
\\[1mm]
Q^R_{IJ} \, \cJ_k &= {\tt Cont}^R_J (k) \, Q^R_{IJ} \,.
\end{aligned}
\label{Eigen AW to MU}
\end{equation}
Conversely, we can reconstruct the matrix unit by solving this eigensystem up to an overall normalisation, which is fixed by \eqref{alg for YSN basis}.
Since the matrix unit should be expanded as
\begin{equation}
Q^{R}_{I J} 
= \frac{d_{R}}{|S_L|} \sum_{g \in S_L} D^{R}_{JI} (g) \, g^{-1}
\label{Expand QRIJ as DRIJ}
\end{equation}
we can also reconstruct the matrix of the Young seminormal representations $D^R_{JI} (g)$ by reading off the coefficients in \eqref{Expand QRIJ as DRIJ}.

We have seen that a complete basis of $\bb{C}[S_L]$ is given by the matrix units $\{ Q^R_{IJ} \}$, and the matrix units can be reconstructed as the solution of the eigensystem \eqref{Eigen AW to MU}. This means that the following permutations
\begin{equation}
    \Big\{ m^{\acting{L}} [T_p] = m^{\acting{R}} [T_p] \,, m^{\acting{L}} [\cJ_k] \,, m^{\acting{R}} [\cJ_k] \, \Big| \, 
    p \vdash L, k \in \{ 2,3, \dots, L \} \Bigr\} 
\label{complete set of AW SL}
\end{equation}
form a complete set of commuting operators in ${\rm End} \, (\cV_{\rm reg})$.

Here are some remarks.
The eigenvalue method for modules was first developed by Murphy \cite{Murphy1981y} for constructing the Young seminormal representation, which was further refined in \cite{vershik2005new}.
We extended this method to construct complete bases of $\bb{C}[S_L]$, corresponding to Artin-Wedderburn and Kronecker decompositions. 

The matrix units $Q^{R}_{I J}$ is called the Young seminormal units, when the subscripts $I,J$ diagonalise the action of YJM elements.\footnote{In this sense, the Young seminormal units can be written as $Q_{IJ}^R = \ket{\atop{R}{I}} \bra{\atop{R}{J}}$, which automatically satisfy \eqref{alg for YSN basis}.}
The Young seminormal units can also be constructed by a recursive method in \cite{GEBook20}. However, the recursive method is not as efficient as our eigenvalue method, because it takes a long time to execute the iterative steps.

The efficiency of our algorithm comes from the fact that we only need the minimal generating set of the centre instead of $T_p$ for all $p$ in \eqref{complete set of AW SL}. This should be contrasted with Murphy's approach where he used to distinguish all basis elements of irreducible representations.

\subsection{Eigenvalue system for the Kronecker decomposition}

The Kronecker basis $\cQ^{R,\Lambda, \tau}_{\ \ K}$ satisfies the following set of eigenvalue system,
\begin{equation}
\begin{aligned}
T_p \, \cQ^{R,\Lambda, \tau}_{\ \ K} = \cQ^{R,\Lambda, \tau}_{\ \ K} \, T_p &= \frac{\chi^R (T_p)}{d_R} \, \cQ^{R,\Lambda, \tau}_{\ \ K} 
\\
m^{\acting{ad}}[{T_q}] ( \cQ^{R,\Lambda, \tau}_{\ \ K} ) &= \frac{\chi^\Lambda (T_1)}{d_\Lambda} \, \cQ^{R,\Lambda, \tau}_{\ \ K} 
\\[1mm]
m^{\acting{ad}}[{\cJ_k}]( \cQ^{R,\Lambda, \tau}_{\ \ K} ) &= {\tt Cont}^\Lambda_K (k) \, \cQ^{R,\Lambda, \tau}_{\ \ K} 
\end{aligned}
\label{Eigen Kron}
\end{equation}
where the last equation follows from \eqref{adjoint action on Kr basis-1} and \eqref{def:JM eigenval}.

Conversely, we take a solution of the eigenvalue system \eqref{Eigen Kron}. We fix the overall normalisation, which is fixed by the $\delta$-function inner product \eqref{Kron basis norm}. For each $R,\Lambda)$, the dimension of the solution space should be equal to $C(R,R,\Lambda)$. We take a real orthogonal basis with respect to the inner product, and use $\tau$ to label its components. In this way, we can reconstruct the Kronecker basis.
Since the Kronecker basis should be expanded as
\begin{equation}
\cQ^{R,\Lambda, \tau}_{\ \ K} 
= \frac{d_{R}}{|S_L|} \, \sum_{g \in S_L} \sum_{I,J=1}^{d_{R}} 
\CGs{R}{R}{\Lambda}{\tau}{I}{J}{K} \, D^{R}_{JI} (g) \, g^{-1}
= \sum_{I,J=1}^{d_{R}} 
\CGs{R}{R}{\Lambda}{\tau}{I}{J}{K} \, Q^{R}_{IJ}
\label{Eigen Kron to Q}
\end{equation}
we can determine the Clebsch-Gordan coefficients as
\begin{equation}
\CGs{R}{R}{\Lambda}{\tau}{I}{J}{K} = \delta \Big( \invs{Q^{R}_{IJ} } \,  \cQ^{R,\Lambda, \tau}_{\ \ K} \Big) .
\end{equation}

Related to this is the fact that in general the set 
\begin{equation}
    \Big\{ m^{\acting{L}} [T_p] = m^{\acting{R}} [T_p]  \,, 
    m^{\acting{ad}} [T_q] \,, m^{\acting{ad}} [\cJ_k] \, \Big| \, p \vdash L, q \vdash L, k \in \{ 2,3, \dots, L \} \Bigr\} 
\end{equation}
forms an {\it incomplete} set of commuting operators in ${\rm End} \, (\cV_{\rm reg})$ due to the extra multiplicity label $\tau$.

\section{Integer eigenvalue system for general rational finite groups}\label{sec:general fg}

It is known that characters and normalised characters of finite groups are algebraic integers \cite[Chapter 6.5]{Serre1977}.
A finite group $G$ is called rational if all irreducible representations have rational-valued characters.
Since rational algebraic integers are ordinary integers, it follows that rational groups have integer characters and integer normalised characters.
In this section, we will generalise the above integer constructions to any pair of rational groups $(G,H)$ where $H $ is a subgroup of $G$ and can be used to define  a corresponding centraliser algebra
\begin{equation}
    \mathbb{C}[G]^H = \{ \sigma \in \mathbb{C}[G] \, : \, h\sigma = \sigma h \, \quad \forall h \in H \}.
\end{equation}
We will start the presentation with a discussion of general pairs $(G,H)$ of group and subgroup, and subsequently specialise to the case where these are both rational groups.

\subsection{Two decompositions of $\mathbb{C}[G]$}
The Artin-Wedderburn decomposition holds for general group algebras (see \cite[Chapter 6.2]{Serre1977} or \cite[Chapter 3-17]{hamermesh2012group}).
Let $\Rep{G}$ be a complete set of unitary irreducible representations of $G$ and for every $R \in \Rep{G}$ let $d_R$ be the dimension.
The elements
\begin{equation}
    Q_{IJ}^R = \frac{d_R}{|G|} \sum_{g \in G} D_{JI}^R(g^{-1}) g, \quad R \in \Rep{G}, \, I,J \in \{ 1,\dots,d_R \}
\end{equation}
form a basis of matrix units of $\mathbb{C}[G]$.
As before, we have
\begin{equation}
\begin{aligned}
    &g_1 \, Q_{IJ}^R = \sum_{K=1}^{d_R} D^R_{KI}(g_1) Q_{KJ}^R, \quad 
    &Q_{IJ}^R \, g_2^{-1} = \sum_{K=1}^{d_R} D^R_{JK}(g_2^{-1}) \, Q_{IK}^R = \sum_{K=1}^{d_R} D^{\overline{R}}_{KJ}(g_2) \, Q_{IK}^R.
\end{aligned}
\label{gen GxG action}
\end{equation}
where we have used unitarity in the last line of the second equation and $\overline{R}$ is the complex conjugate representation of $R$.
Therefore, as a representation of $G \times G$, 
\begin{equation}
    \mathbb{C}[G] \cong \bigoplus_{R \in \Rep{G}} V_R^G \otimes \overline{V}_{\! R}^{\, G},
\label{gen WA decomp CG}
\end{equation}
This implies the famous counting formula
\begin{equation}
    |G| = \sum_{R \in \Rep{G}} d_R^2.
\end{equation}

Restricting $G \times G$ to the diagonal subgroup of elements $(g_1,g_2) = (g,g)$  and decomposing into irreducible representations of this diagonal subgroup gives 
the Kronecker decomposition of $\mathbb{C}[G]$, generalising our discussion for $ \mC [ S_L]$. 
As a representation of the diagonal subgroup we have
\begin{equation}
    V_R^G \otimes \overline{V}_{\! R}^{\, G} \cong \bigoplus_{\Lambda \in \Rep{G}} V_\Lambda^G \otimes V_{R,\overline{R}, \Lambda},
\end{equation}
where the multiplicity space 
\begin{equation}
    V_{R,\overline{R},\Lambda} = \mathrm{Hom}_G(V_\Lambda^G, V_R^G \otimes \overline{V}_R^G ),\quad \dim V_{R,R, \Lambda}  = C(R,\overline{R},\Lambda)
\end{equation}
has dimension given by Clebsch-Gordan multiplicities. 
Therefore, we get
\begin{equation}
    \mathbb{C}[G] \cong \bigoplus_{R,\Lambda \in \Rep{G}} V_\Lambda^G \otimes V_{R,\overline{R},\Lambda}
\end{equation}
which implies the counting formula
\begin{equation}
    |G| = \sum_{R,\Lambda \in \Rep{G}} d_\Lambda \, C(R,\overline{R},\Lambda).
\end{equation}
The corresponding basis has the form
\begin{equation}
    \mathbb{C}[G] = \mathrm{Span}( \mathcal{Q}^{R, \Lambda, \tau}_{\phantom{R,}K} \, \vert \, R,\Lambda \in \Rep{G}, K \in \{ 1,\dots, d_\Lambda \} , \tau \in \{  1,\dots, C(R,R,\Lambda)) \} .
\end{equation}

The centre of a general group algebra $\mathcal{Z}(\mathbb{C}[G])$ has a basis labelled by conjugacy classes.
Let $CL(G)$ be the set of conjugacy classes of $G$ and for every $C \in CL(G)$ define
\begin{equation}
    T_C = \sum_{g \in C} g \, .
\label{def:gen centre}
\end{equation}
The elements form a basis for the centre
\begin{equation}
    \mathcal{Z}(\mathbb{C}[G]) = \mathrm{Span} \Big( T_C \, \Big| \, C \in CL(G) \Big).
\end{equation}
The matrix units diagonalise the left/right action of the centre
\begin{equation}
    m^{\acting{L}}[T_C](Q_{IJ}^R) = \frac{\chi^R(T_C)}{d_R} \, Q^R_{IJ}.
\end{equation}
and the "Kronecker basis", in addition, diagonalises the adjoint action of the centre,
\begin{equation}
    m^{\acting{ad}}[{T_C}](\mathcal{Q}^{R,\Lambda,\tau}_{\phantom{R,}K}) = \frac{\chi^\Lambda(T_C)}{d_\Lambda} \, \mathcal{Q}^{R,\Lambda,\tau}_{\phantom{R,}K}.
\end{equation}
As previously mentioned, the above eigenvalues are integers for all rational groups $G$.

\subsection{Two decompositions of $\bb{C}[G]^H$}
The construction of orbit bases of $\mathbb{C}[G]^H$ is completely analogous to $\cA(\mu)$ and follows by replacing
\begin{equation}
    S_L  \ \to \ G, \qquad
    S_\mu \ \to \ H, \qquad
    \cA(\mu) \ \to \ \bb{C}[G]^H 
\end{equation}
where $H$ is any subgroup of $G$.
An orbit basis of $\mathbb{C}[G]^H$ is defined by 
\begin{equation}    
    P_H(g_i) = \frac{1}{\abs{\mathrm{Stab}(g_i)}}\sum_{h \in H} h g_i h^{-1}
\end{equation}
for a selection of orbit representatives $\{g_i\}$, as in Section \ref{sec:PCA}.
The matrix elements of the left action are defined by 
\begin{equation}
    \sum_{J=1}^l m^{\acting{L}}_{ji} \, [\sigma] \, P_H (g_j) = m^{\acting{L}}[\sigma](P_H(g_i)) = \sigma P_H (g_i )
\end{equation}
as in Section \ref{sec:regular rep Amn}.
 $\sigma$ is an integer linear combination of the basis elements $P_H (g_i)$, the corresponding representation matrices $m^{\acting{L}}_{ji} \, [\sigma], m^{\acting{ad}}_{ji} [\sigma]$ are all integer matrices.

\subsubsection{Restriction basis}

Restricting $G$ to $H$ gives a decomposition of $V_R^G$ into irreducible representations $V_r^H$ of $H$ ,
\begin{equation}
    V_R^G \cong \bigoplus_{r \in \Rep{H}} V_r^H \otimes V_{R \rightarrow r} 
\label{def:gen VG to VH}
\end{equation}
where $V_{R \rightarrow r} = \mathrm{Hom}_H(V_r, V_R)$ is the multiplicity space of dimension
\begin{equation}
    \dim V_{R \rightarrow r} \equiv g(r;R).
\end{equation}
This restriction defines a set of branching coefficients
\begin{equation}
    B^{R \rightarrow r}_{I \rightarrow i, \nu} \,,
 \quad I  \in \{ 1,\dots,d_R  \}    \quad i \in \{ 1,\dots,d_r \} , \quad \nu \in \{ 1,\dots, g(r;R) \} .
\end{equation}

Let us apply the restriction \eqref{def:gen VG to VH} to the Artin-Wedderburn decomposition of $\mathbb{C}[G]$ in \eqref{gen WA decomp CG}.
We obtain a Artin-Wedderburn decomposition of $\mathbb{C}[G]^H$
\begin{equation}
   \mathbb{C}[G]^H \cong \bigoplus_{\substack{R \in \Rep{G} \\ r \in \Rep{H}}} 
   V_{R \rightarrow r} \otimes \overline{V}_{\! R \rightarrow r}
\end{equation}
which implies
\begin{equation}
    \dim \mathbb{C}[G]^H = \sum_{\substack{R \in \Rep{G} \\ r \in \Rep{H}}} g(r;R)^2.
\end{equation}
Following the construction of the restricted Schur basis, we combine the matrix units of $\mathbb{C}[G]$ and branching coefficients to define a set of elements
\begin{equation}
    Q_{\nu_+ \nu_-}^{R, r} = \sum_{I,J} \sum_{i} 
    B^{R \rightarrow r}_{I \rightarrow i, \nu_-} Q_{IJ}^R B^{\overline{R} \rightarrow \overline{r}}_{J \rightarrow i,\nu_+}.
\end{equation}
These elements satisfy
\begin{equation}
    h \, Q^{R, r}_{\nu_+ \nu_-} =  
    Q^{R, r}_{\nu_+ \nu_-} h, \quad \forall h \in H .
\end{equation}

It is known that the centre of a centraliser algebra $\mathcal{Z}(\mathbb{C}[G]^H)$ is generated by $\mathcal{Z}(\mathbb{C}[G])$ together with $\mathcal{Z}(\mathbb{C}[H])$ \cite{Danz2012}.
Let $C \in CL(G)$ and $d \in CL(H)$, then we have the eigensystems
\begin{equation}
\begin{aligned}
    m^{\acting{L}}[T_C](Q^{R, r}_{\nu_+ \nu_-} ) = \frac{\chi^R(T_C) }{d_R} \, Q^{R, r}_{\nu_+ \nu_-} \\
    m^{\acting{L}}[T_d](Q^{R, r}_{\nu_+ \nu_-}) = \frac{\chi^r(T_d) }{d_r} \, Q^{R, r}_{\nu_+ \nu_-}.
\end{aligned}
\end{equation}

\subsubsection{$H$-invariant Kronecker basis}

As before, we can start from the ``Kronecker basis'' of $\mathbb{C}[G]$ to find a basis for $\mathbb{C}[G]^H$.
Here, we restrict to the diagonal subgroup of elements $(h,h) \in H \subset G \times G$.
Under this restriction, we have decompositions 
\begin{equation}
    V^G_{\Lambda} \cong \bigoplus_{ \substack{r \in \Rep{H}}   }  V_r^H \otimes V_{\Lambda \rightarrow r}.
\end{equation}
In particular, we are interested in the trivial representation $r_0$ of $H$ appearing in this decomposition since
\begin{equation}
    \mathbb{C}[G]^H \cong \bigoplus_{R,\Lambda \in \Rep{G}}  V_{R,\overline{R},\Lambda}  \otimes V_{\Lambda \rightarrow r_0}.
\end{equation}
This corresponds to a basis of elements
\begin{equation}
    \mathcal{Q}^{R,\Lambda,H, \tau}_{\beta}, \quad H\subset G, \quad \tau \in \{  1,\dots,C(R,\overline{R},\Lambda) \} , \quad \beta \in \{  1,\dots,
    g(r_0; \Lambda) \} 
\end{equation}
and gives the equality
\begin{equation}
    \dim \mathbb{C}[G]^H = \sum_{R,\Lambda \in \Rep{G}} C(R,\overline{R},\Lambda) \, g(r_0; \Lambda) 
    = \sum_{\substack{R \in \Rep{G} \\ r \in \Rep{H}}} g(r;R)^2.
\end{equation}

\subsection{Integer orthogonal bases for rational groups}

We now restrict to $G, H$ rational groups with $H \subset G$.
As mentioned before, normalised characters of rational groups are integers and the central elements \eqref{def:gen centre} are integer combinations of orbit basis elements.
Therefore, the matrices $m^{\acting{L}}_{ji} \, [T_C], m^{\acting{ad}}_{ji} \, [T_C]$ are integer matrices with integer eigenvalues.
Following Section \ref{sec:EV Amn} we obtain the integer eigenvalue system
\begin{equation}
(\mathbb{C}[G]^H)^{R,r} \cong \, \bigcap_{C,d} \, \Bigl\{ 
\mathrm{Ker}\qty(m^{\acting{L}} [T_{C}]-\widehat{\chi}^{R}_{C})\cap
\mathrm{Ker}\qty(m^{\acting{L}} [T_{d}]-\widehat{\chi}^{r}_{C})
\Bigr\}
\label{def:gen CGH Rr kernel}
\end{equation}
and
\begin{equation}
(\mathbb{C}[G]^H)^{R,\Lambda} \cong \, \bigcap_{C,C'} \, \Bigl\{ 
\mathrm{Ker}\qty(m^{\acting{L}} [T_{C}]-\widehat{\chi}^{R}_{C})\cap 
\mathrm{Ker}\qty(m^{\acting{ad}} [T_{C'}] -\widehat{\chi}^{\Lambda}_{C'}) 
\Bigr\}.
\label{def:gen CGH RLam kernel}
\end{equation}
It suffices to use a subset of conjugacy classes $C,C'$ in $ CL(G)$ with the property that their normalised characters uniquely identify the irreducible representations $ \Rep {G}$. Similarly it suffices to use a subset of conjugacy classes $d \in CL(H) $ which uniquely identify the irreducible representations in $ \Rep{H}$.
The integer eigenvectors can be found following the discussion in Section \ref{sec:integrality}.
Because all characters are rational, $T_C = \mathcal{S}(T_C)$ and it follows that the operators $m^{\mathfrak{L}}[T_C]\,, m^{\mathfrak{L}}[T_d]$, $m^{\acting{ad}}[{T_C}]$, $m^{\acting{ad}}[{T_d}]$ are self-adjoint with respect to the $\delta$-function inner product.
This argument is identical to the proof in equation \eqref{eq: self-adjoint T2}.
Therefore the eigenvectors corresponding to different eigenvalues are orthogonal.

\section{Conclusions}\label{sec:conclusions}

In this paper we developed the theory of multi-matrix invariants and constructions of bases of operators in $\cN=4$ SYM through their connection to permutation centraliser algebras $\cA(\mu)$, which generalise the symmetric group algebras $\mathbb{C}[S_L]$.
We discussed two decompositions of $\mathbb{C}[S_L]$, the Artin-Wedderburn and Kronecker decompositions, and reviewed the correspondence between elements of $\cA(\mu)$ and multi-matrix invariants in Section \ref{sec:PCA}.
Two orthogonal bases of multi-matrix invariants were well-studied in the literature, the restricted Schur and covariant bases. 
The restricted Schur basis decomposes $\cA(\mu)$ into a direct sum of subspaces $\cA^{R, r_1, \dots, r_M}(\mu)$, 
, while the covariant basis gives a direct sum of subspaces $\cA^{R, \Lambda}(\mu)$.
Section \ref{sec: res basis} explained how the restricted Schur basis \eqref{eq: AW Amu} can be understood as descending from the Artin-Wedderburn decomposition of $\mathbb{C}[S_L]$.
Analogously, section \ref{sec: cov basis} explained how the covariant basis \eqref{eq: cov Amu} comes from the Kronecker decomposition.
In section \ref{sec:EV Amn}, we derived an efficient algorithm for constructing finite $N$ orthogonal bases of $\cN=4$ SYM operators which are integer linear combinatons of multi-traces.
We set up integer eigensystems whose solutions give a basis for $\cA^{R, r_1, r_2}(\mu)$ and $\cA^{R, \Lambda}(\mu)$, which are realised as eigenspaces.
The eigenspaces $\cA^{R, r_1, r_2}(\mu)$ have dimensions equal to $g(r_1, r_2 ; R)^2$, while the eigenspaces $\cA^{R, \Lambda}(\mu)$ have dimensions equal to $C(R,R,\Lambda) K_{\Lambda \mu}$.
It was emphasised that these equations, summarised in \eqref{eq:matrix kernels1},\eqref{eq:matrix kernels2}, can be solved by finding kernels of integer matrices using Hermite normal forms.
In section \ref{sec:EV CSL}, we considered generalised eigenvalue systems derived from the Artin-Wedderburn and Kronecker decompositions of $\mathbb{C}[S_L]$.
In the case of the Artin-Wedderburn decomposition, we constructed the matrix units of $\mathbb{C}[S_L]$ in the Young seminormal representation.
For the case of the Kronecker decomposition, we obtained the Kronecker basis and associated Clebsch-Gordan coefficients.
The implementation in {\tt SageMath} is explained in appendix \ref{apx: algo}, and many examples are given in appendices \ref{app:data Amu} and \ref{sec:data}.

The integer orthogonal bases up to $L=10$ can be easily determined by running our code in a laptop, and our code is attached as ancillary files to this paper.
The code needs minor modification beyond $L=14$ because more Casimir operators $T_4 \,, T_5 \,, \dots$ should be added to the eigensystem from the discussion of $k_*$ in section \eqref{sec:centre}.

It is an open problem to systematically determine the multiplicity indices, $\nu_\mp$ in \eqref{eq: R r1 r2} and $\tau, \beta$ in \eqref{def:cA RLam mu}.
Resolving $\nu_\mp$ involves finding a subset of elements in $\cA(\mu)$ that generate a maximal commutative subalgebra of $\cA(\mu)$. 
It is interesting that the $\tau$ multiplicity in the covariant basis has already appeared in the Kronecker decomposition of $\mathbb{C}[S_L]$.
This Clebsch-Gordan multiplicity should be explained from the structural properties of another PCA called $\cK( n )$, which has appeared in the study of tensor models.
The multiplicity problem can be addressed by developing results in \cite{Kimura:2008ac,Mattioli:2016eyp,Geloun_2017,Geloun:2022kma}.
It is straightforward to generalise our algorithm to $M$-matrix models with $M>2$.
We will consider these problems in more detail in upcoming work \cite{Mult_paper}.

The representation theoretic approach we have used in this paper provides a comprehensive characterization of the trace relations for finite $N$ by imposing the constraint $\ell(R) \le N$ for a Young diagram label $R$.
We have focused on the 2-matrix problem here, but the approach also works with multiple fields \cite{Brown:2007xh, Brown:2008ij, deMelloKoch:2012sie}.
An interesting avenue to explore is the application of this systematic description of finite $N$ relations in the context of the construction of $\frac{1}{16}$\,-BPS operators (for related comments and steps in this direction see \cite{Chang:2022mjp,Chang:2024zqi,deMelloKoch:2024pcs}).

{\centering
\subsubsection*{Acknowledgements}
}

\noindent
We thank Gernot Akemann, Joseph Ben Geloun, and Robert de Mello Koch for discussions.
SR is supported by the Science and Technology Facilities Council
(STFC) Consolidated Grant ST/T000686/1 
``Amplitudes, strings and duality'' and a Visiting Professorship at Dublin Institute for Advanced Studies.
The work of RS is supported by NSFC grant no.~12050410255. The work of AP was partly funded by the Deutsche Forschungs-gemeinschaft (DFG) grant SFB 1283/2 2021 E317210226.

\bigskip
\appendix

\section{Notation}\label{app:notation}
In this appendix we define the notation used for branching coefficients and Clebsch-Gordan coefficients for $S_L$ and $U(M)$. They are used to define the decompositions in sections \ref{subsubsec: Kronecker decomp}, \ref{sec: cov basis}, \ref{sec: res basis} and \ref{subsec: gen cov basis}.

\subsection{Branching coefficients}\label{sec:branching coeff}

We review branching coefficients for $S_L$ to its subgroup, which are used to define the restricted Schur basis in section \ref{sec: res basis} and the covariant basis in section \ref{sec: cov basis}.
A Young subgroup of $S_L$ is defined by a list of positive integers $\mu = (\mu_1, \dots, \mu_M)$ such that
\begin{equation}
	\sum_k^M \mu_k = L.
\end{equation}
The Young subgroup corresponding to $\mu$ is defined as
\begin{equation}
    S_\mu = S_{\mu_1} \times S_{\mu_2} \times \dots \times S_{\mu_M} \subseteq S_L
\label{app:Young subgroup}
\end{equation}
Suppose that the restriction of the irreducible representation $R$ under $S_\mu \subset S_L$ decomposes into the direct sum of $(r_1 \otimes \dots \otimes r_M)$, where $r_k \vdash \mu_k$ for $k=1,\dots, M$.
The basis of states in $R$ decomposes as
\begin{equation}
    \ket{\atop{R}{I}} = 
    \sum_{r_1 \vdash \mu_1} \dots \sum_{r_M \vdash \mu_M}
    \sum_{\nu=1}^{g( r_1 , \dots ,  r_M ; R )}
    \sum_{i_1=1}^{d_{r_1}} \dots \sum_{i_M=1}^{d_{r_M}}
    B^{R \to (r_1 \dots  r_M)}_{I \to (i_1  \dots i_M), \nu} \,
    \ket{ \matop{r_1 & \dots & r_M}{i_1 & \dots &i_M}{\nu} }
\label{app:restricted irrep}
\end{equation}
where $B^{R \to (r_1 \dots  r_M)}_{I \to (i_1  \dots i_M), \nu}$ is called the branching coefficient and $\nu$ is called the multiplicity label.
The symbol $g(R \,; r_1, \dots , r_M)$ is the Littlewood-Richardson coefficient defined by
\begin{equation}
g(r_1, \dots , r_M ; R) = \frac{1}{| S_\mu |} 
\sum_{\alpha_1 \in S_{\mu_1}} \dots  \sum_{\alpha_M \in S_{\mu_M}}
\chi^R (\alpha_1 \otimes \dots \otimes \alpha_M) \,
\prod_{k=1}^M \chi^{r_k} (\alpha_k) .
\label{app:Littlewood Richardson}
\end{equation}
We introduce the standard inner product in the restricted basis,
\begin{equation}
\Big\langle \, \matop{r_1 & \dots & r_M}{i_1 & \dots &i_M}{\nu} \, \Big| \, 
\matop{s_1 & \dots & s_M}{j_1 & \dots &j_M}{\xi} \, \Big\rangle 
= \delta_{\nu \xi} \Big( \prod_{k=1}^M \delta^{r_k s_k} \, \delta_{i_k j_k} \Big), \qquad 
(i_k , j_k \in \{ 1, \dots , d_{r_k} \} ).
\end{equation}
The original basis and the restricted basis have the same dimensions,
\begin{equation}
    d_R = \sum_{r_1 \vdash \mu_1} \dots \sum_{r_M \vdash \mu_M}
    \( g( r_1, \dots , r_M ; R) \prod_{k=1}^M d_{r_k} \)
\label{compare rest dim}
\end{equation}
which is consistent with \eqref{eq: smu decomp}. 
This shows that the transformation \eqref{app:restricted irrep} relates two complete orthonormal bases, and therefore unitary. 
Because $S_L$ and $S_\mu$ are real groups, we can choose the branching coefficients to be real, giving
\begin{equation}
    B^{R \to (r_1 \dots  r_M)}_{I \to (i_1  \dots i_M), \nu}
    = \Vev{ \matop{r_1 & \dots & r_M}{i_1 & \dots &i_M}{\nu}  \, \Big| \, \atop{R}{I} } 
    = \Vev{ \atop{R}{I} \, \Big| \, \matop{r_1 & \dots & r_M}{i_1 & \dots &i_M}{\nu} } .
\label{def:branching coeff}
\end{equation}
The inverse transformation of \eqref{app:restricted irrep} is\footnote{Note that the restricted basis depends on $R$ through the multiplicity label, $\nu \in \{ 1, \dots, g(r_1 \dots r_M; R) \}$.}
\begin{equation}
\ket{ \matop{r_1 & \dots & r_M}{i_1 & \dots &i_M}{\nu} } 
= \sum_{I=1}^{d_R} B^{R \to (r_1 \dots  r_M)}_{I \to (i_1  \dots i_M), \nu} \, \ket{\atop{R}{I}} ,
\label{app:restricted irrep inv}
\end{equation}
The restricted basis \eqref{app:restricted irrep} forms an irreducible representation of $S_\mu$ for each multiplicity label $\nu$. As such, a generic element $h \equiv h_1 \otimes \dots \otimes h_M \in S_\mu$ acts diagonally on the multiplicity space as
\begin{equation}
h \, \Big| \matop{r_1 & \dots & r_M}{i_1 & \dots &i_M}{\nu} \, \Big\rangle 
= \sum_{j_1=1}^{d_{r_1}} \dots \sum_{j_M=1}^{d_{r_M}} \ 
D^{r_1}_{j_1 i_1} (h_1) \dots D^{r_M}_{j_M i_M} (h_M) \, 
\Big| \matop{r_1 & \dots & r_M}{j_1 & \dots &j_M}{\nu} \, \Big\rangle  .
\label{app:Drij h}
\end{equation}

The branching coefficients satisfy the completeness relations \cite{hamermesh2012group,Pasukonis:2013ts}
\begin{align}
&\sum_{I} B^{R \to (r_1 \dots  r_M)}_{I \to (i_1  \dots i_M), \nu} \,
B^{R \to (s_1 \dots  s_M)}_{I \to (j_1  \dots j_M), \xi}
= 
\sum_{I} \Vev{ \matop{r_1 & \dots & r_M}{i_1 & \dots &i_M}{\nu}  \, \Big| \, \atop{R}{I} } 
\Vev{ \atop{R}{I} \, \Big| \, \matop{s_1 & \dots & s_M}{j_1 & \dots &j_M}{\xi} }
= \delta_{\nu \xi} \, \Big( \prod_{k=1}^M \delta^{r_k s_k} \, \delta_{i_k j_k} \Big)
\label{product branching} \\[1mm]
&\sum_{r_1 \dots r_M} \sum_{\nu} \sum_{i_1 \dots i_M} 
B^{R \to (r_1 \dots  r_M)}_{J \to (i_1  \dots i_M), \nu} \,
B^{R \to (r_1 \dots  r_M)}_{I \to (i_1  \dots i_M), \nu}
=
\sum_{r_1 \dots r_M} \sum_{\nu} \sum_{i_1 \dots i_M}  
\Vev{ \atop{R}{J} \, \Big| \, \matop{r_1 & \dots & r_M}{i_1 & \dots &i_M}{\nu} }
\Vev{ \matop{r_1 & \dots & r_M}{i_1 & \dots &i_M}{\nu}  \, \Big| \, \atop{R}{I} } 
= \delta_{IJ} \,.
\notag 
\end{align}
They also exhibit an equivariance property
\begin{equation}
\sum_{I} D^R_{JI} ( h ) \, B^{R \to (r_1 \dots r_M) }_{I \to (i_1 \dots i_M), \nu} 
= \sum_{j_1 \dots j_M} B^{R \to (r_1 \dots r_M) }_{J \to (j_1 \dots j_M), \nu} \,
D^{r_1}_{j_1 i_1} (h_1) \dots D^{r_M}_{j_M i_M} (h_M) .
\label{branching equivariance-1} 
\end{equation}

\subsection{CG coefficients of $S_L$}\label{app:CG of SL}

We revire the Clebsch-Gordan (CG) coefficients for $S_L$\,, which are used to define the Kronecker basis in section \ref{subsubsec: Kronecker decomp} and covariant basis in \ref{sec: cov basis}.

Let $V_{R_1}^{S_L} \,, V_{R_2}^{S_L}$ be the irreducible representations of $S_L$\,. We define the tensor product representation by
\begin{equation}
\sigma \ \ket{ \atop{R_1}{I_1} } \otimes \ket{ \atop{R_2}{I_2} }
= 
D^{R_1}_{J_1 I_1} (\sigma) \, D^{R_2}_{J_2 I_2} (\sigma) \,
\ket{ \atop{R_1}{J_1} } \otimes \ket{ \atop{R_2}{J_2} } 
\label{app:g action tensors}
\end{equation}
generalising \eqref{def:g action DRIJ}. This equation implies that the tensor produce representation is decomposed into irreducible representations of $S_L$ as
\begin{equation}
V_{R_1}^{S_L} \otimes V_{R_2}^{S_L} \ \cong \ \bigoplus_{\Lambda \, \vdash L} \ V_\Lambda^{S_L} \otimes V_{R_1 ,R_2 ,\Lambda} \,, \qquad
\dim V_{R_1 ,R_2 ,\Lambda} = C (R_1 , R_2, \Lambda)
\end{equation}
where $C (R_1 , R_2, \Lambda)$ is the Kronecker coefficient defined by
\begin{equation}
C(R_1 ,R_2 ,\Lambda) = \frac{1}{\abs{S_L}} \sum_{g \in S_L} \chi^{R_1} (g) \chi^{R_2} (g) \chi^\Lambda (g) .
\label{app:Kronecker coef}
\end{equation}
By equating the dimensions of the representations, we obtain
\begin{equation}
d_{R_1} \, d_{R_2} = \sum_{\Lambda \, \vdash L} C(\Lambda,R_1,R_2) \, d_{\Lambda} \,.
\end{equation}

By taking a complete basis of $S_L$, we define the CG coefficients from the irreducible decomposition as, 
\begin{equation}
\ket{ \atop{R_1}{I_1} } \otimes \ket{ \atop{R_2}{I_2} }
 = \sum_{\Lambda \, \vdash L} \sum_{\tau =1}^{C(\Lambda,R_1,R_2)} \sum_{K=1}^{d_{\Lambda}}  
\CGs{R_1}{R_2}{\Lambda}{\tau}{I_1}{I_2}{K} \, 
\ket{ \matop{\Lambda}{K}{\tau} } 
\label{app:CG coeff}
\end{equation}
whose inverse relation is
\begin{equation}
\ket{ \matop{\Lambda}{k}{\tau} }
= \sum_{R_1 \, \vdash L} \sum_{R_2 \, \vdash L} \sum_{I_1 =1}^{d_{R_1}} \sum_{I_2 =1}^{d_{R_2}}  
\CGs{R_1}{R_2}{\Lambda}{\tau}{I_1}{I_2}{K} \, 
\ket{ \atop{R_1}{I_1} } \otimes \ket{ \atop{R_2}{I_2} } .
\label{app:CG coeff2}
\end{equation}
Since $S_L$ is a real group, we can choose the CG coefficients to be real.
From \eqref{app:CG coeff} and \eqref{app:CG coeff2} we find the orthogonality relations \cite{hamermesh2012group,Pasukonis:2013ts}
\begin{align}
\sum_{I_1, I_2} 
\CGs{R_1}{R_2}{\Lambda}{\tau}{I_1}{I_2}{K} \, 
\CGs{R_1}{R_2}{\Lambda'}{\tau'}{I_1}{I_2}{K'} 
= \delta^{\Lambda \Lambda'} \, \delta^{\tau \tau'} \delta_{KK'} \,, 
\quad
\sum_{\tau, \Lambda, K} 
\CGs{R_1}{R_2}{\Lambda}{\tau}{I_1}{I_2}{K} \, 
\CGs{R'_1}{R'_2}{\Lambda}{\tau}{J_1}{J_2}{K} \, 
= \delta^{R_1 R'_1} \, \delta^{R_2 R'_2} \, \delta_{I_1 J_1} \, \delta_{I_2 J_2} \,.
\label{BHR_160}
\end{align}
The CG coefficients satisfy the equivariance property 
\begin{align}
\sum_{J_1, J_2}
\Clebsch{\tau}{\Lambda}{R_1}{R_2}{K}{J_1}{J_2} \, 
D^{R_1}_{J_1 I_1} (\sigma) \, D^{R_2}_{J_2 I_2} (\sigma) 
&= \sum_{K'}
D^{\Lambda}_{KK'} (\sigma) \, \Clebsch{\tau}{\Lambda}{R_1}{R_2}{K'}{I_1}{I_2} \,.
\label{app:SDD=SD} 
\end{align}

\subsection{CG coefficients of $U(M)$}\label{app:CG of UM}
The CG coefficients for $U(N)$ are important in defining the general covariant basis in section \ref{subsec: gen cov basis}.
Let $V_M$ be the fundamental representation of $U(M)$, equipped with the standard inner product
\begin{equation}
\langle a  \big| b \rangle_{U(M)} = \delta_{ab} \,, \qquad a, b \in \{ 1, 2, \dots, M \}.
\end{equation}
Under the Schur-Weyl duality, the tensor product $V_M^{\otimes L}$ decomposes as
\begin{equation}
V_M^{\otimes L} = \bigoplus_{\Lambda \, \vdash L} \ ( V^{U(M)}_\Lambda \otimes V^{S_L}_\Lambda ).
\label{app:SW dual SLxU(M)}
\end{equation}
If we write down explicit components, we find
\begin{equation}
\big| \vec a \big\rangle \equiv
\ket{a_1 \,, \dots \,, a_L}_{U(M)} = \sum_{\substack{\Lambda \vdash L \\[.5mm] \ell(\Lambda) \le M}} 
\sum_{M_\Lambda=1}^{\Dim_M (\Lambda)}
\sum_{K=1}^{d_\Lambda}
C^{\Lambda}_K {}^{\Lambda}_{M_\Lambda} (\vec a) \, 
\ket{ \atop{\Lambda}{K} } \ket{ \atop{\Lambda}{M_\Lambda} }_{U(M)}  
\label{app:CG coeff SLxU(M)}
\end{equation}
where $K$ labels a component of the irreducible representation $\Lambda$ of $S_L$\,, and $M_\Lambda$ labels a component of the irreducible representation $\Lambda$ of $U(M)$.
The symbol $C^{\Lambda}_K {}^{\Lambda}_{M_\Lambda} (\vec a)$ in \eqref{app:CG coeff SLxU(M)} is the CG coefficient of $S_L \times U(M)$ used in \eqref{def:general cov basis}.
By comparing the dimensions in \eqref{app:CG coeff SLxU(M)}, we obtain
\begin{equation}
M^L = \sum_{\substack{\Lambda \vdash L \\[.5mm] \ell(\Lambda) \le M}} d_\Lambda \, \Dim_M (\Lambda) .
\end{equation}
The CG coefficients satisfy the orthogonality relations
\begin{equation}
\sum_{\vec a}
C^{\Lambda}_K {}^{\Lambda}_{M_\Lambda} (\vec a) \, 
C^{\Lambda'}_{K'} {}^{\Lambda'}_{M_{\Lambda'}} (\vec a)  
= \delta^{\Lambda \Lambda'} \, \delta_{M_\Lambda M_{\Lambda'}} \, \delta_{K K'} \,, \qquad
\sum_{\Lambda, M_\Lambda , K }
C^{\Lambda}_K {}^{\Lambda}_{M_\Lambda} (\vec a) \, 
C^{\Lambda}_K {}^{\Lambda}_{M_\Lambda} (\vec b) \, 
= \prod_{i=1}^L \delta_{a_i \, b_i} \,.
\label{app:CG orth rels}
\end{equation}

\section{Algorithm: Decomposition of $\cA(\mu_1,\mu_2)$}\label{apx: algo}

In this appendix we outline the algorithm used to construct an integer basis for $\cA^{R,\Lambda}$, $\cA^{R,r_1,r_2}$ using the ingredients introduced in the main text.
We used the {\tt GAP} package in {\tt SageMath} \cite{GAP4,sagemath} to run the code.
The code produces a text file enumerating the integer orthogonal basis elements in terms of linear combinations of multi-traces.
In this appendix we write $\cA(\mu_1, \mu_2) = \cA(m,n)$ in order not to
deviate from the notation in the code.
The code comes in two files:
\begin{itemize}[nosep]
\item {\tt Amn\_Restriction\_Decomposition.sage}
\item {\tt Amn\_Covariant\_Decomposition.sage}
\end{itemize}
and can be run in the terminal by executing, for example, 
\begin{equation}
\text{\tt sage Amn\_Covariant\_Decomposition.sage L n N}
\end{equation}
The first parameter determines $L=m+n$ and the second parameter determines $n$.
The last parameter $N$ divides the output into two files. The first file contains a basis of operators for $L \le N$, and the second file contains a basis for the finite $N$ trace relations. The output is independent of $N$ if $N \ge L$, in which case the second file is empty.

\subsection{Outline of SageMath code}\label{app:sagemath code}
In this part we will give an outline of the {\tt SageMath} code.
Each file/algorithm consists of three parts: (1) Constructing the orbit basis for $\cA(m,n)$.
(2) constructing the representation matrices of central elements $T_2,T_3, \dots$; (3) computing the intersection of kernels in section \ref{sec:EV Amn}.

\begin{enumerate}
    \item 
The first step of the algorithm is to construct $\cA(m,n)$ using the orbit basis described in \ref{sec:PCA}.
The orbits are given by $S_m \times S_n$ acting on $S_L$ by conjugation and can be constructed in {\tt Sage} as follows.
\begin{enumerate}
    \item Construct the groups $S_m \times S_n$ and $S_{m+n}$
    \begin{equation}
        \begin{aligned}
            &{\tt SmSn = libgap.DirectProduct(libgap.SymmetricGroup(m), libgap.SymmetricGroup(n)) }\\
            &{\tt Smn= libgap.SymmetricGroup(m+n)}
        \end{aligned}
    \end{equation}
    \item Compute the orbits
    \begin{equation}
        {\tt orbit\_basis = libgap.OrbitsDomain(SmSn, Smn)}
    \end{equation}
\end{enumerate}
The variable ${\tt orbit\_basis}$ is a set of orbits, each orbit contains elements of $S_L$ related by conjugation of $S_m \times S_n$.
We also collect a set of representatives of each orbit
\begin{equation}
	{\tt basis\_keys = [\,\, b[0] \,\, for \,\, b \,\, in \,\, orbit\_basis \,\,]}
\end{equation}
This gives the orbit basis of $\cA(m,n)$ -- each orbit or representative corresponds to a basis element of $\cA(m,n)$. In the rest of this section we will refer to the representative elements as $g_i$, where $g_i = {\tt basis\_keys[i-1]}$ in the code.

\item The second step is to compute the left and adjoint action of $T_2, T_3 \in \mathcal{Z}(\mathbb{C}[S_L])$ on $\cA(m,n)$. We will compute the matrix elements $m^L_{ji} [T_2]$ given by \eqref{def:m acting on Vmnreg}, and this can be understood in terms of actions on the orbit sets above. Focusing on the left action of $T_2$, we compute the following 
\begin{enumerate}
    \item {\tt GAP} gives the conjugacy classes as a list
    \begin{equation}
        {\tt T2 = libgap.ConjugacyClass(Smn, libgap.eval("(1,2)")).List()}
    \end{equation}
    \item Construct an empty matrix of size $\dim \cA(m,n) = {\tt len(orbits)}$
    \begin{equation}
        {\tt T2\_left\_matrix = zero\_matrix(ZZ,\; len(orbit\_basis))}
    \end{equation}
    The input ${\tt ZZ}$ forces the matrix to have integer entries.
    \item
    The left action on $P_{m,n}(g_i)$ is computed in parallel for every representative.
    We introduce the notation $\{ T_2 g \}$ for the set of elements with non-zero coefficients appearing in the expansion of the product $T_2 g$.
    The action on an orbit basis element is determined by the action on representatives because
    \begin{equation}
    \begin{aligned}
        T_2 P_{m,n}(g_i) &= \frac{1}{|{\rm Stab}(g_i)|}\sum_{h \in S_m \times S_n} h T_2 g_i h^{-1} \\
        &= \frac{1}{|{\rm Stab} (g_i)|}\sum_{h \in S_m \times S_n} \sum_{g \in \{T_2 g_i \}} h g h^{-1}\\
        &= \frac{1}{|{\rm Stab} (g_i)|}\sum_{g \in \{T_2 g_i \}} |{\rm Stab}(g)| P_{m,n}(g)\\
        &= \sum_{g \in \{T_2 g_i \}} \frac{|{\rm Orb} (g_i)|}{|{\rm Orb}(g)|}P_{m,n}(g)
    \end{aligned}
    \end{equation}
    where the last step uses the orbit-stabilizer theorem $|{\rm Stab}(g)||{\rm Orb}(g)| = |S_m \times S_n|$.
    Therefore, as an intermediate step, it is useful to compute 
    \begin{equation}
    {\tt
        t2\_orb = tuple(t2*g \; for \; t2 \; in \; T2) ,
    }
    \end{equation}
    {\tt t2\_orb} corresponds to the set $\{ T_2 g_i \}$.
    Now, for every $h \in \{T_2 g_i \}$ that is also in the orbit $\{g_j \}_{m,n}$ we get a contribution to the coefficient in front of $P_{m,n}(g_j)$ in the expansion of $T_2 P_{m,n}(g_i)$ in the orbit basis.
    It follows that the expansion is given by the vector
    \begin{equation}
    {\tt
            t2\_left\_vec = \begin{aligned}[t]
            {\tt list(}&\\
                    &{\tt len(orbit)/len(orb)*sum(1 \; for \; x \; in \; t2\_orb \; if \; x \; in \; orb)} \\
                    &{\tt for \; orb \; in \; orbits}\\
                &{\tt ) }
            \end{aligned}
    }
    \end{equation}
    where ${\tt \\len(orbit)}$ is the length of the orbit $\{g_i \}_{m,n}$.
    The sum computes the order of the intersection of $\{T_2 g_i \}$ and $\{g_j \}_{m,n}$ for all $j \in \{1, \dots, \dim \cA(m,n)\}$. 
    \begin{equation}
       {\tt sum(1 \; for \; x \; in \; t2\_orb \; if \; x \; in \; orb)  }= \left| \left\{ \text{Elements in {\tt t2\_orb} that are also in {\tt orb}} \right\}\right| 
    \end{equation}
    {\tt t2\_left\_vec} corresponds to the $i$th column in the representation of $T_2$ acting on $\cA(m,n)$.
    We append it to the matrix by setting
    \begin{equation}
        {\tt T2\_left\_matrix[i-1] = t2\_left\_vec}
        \end{equation}
    for every representative.
    \item Lastly, transpose {\tt T2\_left\_matrix}.
\end{enumerate}
Similar procedures can be used to compute the left actions of $T_3$, adjoint actions of $T_2, T_3$ and so on. We call the matrix corresponding to the adjoint action {\tt T2\_adjoint\_matrix} and similarly for $T_3$.

\item 
Having computed the representation matrices we want to compute the kernels in section \ref{sec:EV Amn}.
We will focus on the covariant case, the restricted Schur case is analogous.
For every $R, \Lambda$ we do the following
\begin{enumerate}
    \item Compute the normalised characters/eigenvalues $\widehat{\chi}^R_2, \widehat{\chi}^\Lambda_2$. This is done using the functions {\tt T2eigval(), T3eigval()} which simply computes \eqref{SL Casimir T2}, \eqref{SL Casimir T3},
    \begin{equation}
        \begin{aligned}
            &{\tt R1 = T2eigval(R)}\\
            &{\tt R2 = T3eigval(R)}\\
            &{\tt Lambda1 = T2eigval(Lambda)}\\
            &{\tt Lambda2 = T3eigval(Lambda)}
        \end{aligned}
    \end{equation}
    \item Construct a tuple of matrices 
    \begin{equation}
    \begin{aligned}
    	{\tt EV\_matrices = tuple(}&{\tt R1 - T2\_left\_matrix,}\\
    	&{\tt R2 - T3\_left\_matrix,}\\
    	&{\tt Lambda1 - T2\_adjoint\_matrix,}\\
    	&{\tt Lambda2 - T3\_adjoint\_matrix)}
    \end{aligned}
    \label{code:EV matrices}
    \end{equation}
    \item Construct a new matrix made out of blocks, that is a 4-by-1 block matrix 
    \begin{equation}
        {\tt simul\_kernel\_matrix = block\_matrix(ZZ, len(EV\_matrices), 1,  EV\_matrices)}
    \end{equation}
    where ${\tt ZZ}$ defines the block matrix to be over the integers. 
    \item Compute a basis for the kernel
    \begin{equation}
        {\tt kernel\_basis = simul\_kernel\_matrix.right\_kernel\_matrix()}
    \label{code:right kernel}
    \end{equation}
\end{enumerate}
\end{enumerate}
The matrix {\tt kernel\_basis} is a matrix with rows giving a basis for the kernel of {\tt simul\_kernel\_matrix} or equivalently a basis for $\cA^{R,\Lambda}(m,n)$.
For the benefit of the reader, the next subsection contains a review of a simple algorithm for computing this matrix.
Lastly, we run the Gram-Schmidt process on the matrix {\tt kernel\_basis} and clear the denominators of the vectors to get integer orthogonal bases for $\cA^{R,\Lambda}(m,n)$.
The procedure is very similar for the space $\cA^{R,r_1,r_2}(m,n)$.

Note that the kernel \eqref{code:right kernel} contains a non-trivial solution if and only if\footnote{In the restricted Schur basis, the eigenspace has non-zero dimension if and only if $g(R; r_1, r_2) > 0$.}
\begin{equation}
	C(R,R,\Lambda) > 0, \qquad  \ell(\Lambda) \leq 2, \qquad K_{\Lambda \mu} > 0 .
\end{equation}
Therefore, it is sufficient to compute the eigenvalues and kernels in those cases.
In the code, the set of non-zero pairs $(R, \Lambda)$ are computed using
\begin{equation}
\begin{aligned}
	{\tt sectors = tuple(}&{\tt (R,Lambda) \, \, for \,\, R \,\, in \,\, Partitions(L) \,\, for \,\, Lambda \,\, in \,\, Partitions(L, max\_length=2)} \\
	 &{\tt if \,\, len(R) <= N \,\, and \,\, kroncoeff(R,R,Lambda) > 0 \,\, and} \\
	 &{\tt  symmetrica.kostka\_number(Lambda, [L-i,i]) > 0)}
\end{aligned}
\end{equation}
here {\tt kroncoeff} is a function that computes the Kronecker multiplicity $C(R,R,\Lambda)$ and {\tt kostka\_number} computes the Kostka number $K_{\Lambda \mu}$.

We encounter some problems for $L > 10$. First, output files are huge, typically larger than 1MB. Second, we need to solve memory shortage to run the code. A workstation with 1TB of memory can reach $L=12$; computing $L>12$ will require further refinement of the algorithm or some tricks for the memory management for parallel evaluation.

\subsection{Integer kernels}\label{app:hermite algorithm}

The following algorithm/procedure gives a basis for the kernel of a $p \times q$ matrix $M$ \cite[Algorithm 2.4.5]{Cohen}:
\begin{enumerate}
    \item Set $i=p, j=q, k=q$, $U = I_q$ the $q\times q$ identity matrix. If $p \leq q$ set $l=1$ else $l=p-q+1$.
    \item If $j \neq 1$ go top step 3. Otherwise, go top step 4.
    \item Reduce $j$ by one. If $M_{ij} = 0$ go to step 2. Otherwise do the following:
        \begin{enumerate}
            \item Set $a = M_{ik}, b = M_{ij}$, compute $d,u,v$ such that
            \begin{equation}
                d = \mathrm{gcd}(a,b) = u a + v b.
            \end{equation}
            It is technically necessary for $\abs{u}, \abs{v}$ to be minimal, see the remark below \cite[Algorithm 2.4.5]{Cohen} for more details.
            \item We introduce an auxiliary column vector $B$. For every $r \in \{ 1,\dots,p \}$ set
            \begin{align}
                &B_r = u M_{rk} + v M_{r j} \\
                &M_{rj} = \frac{a}{d} M_{rj} - \frac{b}{d} M_{r k} \\
                &M_{rk} = B_r
            \end{align}
            and analogously for $r=1,\dots,q$
            \begin{align}
                &B_r = u U_{rk} + v U_{r j} \\
                &U_{rj} = \frac{a}{d} U_{rj} - \frac{b}{d} U_{r k} \\
                &U_{rk} = B_r
            \end{align}
        \end{enumerate}
    \item Set $a = M_{ik}$ and do the following steps:
    \begin{enumerate}
        \item If $a < 0$, then for every $r=1,\dots,p$ set
            \begin{equation}
                M_{rk} = - M_{rk}, \quad U_{rk} = - U_{rk}, \quad a = -a.
            \end{equation}
        \item If $a = 0$, increase $k$ by one. Otherwise, each $r = k+1, \dots, q$ 
        \begin{align}
            &q = \lfloor M_{ir}/a \rfloor \\
            &M_{sr} = M_{sr} - qM_{sk}, \quad \text{for every $s\in \{ 1,\dots,p \}$}\\
            &U_{sr} = U_{sr} - qU_{sk}, \quad \text{for every $s\in \{ 1,\dots,q \}$}\\
        \end{align}
        \item If $i = l$ output the $k-1$ first column of $U$ and terminate the algorithm. Otherwise, set $i=i-1, k=k-1, j=k$ and go to step 2.
    \end{enumerate}
\end{enumerate}
The above procedure is not the one used in practice, for reasons of efficiency, {\tt SageMath} will choose an appropriate algorithm based on the input matrix.

\subsection{Simple example: Integer kernel}
\label{apx:example kernel}
In section \ref{sec:simple example} we illustrated a simple example of applying the eigenvalue method to $\mathcal{A}(3,0) = \mathcal{Z}(\mathbb{C}[S_3])$ and $R=[3]$.
This appendix contains the details of computing the integer eigenvectors using the algorithm in the previous subsection. 

We apply the algorithm on the augmented matrix
\begin{equation}
    m^{\mathfrak{L}} [T_2^{(3)}] - 3 I_3= 
    \left(
    \begin{matrix}
    -3 & 3 & 0 \\
    1 & -3 & 2 \\
    0 & 3 & -3 \\
    \hline
    1 & 0 & 0 \\
    0 & 1 & 0 \\
    0 & 0 & 1
    \end{matrix}
    \right)
\label{calc:example1}
\end{equation}
Starting at the third row, we want turn all columns to the left of $-3$ to zero.
We do this by adding the third column to the second column giving
\begin{equation}
    \begin{pmatrix}
    -3 & 3 & 0 \\
    1 & -1 & 2 \\
    0 & 0 & -3 \\
    \hline
    1 & 0 & 0 \\
    0 & 1 & 0 \\
    0 & 1 & 1
    \end{pmatrix}
\label{calc:example2}
\end{equation}
Note that multiplying the upper matrix in \eqref{calc:example1} by the lower matrix of \eqref{calc:example2} on the right, gives the upper matrix of \eqref{calc:example2}.
There is nothing left to clear to the left of $-3$, but we switch the sign by multiplying the third column by $-1$ giving
\begin{equation}
    \begin{pmatrix}
    -3 & 3 & 0 \\
    1 & -1 & -2 \\
    0 & 0 & 3 \\
    \hline
    1 & 0 & 0 \\
    0 & 1 & 0 \\
    0 & 1 & -1
    \end{pmatrix}
\end{equation}
There is nothing to the right of the third column and so no further reduction is needed.

Now consider the second row.
We want to clear all columns to the left of $-1$.
We do this by adding the second column to the first giving
\begin{equation}
    \begin{pmatrix}
    0 & 3 & 0 \\
    0 & -1 & -2 \\
    0 & 0 & 3 \\
    \hline
    1 & 0 & 0 \\
    1 & 1 & 0 \\
    1 & 1 & -1
    \end{pmatrix}
\end{equation}
Since $-1$ is not positive, we multiply the second column by $-1$
\begin{equation}
    \begin{pmatrix}
    0 & -3 & 0 \\
    0 & 1 & -2 \\
    0 & 0 & 3 \\
    \hline
    1 & 0 & 0 \\
    1 & -1 & 0 \\
    1 & -1 & -1
    \end{pmatrix}
\end{equation}
We have already found a basis for the kernel here, the vector $(1,1,1)$.
But for completeness we carry out the full algorithm which says to further reduce to the right of the second column.
Let $q = \lfloor -2/1 \rfloor = -2$, we remove $q=-2$ times the second column from the third column to get
\begin{equation}
    \begin{pmatrix}
    0 & -3 & 6 \\
    0 & 1 & 0 \\
    0 & 0 & 3 \\
    \hline
    1 & 0 & 0 \\
    1 & -1 & -2 \\
    1 & -1 & -3
    \end{pmatrix}
\end{equation}
We now consider the first row and column, and the algorithm says to go to step 4.
Because the entry is already zero we do not have to do anything and the algorithm terminates. As previously mentioned, a basis for the kernel is given by the first column of the augmented matrix.
That is, the vector $(1,1,1)$.

\section{Data: Integer orthogonal bases for $\cA(\mu_1,\mu_2)$}\label{app:data Amu}

In this appendix we give examples of solutions to the eigensystems in section \ref{sec:EV Amn}, giving integer orthogonal bases for two-matrix invariants in $\mathcal{N}=4$ SYM.
\subsection{Half-BPS operators}\label{data:half BPS}

The multi-matrix invariants for $\cA(L,0)$ describe the half-BPS operators of $\cN=4$ SYM.
These operators are labelled by the Young diagram $R$ only, and the restricted Schur and covariant operators become identical.\footnote{The label $\Lambda$ is always given by the totally symmetric representation $[L]$.}
Following the methods in Section \ref{sec:EV Amn} we obtain the integer basis of operators of $\cA(L,0)$, which can be denoted by
\begin{equation}
{\sf O}^R [Z] = \tr_{V_N^{\otimes L}} \Bigl( \mathcal{L} ( \sfQ^R ) \, Z^{\otimes L} \Bigr) 
= \sum_{g \in T_p} \abs{ T_p } \, \tilde c^R (g) \, \cO_g [Z]
\label{app:ORZ bps}
\end{equation}
where $T_p$ is a sum over the conjugacy class of $S_L$ having the cycle type $p \vdash L$. We can view $\sfQ^R$ as the restricted Schur or covariant basis whose multiplicity labels are all trivial,
\begin{equation}
\sfQ^R \ \propto \ 
\begin{cases}
Q^{R , (r_1 , r_2)}_{\nu_+ \nu_-} \,, &\qquad (r_1,r_2, \nu_+, \nu_-)=(R, \emptyset, 1,1)
\\[2mm]
\cQ^{R, \Lambda, (L,0), \tau}_{\beta} \,, &\qquad (\Lambda, \tau, \beta) = ( [L] , 1, 1) .
\end{cases}
\end{equation}

Below we will present the explicit form of ${\sf O}^R [Z]$.
These data suggest that the coefficients in \eqref{app:ORZ bps} are equal to the $S_L$ characters,
\begin{equation}
\abs{ T_p } \, \tilde c^R (g) = \chi^R (T_p) .
\end{equation}
Since the normalised characters $\chi^R (T_p) / d_R$ are known to be integers \cite{BenGeloun:2020yau}, the coefficients have the common factor equal to the dimensions of the $S_L$ representations,
\begin{equation}
{\rm GCD}_{g \in T_p} \{ \tilde c^R (g) \} = d_R \,.
\end{equation}

\subsubsection{$\cA(2,0)$}
$
\begin{alignedat}{9}
R =& [2], &\qquad  
&\Tr(Z)\Tr(Z) + \Tr(Z^2) 
\\
R =& [1,1], &\qquad  
&\Tr(Z)\Tr(Z) - \Tr(Z^2)
\end{alignedat}
$

\subsubsection{$\cA(3,0)$}
$
\begin{alignedat}{9}
    R =& [3], &\qquad
    &\Tr(Z)^3 + 3\Tr(Z)\Tr(Z^2) + 2\Tr(Z^3)
\\
    R =& [2,1], &\qquad
    &2 \, \Bigl\{ \Tr(Z)^3 - \Tr(Z^3) \Bigr\}
\\
    R =& [1,1,1], &\qquad
    &\Tr(Z)^3 - 3\Tr(Z)\Tr(Z^2) + 2\Tr(Z^3)
\end{alignedat}
$

\subsubsection{$\cA(4,0)$}
\resizebox{\linewidth}{!}{%
$
\begin{alignedat}{9}
    R =& [4], &\qquad
    &\Tr(Z)^4+ 6\Tr(Z)^2\Tr(Z^2) + 8\Tr(Z)\Tr(Z^3) + 3\Tr(Z^2)\Tr(Z^2) + 6\Tr(Z^4)
\\
    R =& [3,1], &\qquad
    &3 \, \Bigl\{ \Tr(Z)^4 + 2\Tr(Z)^2\Tr(Z^2) - \Tr(Z^2)\Tr(Z^2) - 2\Tr(Z^4) \Bigr\}
\\
    R =& [2,2], &\qquad
    &2 \, \Bigl\{ \Tr(Z)^4 - 4\Tr(Z)\Tr(Z^3) + 3\Tr(Z^2)\Tr(Z^2) \Bigr\}
\\
    R =& [2,1,1], &\qquad
    &3 \, \Bigl\{ \Tr(Z)^4 - 2\Tr(Z)^2\Tr(Z^2) - \Tr(Z^2)\Tr(Z^2) + 2\Tr(Z^4) \Bigr\}
\\
    R =& [1,1,1,1], &\qquad
    &\Tr(Z)^4 - 6\Tr(Z)^2\Tr(Z^2) + 8\Tr(Z)\Tr(ZZZ) + 3\Tr(Z^2)\Tr(Z^2) - 6\Tr(Z^4)
\end{alignedat}
$}

\subsection{Restricted basis}\label{data:Res Schur}

We will present the operators which are proportional to the restricted Schur operators \eqref{def:res Schur op mu},
\begin{equation}
\begin{aligned}
{\sf O}^{R, r_1, r_2}_{A} (\vec a_{{\mu_1},{\mu_2}}) 
&= \tr_{V_N^{\otimes L}} \Bigl( \mathcal{L} ( \sfQ^{R, r_1, r_2}_{A} ) \, Z^{\otimes {\mu_1}} \otimes W^{\otimes {\mu_2}}  \Bigr) 
\\[1mm]
\sfQ^{R , r_1, r_2}_{A} 
&= \sum_{\nu_+, \nu_-=1}^{g(R;r_1,r_2)} \sfN^{R , (r_1 , r_2)}_{A, \nu_+ \nu_-} \, Q^{R , (r_1 , r_2)}_{\nu_+ \nu_-} \,.
\end{aligned}
\label{app:res Schur op mu}
\end{equation}
where $\sfQ^{R , r_1, r_2}_{A}$ is an integer eigenvector for the eigenvalue system of the restricted Schur basis \eqref{eq: A R r1 r2 kernel}, obtained according to the methods described in Section \ref{sec:integrality}.

\subsubsection{$\cA(1,1)$}
$
\begin{alignedat}{9}
    (R, r_1, r_2) &= ([2], [1], [1]) , &\qquad
    &\Tr(Z)\Tr(W) + \Tr(ZW)
\\
    (R, r_1, r_2) &= ([1,1], [1], [1]), &\qquad
    &\Tr(Z)\Tr(W) - \Tr(ZW)
\end{alignedat}
$

\subsubsection{$\cA(2,1)$}
\resizebox{\linewidth}{!}{%
$
\begin{aligned}
(R,r_1,r_2)&=([3],[2],[1]), &\qquad
    &\Tr(Z)^2 \Tr(W) + 2\Tr(Z)\Tr(ZW) + \Tr(Z^2)\Tr(W) + 2\Tr(Z^2W) 
    \\
(R,r_1,r_2)&=([2,1],[2],[1]),&\qquad
    &2 \, \Bigl\{ \Tr(Z)^2 \Tr(W) - \Tr(Z)\Tr(ZW) + \Tr(Z^2)\Tr(W) - \Tr(Z^2W) \Bigr\}
    \\
(R,r_1,r_2)&=([2,1],[1,1],[1]),&\qquad
    &2 \, \Bigl\{ \Tr(Z)^2 \Tr(W) + \Tr(Z)\Tr(ZW) - \Tr(Z^2)\Tr(W) - \Tr(Z^2W) \Bigr\}
     \\
(R,r_1,r_2)&=([1,1,1],[1,1],[1]),&\qquad
    &\Tr(Z)^2 \Tr(W) - 2\Tr(Z)\Tr(ZW) - \Tr(Z^2)\Tr(W) + 2\Tr(Z^2W)
\end{aligned}
$}

\subsubsection{$\cA(3,1)$}
\resizebox{\linewidth}{!}{%
$
\begin{aligned}
&(R,r_1,r_2)= ([4], [3], [1]), \\
&\quad 6 \Tr(WZZZ) + 6 \Tr(WZZ) \Tr(Z) + 3 \Tr(WZ) \Tr(Z)^2 + \Tr(W) \Tr(Z)^3 + 3 \Tr(WZ) \Tr(ZZ) \\
&\quad + 3 \Tr(W) \Tr(Z) \Tr(ZZ) + 2 \Tr(W) \Tr(ZZZ) \\
&(R,r_1,r_2)= ([3,1], [3], [1]), \\
&\quad 3 \, \Bigl\{ -2 \Tr(WZZZ) - 2 \Tr(WZZ) \Tr(Z) - \Tr(WZ) \Tr(Z)^2 + \Tr(W) \Tr(Z)^3 - \Tr(WZ) \Tr(ZZ) \\
&\quad + 3 \Tr(W) \Tr(Z) \Tr(ZZ) + 2 \Tr(W) \Tr(ZZZ) \Bigr\}
\\
&(R,r_1,r_2)= ([3,1], [2,1], [1]), \\
&\quad 6 \, \Bigl\{ -2 \Tr(WZZZ) + \Tr(WZZ) \Tr(Z) + 2 \Tr(WZ) \Tr(Z)^2 + \Tr(W) \Tr(Z)^3 - \Tr(WZ) \Tr(ZZ) \\
&\quad- \Tr(W) \Tr(ZZZ) \Bigr\}
\\
&(R,r_1,r_2)= ([2,2], [2,1], [1]), \\
&\quad 2 \, \Bigl\{ -3 \Tr(WZZ) \Tr(Z) + \Tr(W) \Tr(Z)^3 + 3 \Tr(WZ) \Tr(ZZ) - \Tr(W) \Tr(ZZZ) \Bigr\},   
\\
&(R,r_1,r_2)= ([2,1,1], [2,1], [1]), \\
&\quad 6 \, \Bigl\{ 2 \Tr(WZZZ) + \Tr(WZZ) \Tr(Z) - 2 \Tr(WZ) \Tr(Z)^2 + \Tr(W) \Tr(Z)^3 - \Tr(WZ) \Tr(ZZ) \\
&\quad- \Tr(W) \Tr(ZZZ) \Bigr\}
\\
&(R,r_1,r_2)= ([2,1,1], [1,1,1], [1]), \\
&\quad 3 \, \Bigl\{ 2 \Tr(WZZZ) - 2 \Tr(WZZ) \Tr(Z) + \Tr(WZ) \Tr(Z)^2 + \Tr(W) \Tr(Z)^3 - \Tr(WZ) \Tr(ZZ) \\
&\quad - 3 \Tr(W) \Tr(Z) \Tr(ZZ) + 2 \Tr(W) \Tr(ZZZ)\Bigr\}
\\
&(R,r_1,r_2)= ([1,1,1,1], [1,1,1], [1]), \\
&\quad -6 \Tr(WZZZ) + 6 \Tr(WZZ) \Tr(Z) - 3 \Tr(WZ) \Tr(Z)^2 + \Tr(W) \Tr(Z)^3 + 3 \Tr(WZ) \Tr(ZZ) \\
&\quad - 3 \Tr(W) \Tr(Z) \Tr(ZZ) + 2 \Tr(W) \Tr(ZZZ)
\end{aligned}
$}

\subsubsection{$\cA(2,2)$}
\resizebox{\linewidth}{!}{%
$
\begin{alignedat}{9}
&(R,r_1,r_2)= ([4], [2], [2]), \\
&\quad 4 \Tr(WWZZ)+ 2 \Tr(WZ)^2 + 2 \Tr(WZWZ)+ 4 \Tr(W)\Tr(WZZ)+ 4 \Tr(WWZ)\Tr(Z)
\\
&\quad + 4 \Tr(W)\Tr(WZ)\Tr(Z)+ \Tr(W)^2 \Tr(Z)^2 + \Tr(WW)\Tr(Z)^2 + \Tr(W)^2 \Tr(ZZ)+ \Tr(WW)\Tr(ZZ),
\\[1mm]
 &(R,r_1,r_2)= ([3,1], [2], [2]), \\
&\quad -2 \Tr(WZ)^2 - 2 \Tr(WZWZ)+ \Tr(W)^2 \Tr(Z)^2 + \Tr(WW)\Tr(Z)^2 + \Tr(W)^2 \Tr(ZZ)+ \Tr(WW)\Tr(ZZ),
\\[1mm]
 &(R,r_1,r_2)= ([3,1], [2], [1,1]), \\
&\quad 2 \, \Bigl\{ -2 \Tr(WWZZ)+ 2 \Tr(W)\Tr(WZZ)- 2 \Tr(WWZ)\Tr(Z)+ 2 \Tr(W)\Tr(WZ)\Tr(Z)+ \Tr(W)^2 \Tr(Z)^2 
\\
&\quad - \Tr(WW)\Tr(Z)^2 + \Tr(W)^2 \Tr(ZZ) - \Tr(WW)\Tr(ZZ) \Bigr\}, 
\\
 &(R,r_1,r_2)= ([3,1], [1,1], [2]), \\
&\quad  2 \, \Bigl\{ -2 \Tr(WWZZ)- 2 \Tr(W)\Tr(WZZ)+ 2 \Tr(WWZ)\Tr(Z)+ 2 \Tr(W)\Tr(WZ)\Tr(Z)+  \Tr(W)^2 \Tr(Z)^2 
\\
&\quad + \Tr(WW)\Tr(Z)^2 - \Tr(W)^2 \Tr(ZZ)- \Tr(WW)\Tr(ZZ) \Bigr\}, 
\\
 &(R,r_1,r_2)= ([2,2], [2], [2]), \\
&\quad 2 \, \Bigl\{ -2 \Tr(WWZZ)+ 2 \Tr(WZ)^2 + 2 \Tr(WZWZ)- 2 \Tr(W)\Tr(WZZ)- 2 \Tr(WWZ)\Tr(Z)
\\
&\quad - 2 \Tr(W)\Tr(WZ)\Tr(Z) + \Tr(W)^2 \Tr(Z)^2 + \Tr(WW)\Tr(Z)^2 + \Tr(W)^2 \Tr(ZZ) + \Tr(WW)\Tr(ZZ) \Bigr\},  
\\
 &(R,r_1,r_2)= ([2,2], [1,1], [1,1]), \\
&\quad 2 \, \Bigl\{ 2 \Tr(WWZZ)+ 2 \Tr(WZ)^2 - 2 \Tr(WZWZ)- 2 \Tr(W)\Tr(WZZ)- 2 \Tr(WWZ)\Tr(Z)
\\
&\quad + 2 \Tr(W)\Tr(WZ)\Tr(Z) + \Tr(W)^2 \Tr(Z)^2 - \Tr(WW)\Tr(Z)^2 - \Tr(W)^2 \Tr(ZZ)+ \Tr(WW)\Tr(ZZ) \Bigr\},   
\\
 &(R,r_1,r_2)= ([2,1,1], [2], [1,1]), \\
&\quad 2 \, \Bigl\{ 2 \Tr(WWZZ)- 2 \Tr(W)\Tr(WZZ)+ 2 \Tr(WWZ)\Tr(Z)- 2 \Tr(W)\Tr(WZ)\Tr(Z)+ \Tr(W)^2 \Tr(Z)^2 
\\
&\quad - \Tr(WW)\Tr(Z)^2 + \Tr(W)^2 \Tr(ZZ)- \Tr(WW)\Tr(ZZ) \Bigr\},  
\\
 &(R,r_1,r_2)= ([2,1,1], [1,1], [2]), \\
&\quad 2 \, \Bigl\{ 2 \Tr(WWZZ)+ 2 \Tr(W)\Tr(WZZ)- 2 \Tr(WWZ)\Tr(Z)- 2 \Tr(W)\Tr(WZ)\Tr(Z)+ \Tr(W)^2 \Tr(Z)^2 
\\
&\quad + \Tr(WW)\Tr(Z)^2 - \Tr(W)^2 \Tr(ZZ)- \Tr(WW)\Tr(ZZ) \Bigr\},   
\\
 &(R,r_1,r_2)= ([2,1,1], [1,1], [1,1]), \\
&\quad -2 \Tr(WZ)^2 + 2 \Tr(WZWZ)+ \Tr(W)^2 \Tr(Z)^2 - \Tr(WW)\Tr(Z)^2 - \Tr(W)^2 \Tr(ZZ)+ \Tr(WW)\Tr(ZZ), 
\\[1mm]
 &(R,r_1,r_2)= ([1,1,1,1], [1,1], [1,1]), \\
&\quad -4 \Tr(WWZZ)+ 2 \Tr(WZ)^2 - 2 \Tr(WZWZ)+ 4 \Tr(W)\Tr(WZZ)+ 4 \Tr(WWZ)\Tr(Z)
\\
&\quad - 4 \Tr(W)\Tr(WZ)\Tr(Z) + \Tr(W)^2 \Tr(Z)^2 - \Tr(WW)\Tr(Z)^2 - \Tr(W)^2 \Tr(ZZ)+ \Tr(WW)\Tr(ZZ)
\end{alignedat}
$}

\subsubsection{$\cA(4,1)$}
\resizebox{\linewidth}{!}{%
$
\begin{alignedat}{9}
 &(R,r_1,r_2)= ([5], [4], [1]), \\
&\quad   24 \Tr(WZZZZ) + 24 \Tr(WZZZ) \Tr(Z) + 12 \Tr(WZZ) \Tr(Z)^2 +   4 \Tr(WZ) \Tr(Z)^3 + \Tr(W) \Tr(Z)^4 \\
&\quad + 12 \Tr(WZZ) \Tr(ZZ) +   12 \Tr(WZ) \Tr(Z) \Tr(ZZ) + 6 \Tr(W) \Tr(Z)^2 \Tr(ZZ) +   3 \Tr(W) \Tr(ZZ)^2 \\
&\quad + 8 \Tr(WZ) \Tr(ZZZ) + 8 \Tr(W) \Tr(Z) \Tr(ZZZ) +   6 \Tr(W) \Tr(ZZZZ) 
\\
 &(R,r_1,r_2)= ([4,1], [4], [1]), \\
 &\quad 4 \, \Bigl\{ -6 \Tr(WZZZZ) -   6 \Tr(WZZZ) \Tr(Z) - 3 \Tr(WZZ) \Tr(Z)^2 - \Tr(WZ) \Tr(Z)^3 +   \Tr(W) \Tr(Z)^4 \\
&\quad - 3 \Tr(WZZ) \Tr(ZZ) - 3 \Tr(WZ) \Tr(Z) \Tr(ZZ) +   6 \Tr(W) \Tr(Z)^2 \Tr(ZZ) + 3 \Tr(W) \Tr(ZZ)^2 \\
&\quad - 2 \Tr(WZ) \Tr(ZZZ) +   8 \Tr(W) \Tr(Z) \Tr(ZZZ) + 6 \Tr(W) \Tr(ZZZZ) \Bigr\}
\\
 &(R,r_1,r_2)= ([4,1], [3,1],   [1]), \\
 &\quad 12 \, \Bigl\{-6 \Tr(WZZZZ) + 2 \Tr(WZZZ) \Tr(Z) + 5 \Tr(WZZ) \Tr(Z)^2 +   3 \Tr(WZ) \Tr(Z)^3 + \Tr(W) \Tr(Z)^4 \\
&\quad - 3 \Tr(WZZ) \Tr(ZZ) +   \Tr(WZ) \Tr(Z) \Tr(ZZ) + 2 \Tr(W) \Tr(Z)^2 \Tr(ZZ) - \Tr(W) \Tr(ZZ)^2 \\
&\quad -   2 \Tr(WZ) \Tr(ZZZ) - 2 \Tr(W) \Tr(ZZZZ) \Bigr\}
\\
 &(R,r_1,r_2)= ([3,2], [3,1],   [1]), \\
&\quad  6 \, \Bigl\{-4 \Tr(WZZZ) \Tr(Z) - 4 \Tr(WZZ) \Tr(Z)^2 + \Tr(W) \Tr(Z)^4 +   4 \Tr(WZ) \Tr(Z) \Tr(ZZ) \\
&\quad + 2 \Tr(W) \Tr(Z)^2 \Tr(ZZ) - \Tr(W) \Tr(ZZ)^2 +   4 \Tr(WZ) \Tr(ZZZ) - 2 \Tr(W) \Tr(ZZZZ) \Bigr\}
\\
 &(R,r_1,r_2)= ([3,2], [2,2],   [1]), \\
&\quad  4 \, \Bigl\{ -6 \Tr(WZZZ) \Tr(Z) + 2 \Tr(WZ) \Tr(Z)^3 + \Tr(W) \Tr(Z)^4 +   6 \Tr(WZZ) \Tr(ZZ) \\
&\quad + 3 \Tr(W) \Tr(ZZ)^2  - 2 \Tr(WZ) \Tr(ZZZ) -   4 \Tr(W) \Tr(Z) \Tr(ZZZ) \Bigr\}
\\
 &(R,r_1,r_2)= ([3,1,1], [3,1], [1]), \\
&\quad  12 \, \Bigl\{ 4 \Tr(WZZZZ) + 2 \Tr(WZZZ) \Tr(Z) - 2 \Tr(WZ) \Tr(Z)^3 + \Tr(W) \Tr(Z)^4 +   2 \Tr(WZZ) \Tr(ZZ) \\
&\quad - 4 \Tr(WZ) \Tr(Z) \Tr(ZZ) + 2 \Tr(W) \Tr(Z)^2 \Tr(ZZ) -    \Tr(W) \Tr(ZZ)^2 - 2 \Tr(WZ) \Tr(ZZZ) \\
&\quad -   2 \Tr(W) \Tr(ZZZZ) \Bigr\}
\\
 &(R,r_1,r_2)= ([3,1,1], [2,1,1], [1]), \\
&\quad  12 \, \Bigl\{ 4 \Tr(WZZZZ) - 2 \Tr(WZZZ) \Tr(Z) + 2 \Tr(WZ) \Tr(Z)^3 + \Tr(W) \Tr(Z)^4 -   2 \Tr(WZZ) \Tr(ZZ) \\
&\quad - 4 \Tr(WZ) \Tr(Z) \Tr(ZZ) - 2 \Tr(W) \Tr(Z)^2 \Tr(ZZ) -    \Tr(W) \Tr(ZZ)^2 + 2 \Tr(WZ) \Tr(ZZZ) \\
&\quad +   2 \Tr(W) \Tr(ZZZZ) \Bigr\}
\\
 &(R,r_1,r_2)= ([2,2,1], [2,2], [1]), \\
&\quad  4 \, \Bigl\{ 6 \Tr(WZZZ) \Tr(Z) - 2 \Tr(WZ) \Tr(Z)^3 + \Tr(W) \Tr(Z)^4 -   6 \Tr(WZZ) \Tr(ZZ) + 3 \Tr(W) \Tr(ZZ)^2 \\
&\quad + 2 \Tr(WZ) \Tr(ZZZ) -   4 \Tr(W) \Tr(Z) \Tr(ZZZ) \Bigr\}
\\
 &(R,r_1,r_2)= ([2,2,1], [2,1,1], [1]), \\
&\quad  6 \, \Bigl\{ 4 \Tr(WZZZ) \Tr(Z) - 4 \Tr(WZZ) \Tr(Z)^2 + \Tr(W) \Tr(Z)^4 +   4 \Tr(WZ) \Tr(Z) \Tr(ZZ) \\
&\quad- 2 \Tr(W) \Tr(Z)^2 \Tr(ZZ) - \Tr(W) \Tr(ZZ)^2 -   4 \Tr(WZ) \Tr(ZZZ) + 2 \Tr(W) \Tr(ZZZZ) \Bigr\}
\end{alignedat}
$}

\resizebox{\linewidth}{!}{%
$
\begin{alignedat}{9}
 &(R,r_1,r_2)= ([2,1,1,1], [2,1,1],   [1]), \\
&\quad 12 \, \Bigl\{ -6 \Tr(WZZZZ) - 2 \Tr(WZZZ) \Tr(Z) + 5 \Tr(WZZ) \Tr(Z)^2 -   3 \Tr(WZ) \Tr(Z)^3 + \Tr(W) \Tr(Z)^4 \\
&\quad + 3 \Tr(WZZ) \Tr(ZZ) +   \Tr(WZ) \Tr(Z) \Tr(ZZ) - 2 \Tr(W) \Tr(Z)^2 \Tr(ZZ) - \Tr(W) \Tr(ZZ)^2 \\
&\quad +  2 \Tr(WZ) \Tr(ZZZ) + 2 \Tr(W) \Tr(ZZZZ) \Bigr\}
\\
 &(R,r_1,r_2)= ([2,1,1,1], [1,1,1,1],   [1]), \\
&\quad  4 \, \Bigl\{ -6 \Tr(WZZZZ) + 6 \Tr(WZZZ) \Tr(Z) - 3 \Tr(WZZ) \Tr(Z)^2 +   \Tr(WZ) \Tr(Z)^3 + \Tr(W) \Tr(Z)^4 \\
&\quad + 3 \Tr(WZZ) \Tr(ZZ) -   3 \Tr(WZ) \Tr(Z) \Tr(ZZ) - 6 \Tr(W) \Tr(Z)^2 \Tr(ZZ) +   3 \Tr(W) \Tr(ZZ)^2 \\
&\quad + 2 \Tr(WZ) \Tr(ZZZ) + 8 \Tr(W) \Tr(Z) \Tr(ZZZ) -   6 \Tr(W) \Tr(ZZZZ) \Bigr\}
\\
 &(R,r_1,r_2)= ([1,1,1,1,1], [1,1,1,1], [1]), \\
&\quad   24 \Tr(WZZZZ) - 24 \Tr(WZZZ) \Tr(Z) + 12 \Tr(WZZ) \Tr(Z)^2 -   4 \Tr(WZ) \Tr(Z)^3 + \Tr(W) \Tr(Z)^4 \\
&\quad - 12 \Tr(WZZ) \Tr(ZZ) +   12 \Tr(WZ) \Tr(Z) \Tr(ZZ) - 6 \Tr(W) \Tr(Z)^2 \Tr(ZZ) +   3 \Tr(W) \Tr(ZZ)^2 \\
&\quad - 8 \Tr(WZ) \Tr(ZZZ) + 8 \Tr(W) \Tr(Z) \Tr(ZZZ) -   6 \Tr(W) \Tr(ZZZZ)
\end{alignedat}
$}

\subsubsection{$\cA(3,2)$}
\resizebox{\linewidth}{!}{%
$
\begin{alignedat}{9}
&(R,r_1,r_2)=([5], [3], [2]), \\
&\quad 12 \Tr(WWZZZ) + 12 \Tr(WZWZZ) + 12 \Tr(WZ) \Tr(WZZ) +   12 \Tr(W) \Tr(WZZZ) + 12 \Tr(WWZZ) \Tr(Z) \\
&\quad + 6 \Tr(WZ)^2 \Tr(Z) +   6 \Tr(WZWZ) \Tr(Z) + 12 \Tr(W) \Tr(WZZ) \Tr(Z) + 6 \Tr(WWZ) \Tr(Z)^2 \\
&\quad +   6 \Tr(W) \Tr(WZ) \Tr(Z)^2 + \Tr(W)^2 \Tr(Z)^3 + \Tr(WW) \Tr(Z)^3 +   6 \Tr(WWZ) \Tr(ZZ) + 6 \Tr(W) \Tr(WZ) \Tr(ZZ) \\
&\quad + 3 \Tr(W)^2 \Tr(Z) \Tr(ZZ) + 3 \Tr(WW) \Tr(Z) \Tr(ZZ) +   2 \Tr(W)^2 \Tr(ZZZ) + 2 \Tr(WW) \Tr(ZZZ) 
\\
&(R,r_1,r_2)=([4,1], [3], [2]), \\
&\quad 6 \, \Bigl\{   2 \Tr(WWZZZ) - 8 \Tr(WZWZZ) - 8 \Tr(WZ) \Tr(WZZ) + 2 \Tr(W) \Tr(WZZZ) +   2 \Tr(WWZZ) \Tr(Z) \\
&\quad - 4 \Tr(WZ)^2 \Tr(Z) - 4 \Tr(WZWZ) \Tr(Z) +   2 \Tr(W) \Tr(WZZ) \Tr(Z) + \Tr(WWZ) \Tr(Z)^2 + \Tr(W) \Tr(WZ) \Tr(Z)^2 \\
&\quad  + \Tr(W)^2 \Tr(Z)^3 + \Tr(WW) \Tr(Z)^3 + \Tr(WWZ) \Tr(ZZ) + \Tr(W) \Tr(WZ) \Tr(ZZ) + 3 \Tr(W)^2 \Tr(Z) \Tr(ZZ) \\
&\quad +   3 \Tr(WW) \Tr(Z) \Tr(ZZ) + 2 \Tr(W)^2 \Tr(ZZZ) +   2 \Tr(WW) \Tr(ZZZ) \Bigr\}
\\
&(R,r_1,r_2)=([4,1], [3], [1,1]), \\
&\quad  2 \, \Bigl\{  -6 \Tr(WWZZZ) + 6 \Tr(W) \Tr(WZZZ) - 6 \Tr(WWZZ) \Tr(Z) +   6 \Tr(W) \Tr(WZZ) \Tr(Z) - 3 \Tr(WWZ) \Tr(Z)^2 \\
&\quad +   3 \Tr(W) \Tr(WZ) \Tr(Z)^2 + \Tr(W)^2 \Tr(Z)^3 - \Tr(WW) \Tr(Z)^3 -   3 \Tr(WWZ) \Tr(ZZ) + 3 \Tr(W) \Tr(WZ) \Tr(ZZ) \\
&\quad +   3 \Tr(W)^2 \Tr(Z) \Tr(ZZ) - 3 \Tr(WW) \Tr(Z) \Tr(ZZ) +   2 \Tr(W)^2 \Tr(ZZZ) - 2 \Tr(WW) \Tr(ZZZ) \Bigr\}
\\
&(R,r_1,r_2)=([4,1], [2,1], [2]), \\
&\quad  6 \, \Bigl\{  -4 \Tr(WWZZZ) - 2 \Tr(WZWZZ) - 2 \Tr(WZ) \Tr(WZZ) -   4 \Tr(W) \Tr(WZZZ) + 2 \Tr(WWZZ) \Tr(Z) \\
&\quad + 2 \Tr(WZ)^2 \Tr(Z) +   2 \Tr(WZWZ) \Tr(Z) + 2 \Tr(W) \Tr(WZZ) \Tr(Z) + 4 \Tr(WWZ) \Tr(Z)^2  \\
&\quad +   4 \Tr(W) \Tr(WZ) \Tr(Z)^2 + \Tr(W)^2 \Tr(Z)^3 + \Tr(WW) \Tr(Z)^3 -   2 \Tr(WWZ) \Tr(ZZ) - 2 \Tr(W) \Tr(WZ) \Tr(ZZ)  \\
&\quad - \Tr(W)^2 \Tr(ZZZ) - \Tr(WW) \Tr(ZZZ) \Bigr\}
\end{alignedat}
$}

\resizebox{\linewidth}{!}{%
$
\begin{alignedat}{9}
&(R,r_1,r_2)=([3,2], [3], [2]), \\
&\quad  3 \, \Bigl\{  -4 \Tr(WWZZZ) + 4 \Tr(WZWZZ) + 4 \Tr(WZ) \Tr(WZZ) -   4 \Tr(W) \Tr(WZZZ) - 4 \Tr(WWZZ) \Tr(Z) \\
&\quad + 2 \Tr(WZ)^2 \Tr(Z) +   2 \Tr(WZWZ) \Tr(Z) - 4 \Tr(W) \Tr(WZZ) \Tr(Z) - 2 \Tr(WWZ) \Tr(Z)^2 -   2 \Tr(W) \Tr(WZ) \Tr(Z)^2 \\
&\quad + \Tr(W)^2 \Tr(Z)^3 + \Tr(WW) \Tr(Z)^3 -   2 \Tr(WWZ) \Tr(ZZ) - 2 \Tr(W) \Tr(WZ) \Tr(ZZ) +   3 \Tr(W)^2 \Tr(Z) \Tr(ZZ) \\
&\quad + 3 \Tr(WW) \Tr(Z) \Tr(ZZ) +   2 \Tr(W)^2 \Tr(ZZZ) + 2 \Tr(WW) \Tr(ZZZ) \Bigr\}
\\
&(R,r_1,r_2)=([3,2], [2,1], [2]), \\
&\quad  12 \, \Bigl\{  -\Tr(WWZZZ) + \Tr(WZWZZ) + \Tr(WZ) \Tr(WZZ) - \Tr(W) \Tr(WZZZ) -   4 \Tr(WWZZ) \Tr(Z) \\
&\quad  - \Tr(WZ)^2 \Tr(Z) - \Tr(WZWZ) \Tr(Z) -   4 \Tr(W) \Tr(WZZ) \Tr(Z) + \Tr(WWZ) \Tr(Z)^2 + \Tr(W) \Tr(WZ) \Tr(Z)^2  \\
&\quad +   \Tr(W)^2 \Tr(Z)^3 + \Tr(WW) \Tr(Z)^3 + 4 \Tr(WWZ) \Tr(ZZ) +   4 \Tr(W) \Tr(WZ) \Tr(ZZ) - \Tr(W)^2 \Tr(ZZZ) \\
&\quad -   \Tr(WW) \Tr(ZZZ) \Bigr\}
\\
&(R,r_1,r_2)=([3,2], [2,1], [1,1]), \\
&\quad 4 \, \Bigl\{   3 \Tr(WWZZZ) - 3 \Tr(WZWZZ) + 3 \Tr(WZ) \Tr(WZZ) - 3 \Tr(W) \Tr(WZZZ) +   3 \Tr(WZ)^2 \Tr(Z) \\
&\quad - 3 \Tr(WZWZ) \Tr(Z) - 3 \Tr(WWZ) \Tr(Z)^2 +   3 \Tr(W) \Tr(WZ) \Tr(Z)^2 + \Tr(W)^2 \Tr(Z)^3 - \Tr(WW) \Tr(Z)^3 \\
&\quad - \Tr(W)^2 \Tr(ZZZ) + \Tr(WW) \Tr(ZZZ) \Bigr\}
\\
&(R,r_1,r_2)=([3,1,1], [3], [1,1]), \\
&\quad 3 \, \Bigl\{   4 \Tr(WWZZZ) - 4 \Tr(W) \Tr(WZZZ) + 4 \Tr(WWZZ) \Tr(Z) -   4 \Tr(W) \Tr(WZZ) \Tr(Z) + 2 \Tr(WWZ) \Tr(Z)^2 \\
&\quad -   2 \Tr(W) \Tr(WZ) \Tr(Z)^2 + \Tr(W)^2 \Tr(Z)^3 - \Tr(WW) \Tr(Z)^3 +   2 \Tr(WWZ) \Tr(ZZ) - 2 \Tr(W) \Tr(WZ) \Tr(ZZ) \\
&\quad +   3 \Tr(W)^2 \Tr(Z) \Tr(ZZ) - 3 \Tr(WW) \Tr(Z) \Tr(ZZ) +   2 \Tr(W)^2 \Tr(ZZZ) - 2 \Tr(WW) \Tr(ZZZ) \Bigr\}
\\
&(R,r_1,r_2)=([3,1,1], [2,1],   [2]), \\
&\quad 12 \, \Bigl\{   \Tr(WWZZZ) + 3 \Tr(WZWZZ) + 3 \Tr(WZ) \Tr(WZZ) + \Tr(W) \Tr(WZZZ) +   2 \Tr(WWZZ) \Tr(Z)  \\
&\quad - 3 \Tr(WZ)^2 \Tr(Z) - 3 \Tr(WZWZ) \Tr(Z) +   2 \Tr(W) \Tr(WZZ) \Tr(Z) - \Tr(WWZ) \Tr(Z)^2 - \Tr(W) \Tr(WZ) \Tr(Z)^2  \\
&\quad +   \Tr(W)^2 \Tr(Z)^3 + \Tr(WW) \Tr(Z)^3 - 2 \Tr(WWZ) \Tr(ZZ) -   2 \Tr(W) \Tr(WZ) \Tr(ZZ) - \Tr(W)^2 \Tr(ZZZ) \\
&\quad -  \Tr(WW) \Tr(ZZZ) \Bigr\}
\\
&(R,r_1,r_2)=([3,1,1], [2,1], [1,1]), \\
&\quad 12 \, \Bigl\{   \Tr(WWZZZ) + 3 \Tr(WZWZZ) - 3 \Tr(WZ) \Tr(WZZ) - \Tr(W) \Tr(WZZZ) -   2 \Tr(WWZZ) \Tr(Z) \\
&\quad - 3 \Tr(WZ)^2 \Tr(Z) + 3 \Tr(WZWZ) \Tr(Z) +   2 \Tr(W) \Tr(WZZ) \Tr(Z) - \Tr(WWZ) \Tr(Z)^2 + \Tr(W) \Tr(WZ) \Tr(Z)^2 \\
&\quad +   \Tr(W)^2 \Tr(Z)^3 - \Tr(WW) \Tr(Z)^3 + 2 \Tr(WWZ) \Tr(ZZ) -   2 \Tr(W) \Tr(WZ) \Tr(ZZ) - \Tr(W)^2 \Tr(ZZZ) \\
&\quad +   \Tr(WW) \Tr(ZZZ) \Bigr\}
\\
&(R,r_1,r_2)=([3,1,1], [1,1,1], [2]), \\
&\quad 3 \, \Bigl\{   4 \Tr(WWZZZ) + 4 \Tr(W) \Tr(WZZZ) - 4 \Tr(WWZZ) \Tr(Z) -   4 \Tr(W) \Tr(WZZ) \Tr(Z) + 2 \Tr(WWZ) \Tr(Z)^2 \\
&\quad  +   2 \Tr(W) \Tr(WZ) \Tr(Z)^2 + \Tr(W)^2 \Tr(Z)^3 + \Tr(WW) \Tr(Z)^3 -   2 \Tr(WWZ) \Tr(ZZ) - 2 \Tr(W) \Tr(WZ) \Tr(ZZ) \\
&\quad  -   3 \Tr(W)^2 \Tr(Z) \Tr(ZZ) - 3 \Tr(WW) \Tr(Z) \Tr(ZZ) +   2 \Tr(W)^2 \Tr(ZZZ) + 2 \Tr(WW) \Tr(ZZZ) \Bigr\}
\\
&(R,r_1,r_2)=([2,2,1], [2,1],   [2]), \\
&\quad 4 \, \Bigl\{   3 \Tr(WWZZZ) - 3 \Tr(WZWZZ) - 3 \Tr(WZ) \Tr(WZZ) + 3 \Tr(W) \Tr(WZZZ) +   3 \Tr(WZ)^2 \Tr(Z) \\
&\quad  + 3 \Tr(WZWZ) \Tr(Z) - 3 \Tr(WWZ) \Tr(Z)^2 -   3 \Tr(W) \Tr(WZ) \Tr(Z)^2 + \Tr(W)^2 \Tr(Z)^3 + \Tr(WW) \Tr(Z)^3 \\
&\quad  -   \Tr(W)^2 \Tr(ZZZ) - \Tr(WW) \Tr(ZZZ) \Bigr\}
\end{alignedat}
$}

\resizebox{\linewidth}{!}{%
$
\begin{alignedat}{9}
&(R,r_1,r_2)=([2,2,1], [2,1], [1,1]), \\
&\quad  12 \, \Bigl\{  -\Tr(WWZZZ) + \Tr(WZWZZ) - \Tr(WZ) \Tr(WZZ) + \Tr(W) \Tr(WZZZ) +   4 \Tr(WWZZ) \Tr(Z) \\
&\quad - \Tr(WZ)^2 \Tr(Z) + \Tr(WZWZ) \Tr(Z) -   4 \Tr(W) \Tr(WZZ) \Tr(Z) + \Tr(WWZ) \Tr(Z)^2 - \Tr(W) \Tr(WZ) \Tr(Z)^2 \\
&\quad +   \Tr(W)^2 \Tr(Z)^3 - \Tr(WW) \Tr(Z)^3 - 4 \Tr(WWZ) \Tr(ZZ) +   4 \Tr(W) \Tr(WZ) \Tr(ZZ) - \Tr(W)^2 \Tr(ZZZ) \\
&\quad +   \Tr(WW) \Tr(ZZZ) \Bigr\}
\\
&(R,r_1,r_2)=([2,2,1], [1,1,1], [1,1]), \\
&\quad  3 \, \Bigl\{  -4 \Tr(WWZZZ) + 4 \Tr(WZWZZ) - 4 \Tr(WZ) \Tr(WZZ) +   4 \Tr(W) \Tr(WZZZ) + 4 \Tr(WWZZ) \Tr(Z) \\
&\quad + 2 \Tr(WZ)^2 \Tr(Z) -   2 \Tr(WZWZ) \Tr(Z) - 4 \Tr(W) \Tr(WZZ) \Tr(Z) - 2 \Tr(WWZ) \Tr(Z)^2 +   2 \Tr(W) \Tr(WZ) \Tr(Z)^2 \\
&\quad + \Tr(W)^2 \Tr(Z)^3 - \Tr(WW) \Tr(Z)^3 +   2 \Tr(WWZ) \Tr(ZZ) - 2 \Tr(W) \Tr(WZ) \Tr(ZZ) -   3 \Tr(W)^2 \Tr(Z) \Tr(ZZ) \\
&\quad + 3 \Tr(WW) \Tr(Z) \Tr(ZZ) +   2 \Tr(W)^2 \Tr(ZZZ) - 2 \Tr(WW) \Tr(ZZZ) \Bigr\}
\\
&(R,r_1,r_2)=([2,1,1,1], [2,1],   [1,1]), \\
&\quad  6 \, \Bigl\{  -4 \Tr(WWZZZ) - 2 \Tr(WZWZZ) + 2 \Tr(WZ) \Tr(WZZ) +   4 \Tr(W) \Tr(WZZZ) - 2 \Tr(WWZZ) \Tr(Z) \\
&\quad + 2 \Tr(WZ)^2 \Tr(Z) -   2 \Tr(WZWZ) \Tr(Z) + 2 \Tr(W) \Tr(WZZ) \Tr(Z) + 4 \Tr(WWZ) \Tr(Z)^2 -   4 \Tr(W) \Tr(WZ) \Tr(Z)^2 \\
&\quad + \Tr(W)^2 \Tr(Z)^3 - \Tr(WW) \Tr(Z)^3 +   2 \Tr(WWZ) \Tr(ZZ) - 2 \Tr(W) \Tr(WZ) \Tr(ZZ) - \Tr(W)^2 \Tr(ZZZ) \\
&\quad +   \Tr(WW) \Tr(ZZZ) \Bigr\}
\\
&(R,r_1,r_2)=([2,1,1,1], [1,1,1], [2]), \\
&\quad  2 \, \Bigl\{  -6 \Tr(WWZZZ) - 6 \Tr(W) \Tr(WZZZ) + 6 \Tr(WWZZ) \Tr(Z) +   6 \Tr(W) \Tr(WZZ) \Tr(Z) - 3 \Tr(WWZ) \Tr(Z)^2 \\
&\quad -   3 \Tr(W) \Tr(WZ) \Tr(Z)^2 + \Tr(W)^2 \Tr(Z)^3 + \Tr(WW) \Tr(Z)^3 +   3 \Tr(WWZ) \Tr(ZZ) + 3 \Tr(W) \Tr(WZ) \Tr(ZZ) \\
&\quad -   3 \Tr(W)^2 \Tr(Z) \Tr(ZZ) - 3 \Tr(WW) \Tr(Z) \Tr(ZZ) +   2 \Tr(W)^2 \Tr(ZZZ) + 2 \Tr(WW) \Tr(ZZZ) \Bigr\}
\\
&(R,r_1,r_2)=([2,1,1,1], [1,1,1],   [1,1]), \\
&\quad 6 \, \Bigl\{   2 \Tr(WWZZZ) - 8 \Tr(WZWZZ) + 8 \Tr(WZ) \Tr(WZZ) - 2 \Tr(W) \Tr(WZZZ) -   2 \Tr(WWZZ) \Tr(Z) \\
&\quad - 4 \Tr(WZ)^2 \Tr(Z) + 4 \Tr(WZWZ) \Tr(Z) +   2 \Tr(W) \Tr(WZZ) \Tr(Z) + \Tr(WWZ) \Tr(Z)^2 - \Tr(W) \Tr(WZ) \Tr(Z)^2 \\
&\quad +   \Tr(W)^2 \Tr(Z)^3 - \Tr(WW) \Tr(Z)^3 - \Tr(WWZ) \Tr(ZZ) +   \Tr(W) \Tr(WZ) \Tr(ZZ) - 3 \Tr(W)^2 \Tr(Z) \Tr(ZZ) \\
&\quad +   3 \Tr(WW) \Tr(Z) \Tr(ZZ) + 2 \Tr(W)^2 \Tr(ZZZ) -   2 \Tr(WW) \Tr(ZZZ) \Bigr\}
\\
&(R,r_1,r_2)=([1,1,1,1,1], [1,1,1], [1,1]), \\
&\quad 12 \Tr(WWZZZ) + 12 \Tr(WZWZZ) - 12 \Tr(WZ) \Tr(WZZ) -   12 \Tr(W) \Tr(WZZZ) - 12 \Tr(WWZZ) \Tr(Z) \\
&\quad + 6 \Tr(WZ)^2 \Tr(Z) -   6 \Tr(WZWZ) \Tr(Z) + 12 \Tr(W) \Tr(WZZ) \Tr(Z) + 6 \Tr(WWZ) \Tr(Z)^2  \\
&\quad -  6 \Tr(W) \Tr(WZ) \Tr(Z)^2 + \Tr(W)^2 \Tr(Z)^3 - \Tr(WW) \Tr(Z)^3 -   6 \Tr(WWZ) \Tr(ZZ) + 6 \Tr(W) \Tr(WZ) \Tr(ZZ)  \\
&\quad -  3 \Tr(W)^2 \Tr(Z) \Tr(ZZ) + 3 \Tr(WW) \Tr(Z) \Tr(ZZ) +   2 \Tr(W)^2 \Tr(ZZZ) - 2 \Tr(WW) \Tr(ZZZ) 
\end{alignedat}
$}

\subsubsection{$\cA(3,3)$}

This is the first example of the restricted Schur basis with a non-trivial Littlewood-Richardson coefficient
\begin{equation}
g( [2,1], [2,1] ; [3,2,1])=2.
\end{equation}
Below we show an integer basis of the four states with $(R,r_1,r_2)=([3,2,1], [2,1], [2,1])$, which are orthogonal with respect to the $\delta$-function inner product.

\resizebox{\linewidth}{!}{%
$
\begin{aligned}
&(R,r_1,r_2)=([3,2,1], [2,1], [2,1]), \\
&\begin{aligned}
&12 \left\{ \quad  \Tr(Z)^3 \Tr(W)^3 - \Tr(Z)^3\Tr(WWW) - 3 \Tr(Z)\Tr(ZZW)\Tr(W)^2 \right. \\
&\quad + 3\Tr(Z)\Tr(ZZWWW) - 3\Tr(Z)\Tr(ZW)^2\Tr(W) + 3\Tr(Z)\Tr(ZWZWW) \\
&\quad + 3\Tr(ZZ)\Tr(ZW)\Tr(W)^2 - 3\Tr(ZZ)\Tr(ZWWW) - \Tr(ZZZ)\Tr(W)^3 \\
&\quad + \Tr(ZZZ)\Tr(WWW)- 6\Tr(ZZW)\Tr(ZWW) + 3\Tr(ZZWZW)\Tr(W) \\
&\left.  \quad + 3\Tr(ZZWW)\Tr(ZW)\right\}
\end{aligned}
\\
&(R,r_1,r_2)=([3,2,1], [2,1], [2,1]), \\
&\begin{aligned}
&12\left\{ \quad \Tr(Z)^3 \Tr(W)^3- \Tr(Z)^3 \Tr(WWW) + 9\Tr(Z)^2 \Tr(ZW) \Tr(WW)\right .  \\
&\quad - 9\Tr(Z)^2 \Tr(ZWW)\Tr(W) + 6\Tr(Z)\Tr(ZZW)\Tr(W)^2 - 6\Tr(Z)\Tr(ZZWWW) \\
&\quad - 3\Tr(Z)\Tr(ZW)^2\Tr(W) + 3\Tr(Z)\Tr(ZWZWW) - 6\Tr(ZZ)\Tr(ZW)\Tr(W)^2 \\
&\quad + 6\Tr(ZZ)\Tr(ZWWW) - \Tr(ZZZ)\Tr(W)^3 + \Tr(ZZZ)\Tr(WWW) \\
&\quad - 9\Tr(ZZZW)\Tr(WW) + 9\Tr(ZZZWW)\Tr(W) - 6\Tr(ZZW)\Tr(ZWW) \\
&\left . \quad + 3\Tr(ZZWZW)\Tr(W) + 3\Tr(ZZWW)\Tr(ZW)\right\}
\end{aligned}
\\
&(R,r_1,r_2)=([3,2,1], [2,1], [2,1]), \\
&\begin{aligned}
&36\left\{ \quad   \Tr(Z)\Tr(ZZW)\Tr(WW) -  \Tr(Z)\Tr(ZZWW)\Tr(W)\right .  \\
&\quad -  \Tr(Z)\Tr(ZW)\Tr(ZWW) +  \Tr(Z)\Tr(ZWZW)\Tr(W) \\
&\quad -  \Tr(ZZ)\Tr(ZW)\Tr(WW) +  \Tr(ZZ)\Tr(ZWW)\Tr(W) \\
&\quad -  \Tr(ZZW)\Tr(ZW)\Tr(W) +  \Tr(ZZWWZW) \\
&\left . \quad +  \Tr(ZW)^3 -  \Tr(ZWZWZW)\right\}
\end{aligned}
\\
&(R,r_1,r_2)=([3,2,1], [2,1], [2,1]), \\
& \begin{aligned}
&36\left\{ \quad \Tr(Z)\Tr(ZZW)\Tr(WW) - \Tr(Z)\Tr(ZZWW)\Tr(W) - \Tr(Z)\Tr(ZW)\Tr(ZWW)\right .  \\
&\quad + \Tr(Z)\Tr(ZWZW)\Tr(W) - \Tr(ZZ)\Tr(ZW)\Tr(WW) + \Tr(ZZ)\Tr(ZWW)\Tr(W) \\
&\quad - \Tr(ZZW)\Tr(ZW)\Tr(W) + 23\Tr(ZZWZWW) - 22\Tr(ZZWWZW) \\
&\left . \quad + \Tr(ZW)^3 - \Tr(ZWZWZW) \right\}
\end{aligned} 
\end{aligned}
$}

\subsection{Covariant basis}\label{data:cov op}

We will present the operators which are proportional to the covariant  operators \eqref{def:cov op mu},
\begin{equation}
\begin{aligned}
{\sf O}^{R, \Lambda, \mu, \tau}_{\beta}(\vec a_\mu) 
&= \tr_{V_N^{\otimes L}} \Bigl( \mathcal{L} ( \sfQ^{R, \Lambda}_{A} ) \, Z^{\otimes {\mu_1}} \otimes W^{\otimes {\mu_2}}  \Bigr) 
\\[1mm]
\sfQ^{R, \Lambda}_{A}
&= \sum_{\tau=1}^{C(R,R,\Lambda)} \sum_{\beta = 1}^{K_{\Lambda, ({\mu_1},{\mu_2})} }
\sfN^{R, \Lambda, \tau}_{A, \beta} \, \cQ^{R, \Lambda, ({\mu_1},{\mu_2}), \tau}_{\beta} \,.
\end{aligned}
\label{app:cov op mu}
\end{equation}
where $\sfQ^{R, \Lambda}_{A}$ is an integer eigenvector for the eigenvalue system of the covariant basis \eqref{eq: A R Lambda kernel}, obtained according to the methods described in Section \ref{sec:integrality}. The index $\beta$ may be omitted because $K_{\Lambda \mu}=0$ or $1$ for $\mu=({\mu_1},{\mu_2})$.

\subsubsection{$\cA(1,1)$}
$
\begin{alignedat}{9}
    (R, \Lambda) &= ([2], [2]) , &\qquad
    &\Tr(Z)\Tr(W) + \Tr(ZW)
\\
    (R, \Lambda) &= ([1,1], [2]), &\qquad
    &\Tr(Z)\Tr(W) - \Tr(ZW)
\end{alignedat}
$

\subsubsection{A(2,1)}

\resizebox{\linewidth}{!}{%
$
\begin{alignedat}{9}
(R, \Lambda)&=([3],[3]), &\qquad
        &\Tr(Z)^2\Tr(W) + 2\Tr(Z)\Tr(ZW) + \Tr(Z^2)\Tr(W) + 2\Tr(Z^2W)
    \\
(R, \Lambda)&=([2,1],[3]),&\qquad
    &2 \, \Bigl\{  \Tr(Z)^2\Tr(W) - \Tr(Z^2W) \Bigr\}
    \\
(R, \Lambda)&=([2,1],[2,1]),&\qquad
    &2 \, \Bigl\{ \Tr(Z)\Tr(ZW) - \Tr(Z^2)\Tr(W) \Bigr\}
     \\
(R, \Lambda)&=([1,1,1],[3]),&\qquad
    &\Tr(Z)^2\Tr(W) - 2\Tr(Z)\Tr(ZW) -\Tr(Z^2)\Tr(W) + 2\Tr(Z^2W)
\end{alignedat}
$}

\subsubsection{A(3,1)}
\resizebox{\linewidth}{!}{%
$
\begin{alignedat}{9}
&(R,\Lambda)= ([4], [4]), \\
&\quad 6 \Tr(WZZZ) + 6 \Tr(WZZ) \Tr(Z) + 3 \Tr(WZ) \Tr(Z)^2 + \Tr(W) \Tr(Z)^3 + 3 \Tr(WZ) \Tr(ZZ) \\
&\quad + 3 \Tr(W) \Tr(Z) \Tr(ZZ) + 2 \Tr(W) \Tr(ZZZ)  \\
&(R,\Lambda)= ([3,1], [4]), \\
&\quad 3  \, \Bigl\{ -2 \Tr(WZZZ) + \Tr(WZ) \Tr(Z)^2 + \Tr(W) \Tr(Z)^3 - \Tr(WZ) \Tr(ZZ) + \Tr(W) \Tr(Z) \Tr(ZZ) \Bigr\} \\
&(R,\Lambda)= ([3,1], [3,1]), \\
&\quad 6 \, \Bigl\{ \Tr(WZZ) \Tr(Z) + \Tr(WZ) \Tr(Z)^2 - \Tr(W) \Tr(Z) \Tr(ZZ) - \Tr(W) \Tr(ZZZ) \Bigr\} \\
&(R,\Lambda)= ([2,2], [4]), \\
&\quad 2 \, \Bigl\{ -3 \Tr(WZZ) \Tr(Z) + \Tr(W) \Tr(Z)^3 + 3 \Tr(WZ) \Tr(ZZ) - \Tr(W) \Tr(ZZZ) \Bigr\} \\
&(R,\Lambda)= ([2,1,1], [4]), \\
&\quad 3 \, \Bigl\{ 2 \Tr(WZZZ) - \Tr(WZ) \Tr(Z)^2 + \Tr(W) \Tr(Z)^3 - \Tr(WZ) \Tr(ZZ) - \Tr(W) \Tr(Z) \Tr(ZZ) \Bigr\} \\
&(R,\Lambda)= ([2,1,1], [3,1]), \\
&\quad 6 \, \Bigl\{ -\Tr(WZZ) \Tr(Z) + \Tr(WZ) \Tr(Z)^2 - \Tr(W) \Tr(Z) \Tr(ZZ) + \Tr(W) \Tr(ZZZ) \Bigr\} \\
&(R,\Lambda)= ([1,1,1,1], [4]), \\
&\quad -6 \Tr(WZZZ) + 6 \Tr(WZZ) \Tr(Z) - 3 \Tr(WZ) \Tr(Z)^2 + \Tr(W) \Tr(Z)^3 \\
&\quad + 3 \Tr(WZ) \Tr(ZZ) - 3 \Tr(W) \Tr(Z) \Tr(ZZ) + 2 \Tr(W) \Tr(ZZZ)
\end{alignedat}
$}
\subsubsection{A(2,2)}
\resizebox{\linewidth}{!}{%
$
\begin{alignedat}{9}
&(R,\Lambda)= ([4], [4]), \\
&\quad 4 \Tr(WWZZ) + 2 \Tr(WZ)^2 + 2 \Tr(WZWZ) + 4 \Tr(W) \Tr(WZZ) + 4 \Tr(WWZ) \Tr(Z) \\
&\quad + 4 \Tr(W) \Tr(WZ) \Tr(Z) + \Tr(W)^2 \Tr(Z)^2 + \Tr(WW) \Tr(Z)^2 + \Tr(W)^2 \Tr(ZZ) + \Tr(WW) \Tr(ZZ) \\
&(R,\Lambda)= ([3,1], [4]), \\
&\quad -4 \Tr(WWZZ) - 2 \Tr(WZ)^2 - 2 \Tr(WZWZ) + 4 \Tr(W) \Tr(WZ) \Tr(Z) + 3 \Tr(W)^2 \Tr(Z)^2 \\
&\quad + \Tr(WW) \Tr(Z)^2 + \Tr(W)^2 \Tr(ZZ) - \Tr(WW) \Tr(ZZ)  \\
&(R,\Lambda)= ([3,1], [3,1]), \\
&\quad 2 \, \Bigl\{ -2 \Tr(W) \Tr(WZZ) + 2 \Tr(WWZ) \Tr(Z) + \Tr(WW) \Tr(Z)^2 - \Tr(W)^2 \Tr(ZZ) \Bigr\} \\
&(R,\Lambda)= ([3,1], [2,2]), \\
&\quad 2 \, \Bigl\{ 2 \Tr(WWZZ) - 2 \Tr(WZ)^2 - 2 \Tr(WZWZ) - 2 \Tr(W) \Tr(WZ) \Tr(Z) + \Tr(WW) \Tr(Z)^2 \\
&\quad + \Tr(W)^2 \Tr(ZZ) + 2 \Tr(WW) \Tr(ZZ) \Bigr\} \\
&(R,\Lambda)= ([2,2], [4]), \\
&\quad 2 \, \Bigl\{ 2 \Tr(WZ)^2 - 2 \Tr(W) \Tr(WZZ) - 2 \Tr(WWZ) \Tr(Z) + \Tr(W)^2 \Tr(Z)^2 + \Tr(WW) \Tr(ZZ) \Bigr\} \\
&(R,\Lambda)= ([2,2], [2,2]), \\
&\quad 2 \, \Bigl\{ -2 \Tr(WWZZ) + 2 \Tr(WZWZ) - 2 \Tr(W) \Tr(WZ) \Tr(Z) + \Tr(WW) \Tr(Z)^2 + \Tr(W)^2 \Tr(ZZ) \Bigr\} \\
&(R,\Lambda)= ([2,1,1], [4]), \\
&\quad  4 \Tr(WWZZ) - 2 \Tr(WZ)^2 + 2 \Tr(WZWZ) - 4 \Tr(W) \Tr(WZ) \Tr(Z) + 3 \Tr(W)^2 \Tr(Z)^2 \\
&\quad - \Tr(WW) \Tr(Z)^2 - \Tr(W)^2 \Tr(ZZ) - \Tr(WW) \Tr(ZZ)  \\
&(R,\Lambda)= ([2,1,1], [3,1]), \\
&\quad 2 \, \Bigl\{ 2 \Tr(W) \Tr(WZZ) - 2 \Tr(WWZ) \Tr(Z) + \Tr(WW) \Tr(Z)^2 - \Tr(W)^2 \Tr(ZZ) \Bigr\} \\
&(R,\Lambda)= ([2,1,1], [2,2]), \\
&\quad 2 \, \Bigl\{ 2 \Tr(WWZZ) + 2 \Tr(WZ)^2 - 2 \Tr(WZWZ) - 2 \Tr(W) \Tr(WZ) \Tr(Z) + \Tr(WW) \Tr(Z)^2 \\
&\quad + \Tr(W)^2 \Tr(ZZ) - 2 \Tr(WW) \Tr(ZZ) \Bigr\} \\
&(R,\Lambda)= ([1,1,1,1], [4]), \\
&\quad -4 \Tr(WWZZ) + 2 \Tr(WZ)^2 - 2 \Tr(WZWZ) + 4 \Tr(W) \Tr(WZZ) + 4 \Tr(WWZ) \Tr(Z) \\
&\quad - 4 \Tr(W) \Tr(WZ) \Tr(Z) + \Tr(W)^2 \Tr(Z)^2 - \Tr(WW) \Tr(Z)^2 - \Tr(W)^2 \Tr(ZZ) + \Tr(WW) \Tr(ZZ)
\end{alignedat}
$}

\subsubsection{A(4,1)}
\resizebox{\linewidth}{!}{%
$
\begin{alignedat}{9}
&(R,\Lambda)= ([5], [5]), \\
&\quad 24 \Tr(WZZZZ) + 24 \Tr(WZZZ) \Tr(Z) + 12 \Tr(WZZ) \Tr(Z)^2 + 4 \Tr(WZ) \Tr(Z)^3 + \Tr(W) \Tr(Z)^4 \\
&\quad + 12 \Tr(WZZ) \Tr(ZZ) + 12 \Tr(WZ) \Tr(Z) \Tr(ZZ) + 6 \Tr(W) \Tr(Z)^2 \Tr(ZZ) + 3 \Tr(W) \Tr(ZZ)^2 \\
&\quad + 8 \Tr(WZ) \Tr(ZZZ) + 8 \Tr(W) \Tr(Z) \Tr(ZZZ) + 6 \Tr(W) \Tr(ZZZZ) \\
&(R,\Lambda)= ([4,1], [5]), \\
&\quad 4 \, \Bigl\{ -6 \Tr(WZZZZ) + 3 \Tr(WZZ) \Tr(Z)^2 + 2 \Tr(WZ) \Tr(Z)^3 + \Tr(W) \Tr(Z)^4 - 3 \Tr(WZZ) \Tr(ZZ) \\
&\quad + 3 \Tr(W) \Tr(Z)^2 \Tr(ZZ) - 2 \Tr(WZ) \Tr(ZZZ) + 2 \Tr(W) \Tr(Z) \Tr(ZZZ) \Bigr\} \\
&(R,\Lambda)= ([4,1], [4,1]), \\
&\quad 12 \, \Bigl\{ 2 \Tr(WZZZ) \Tr(Z) + 2 \Tr(WZZ) \Tr(Z)^2 + \Tr(WZ) \Tr(Z)^3 + \Tr(WZ) \Tr(Z) \Tr(ZZ) \\
&\quad - \Tr(W) \Tr(Z)^2 \Tr(ZZ) - \Tr(W) \Tr(ZZ)^2 - 2 \Tr(W) \Tr(Z) \Tr(ZZZ) - 2 \Tr(W) \Tr(ZZZZ) \Bigr\} \\
&(R,\Lambda)= ([3,2], [5]), \\
&\quad -24 \Tr(WZZZ) \Tr(Z) - 12 \Tr(WZZ) \Tr(Z)^2 + 4 \Tr(WZ) \Tr(Z)^3 + 5 \Tr(W) \Tr(Z)^4 \\
&\quad + 12 \Tr(WZZ) \Tr(ZZ) + 12 \Tr(WZ) \Tr(Z) \Tr(ZZ) + 6 \Tr(W) \Tr(Z)^2 \Tr(ZZ) + 3 \Tr(W) \Tr(ZZ)^2 \\
&\quad + 8 \Tr(WZ) \Tr(ZZZ) - 8 \Tr(W) \Tr(Z) \Tr(ZZZ) - 6 \Tr(W) \Tr(ZZZZ) \\
&(R,\Lambda)= ([3,2], [4,1]), \\
&\quad 24 \, \Bigl\{ -\Tr(WZZZ) \Tr(Z) + 2 \Tr(WZZ) \Tr(Z)^2 + \Tr(WZ) \Tr(Z)^3 + 3 \Tr(WZZ) \Tr(ZZ) \\
&\quad - 2 \Tr(WZ) \Tr(Z) \Tr(ZZ) - \Tr(W) \Tr(Z)^2 \Tr(ZZ) + 2 \Tr(W) \Tr(ZZ)^2 - 3 \Tr(WZ) \Tr(ZZZ) \\
&\quad - 2 \Tr(W) \Tr(Z) \Tr(ZZZ) + \Tr(W) \Tr(ZZZZ) \Bigr\} \\
&(R,\Lambda)= ([3,1,1], [5]), \\
&\quad 6 \, \Bigl\{ 4 \Tr(WZZZZ) + \Tr(W) \Tr(Z)^4 - 4 \Tr(WZ) \Tr(Z) \Tr(ZZ) - \Tr(W) \Tr(ZZ)^2 \Bigr\},  \\
&(R,\Lambda)= ([3,1,1], [4,1]), \\
&\quad 24 \, \Bigl\{ -\Tr(WZZZ) \Tr(Z) + \Tr(WZ) \Tr(Z)^3 - \Tr(WZZ) \Tr(ZZ) - \Tr(W) \Tr(Z)^2 \Tr(ZZ) \\
&\quad + \Tr(WZ) \Tr(ZZZ) + \Tr(W) \Tr(ZZZZ) \Bigr\} \\
&(R,\Lambda)= ([2,2,1], [5]), \\
&\quad 24 \Tr(WZZZ) \Tr(Z) - 12 \Tr(WZZ) \Tr(Z)^2 - 4 \Tr(WZ) \Tr(Z)^3 + 5 \Tr(W) \Tr(Z)^4 \\
&\quad - 12 \Tr(WZZ) \Tr(ZZ) + 12 \Tr(WZ) \Tr(Z) \Tr(ZZ) - 6 \Tr(W) \Tr(Z)^2 \Tr(ZZ) + 3 \Tr(W) \Tr(ZZ)^2 \\
&\quad - 8 \Tr(WZ) \Tr(ZZZ) - 8 \Tr(W) \Tr(Z) \Tr(ZZZ) + 6 \Tr(W) \Tr(ZZZZ) \\
&(R,\Lambda)= ([2,2,1], [4,1]), \\
&\quad 24 \, \Bigl\{ -\Tr(WZZZ) \Tr(Z) - 2 \Tr(WZZ) \Tr(Z)^2 + \Tr(WZ) \Tr(Z)^3 + 3 \Tr(WZZ) \Tr(ZZ) \\
&\quad + 2 \Tr(WZ) \Tr(Z) \Tr(ZZ) - \Tr(W) \Tr(Z)^2 \Tr(ZZ) - 2 \Tr(W) \Tr(ZZ)^2 - 3 \Tr(WZ) \Tr(ZZZ) \\
&\quad + 2 \Tr(W) \Tr(Z) \Tr(ZZZ) + \Tr(W) \Tr(ZZZZ) \Bigr\} \\
&(R,\Lambda)= ([2,1,1,1], [5]), \\
&\quad 4 \, \Bigl\{ -6 \Tr(WZZZZ) + 3 \Tr(WZZ) \Tr(Z)^2 - 2 \Tr(WZ) \Tr(Z)^3 + \Tr(W) \Tr(Z)^4 + 3 \Tr(WZZ) \Tr(ZZ) \\
&\quad - 3 \Tr(W) \Tr(Z)^2 \Tr(ZZ) + 2 \Tr(WZ) \Tr(ZZZ) + 2 \Tr(W) \Tr(Z) \Tr(ZZZ) \Bigr\} 
\end{alignedat}
$}

\resizebox{\linewidth}{!}{%
$
\begin{alignedat}{9}
&(R,\Lambda)= ([2,1,1,1], [4,1]), \\
&\quad 12 \, \Bigl\{ 2 \Tr(WZZZ) \Tr(Z) - 2 \Tr(WZZ) \Tr(Z)^2 + \Tr(WZ) \Tr(Z)^3 - \Tr(WZ) \Tr(Z) \Tr(ZZ) \\
&\quad - \Tr(W) \Tr(Z)^2 \Tr(ZZ) + \Tr(W) \Tr(ZZ)^2 + 2 \Tr(W) \Tr(Z) \Tr(ZZZ) - 2 \Tr(W) \Tr(ZZZZ) \Bigr\} \\
&(R,\Lambda)= ([1,1,1,1,1], [5]), \\
&\quad 24 \Tr(WZZZZ) - 24 \Tr(WZZZ) \Tr(Z) + 12 \Tr(WZZ) \Tr(Z)^2 - 4 \Tr(WZ) \Tr(Z)^3 + \Tr(W) \Tr(Z)^4 \\
&\quad - 12 \Tr(WZZ) \Tr(ZZ) + 12 \Tr(WZ) \Tr(Z) \Tr(ZZ) - 6 \Tr(W) \Tr(Z)^2 \Tr(ZZ) + 3 \Tr(W) \Tr(ZZ)^2 \\
&\quad - 8 \Tr(WZ) \Tr(ZZZ) + 8 \Tr(W) \Tr(Z) \Tr(ZZZ) - 6 \Tr(W) \Tr(ZZZZ)
\end{alignedat}
$}

\subsubsection{A(3,2)}
This is the first example of the covariant basis with a non-trivial Kronecker coefficient
\begin{equation}
C([3,1,1], [3,1,1], [3,2]) =2.
\end{equation}
Below we show an integer basis of the two states with $(R,\Lambda)=([3,1,1], [3,2])$, which are orthogonal with respect to the $\delta$-function inner product.

$
\begin{aligned}
&(R,\Lambda)= ([3,1,1], [3,2]), \\
&\begin{aligned}
&3\left\{ \quad  \Tr(Z)^3 \Tr(WW) - 2\Tr(Z)^2 \Tr(ZW) \Tr(W) + \Tr(Z)\Tr(ZZ)\Tr(W)^2 \right . \\
&\quad + 4\Tr(Z)\Tr(ZZWW) - 4\Tr(Z)\Tr(ZWZW) - 2\Tr(ZZ)\Tr(ZWW) \\
&\left. \quad - 2\Tr(ZZZ)\Tr(WW) + 4\Tr(ZZW)\Tr(ZW) \right\}
\end{aligned}
\\
&(R,\Lambda)= ([3,1,1], [3,2]), \\
&\begin{aligned}
&12\left\{ \quad \Tr(Z)^2 \Tr(ZWW) - \Tr(Z)\Tr(ZZ)\Tr(WW) - 2\Tr(Z)\Tr(ZZW)\Tr(W) \right . \\
&\left . \quad + \Tr(Z)\Tr(ZW)^2 + \Tr(ZZZ)\Tr(W)^2 + \Tr(ZZZWW) - \Tr(ZZWZW)\right\} 
\end{aligned}
\end{aligned}
$

The remaining states are given as follows.

\resizebox{\linewidth}{!}{%
$
\begin{alignedat}{9}
&(R,\Lambda)= ([5], [5]), \\
&\quad 12 \Tr(WWZZZ) + 12 \Tr(WZWZZ) + 12 \Tr(WZ) \Tr(WZZ) + 12 \Tr(W) \Tr(WZZZ) + 12 \Tr(WWZZ) \Tr(Z) \\
&\quad + 6 \Tr(WZ)^2 \Tr(Z) + 6 \Tr(WZWZ) \Tr(Z) + 12 \Tr(W) \Tr(WZZ) \Tr(Z) + 6 \Tr(WWZ) \Tr(Z)^2 \\
&\quad + 6 \Tr(W) \Tr(WZ) \Tr(Z)^2 + \Tr(W)^2 \Tr(Z)^3 + \Tr(WW) \Tr(Z)^3 + 6 \Tr(WWZ) \Tr(ZZ) + 6 \Tr(W) \Tr(WZ) \Tr(ZZ) \\
&\quad + 3 \Tr(W)^2 \Tr(Z) \Tr(ZZ) + 3 \Tr(WW) \Tr(Z) \Tr(ZZ) + 2 \Tr(W)^2 \Tr(ZZZ) + 2 \Tr(WW) \Tr(ZZZ)  \\
&(R,\Lambda)= ([4,1], [5]), \\
&\quad 2 \, \Bigl\{ -6 \Tr(WWZZZ) - 6 \Tr(WZWZZ) - 6 \Tr(WZ) \Tr(WZZ) + 6 \Tr(W) \Tr(WZZ) \Tr(Z) + 3 \Tr(WWZ) \Tr(Z)^2 \\
&\quad + 6 \Tr(W) \Tr(WZ) \Tr(Z)^2 + 2 \Tr(W)^2 \Tr(Z)^3 + \Tr(WW) \Tr(Z)^3 - 3 \Tr(WWZ) \Tr(ZZ) + 3 \Tr(W)^2 \Tr(Z) \Tr(ZZ) \\
&\quad + \Tr(W)^2 \Tr(ZZZ) - \Tr(WW) \Tr(ZZZ) \Bigr\} \\
&(R,\Lambda)= ([4,1], [4,1]), \\
&\quad 6 \, \Bigl\{ -6 \Tr(W) \Tr(WZZZ) + 4 \Tr(WWZZ) \Tr(Z) + 2 \Tr(WZ)^2 \Tr(Z) + 2 \Tr(WZWZ) \Tr(Z) \\
&\quad - 2 \Tr(W) \Tr(WZZ) \Tr(Z) + 4 \Tr(WWZ) \Tr(Z)^2 + \Tr(W) \Tr(WZ) \Tr(Z)^2 + \Tr(WW) \Tr(Z)^3 \\
&\quad - 3 \Tr(W) \Tr(WZ) \Tr(ZZ) - 2 \Tr(W)^2 \Tr(Z) \Tr(ZZ) + \Tr(WW) \Tr(Z) \Tr(ZZ) - 2 \Tr(W)^2 \Tr(ZZZ) \Bigr\}
\end{alignedat}
$}

\resizebox{\linewidth}{!}{%
$
\begin{alignedat}{9}
&(R,\Lambda)= ([4,1], [3,2]), \\
&\quad 6 \, \Bigl\{ 6 \Tr(WWZZZ) - 6 \Tr(WZWZZ) - 6 \Tr(WZ) \Tr(WZZ) + 4 \Tr(WWZZ) \Tr(Z) - 4 \Tr(WZ)^2 \Tr(Z) \\
&\quad  - 4 \Tr(WZWZ) \Tr(Z) - 2 \Tr(W) \Tr(WZZ) \Tr(Z) + \Tr(WWZ) \Tr(Z)^2 - 2 \Tr(W) \Tr(WZ) \Tr(Z)^2 \\
&\quad + \Tr(WW) \Tr(Z)^3 + 3 \Tr(WWZ) \Tr(ZZ) + \Tr(W)^2 \Tr(Z) \Tr(ZZ) + 4 \Tr(WW) \Tr(Z) \Tr(ZZ) \\
&\quad + \Tr(W)^2 \Tr(ZZZ) + 3 \Tr(WW) \Tr(ZZZ) \Bigr\} \\
&(R,\Lambda)= ([3,2], [5]), \\
&\quad 12 \Tr(WZ) \Tr(WZZ) - 12 \Tr(W) \Tr(WZZZ) - 12 \Tr(WWZZ) \Tr(Z) + 6 \Tr(WZ)^2 \Tr(Z) - 6 \Tr(WZWZ) \Tr(Z) \\
&\quad - 12 \Tr(W) \Tr(WZZ) \Tr(Z) - 6 \Tr(WWZ) \Tr(Z)^2 + 6 \Tr(W) \Tr(WZ) \Tr(Z)^2 + 5 \Tr(W)^2 \Tr(Z)^3 + \Tr(WW) \Tr(Z)^3 \\
&\quad + 6 \Tr(WWZ) \Tr(ZZ) + 6 \Tr(W) \Tr(WZ) \Tr(ZZ) + 3 \Tr(W)^2 \Tr(Z) \Tr(ZZ) + 3 \Tr(WW) \Tr(Z) \Tr(ZZ) \\
&\quad - 2 \Tr(W)^2 \Tr(ZZZ) + 2 \Tr(WW) \Tr(ZZZ) \\
&(R,\Lambda)= ([3,2], [4,1]), \\
&\quad 12 \, \Bigl\{ -3 \Tr(WZ) \Tr(WZZ) + 3 \Tr(W) \Tr(WZZZ) - 2 \Tr(WWZZ) \Tr(Z) - 4 \Tr(WZ)^2 \Tr(Z) - \Tr(WZWZ) \Tr(Z) \\
&\quad - 2 \Tr(W) \Tr(WZZ) \Tr(Z) + 4 \Tr(WWZ) \Tr(Z)^2 + \Tr(W) \Tr(WZ) \Tr(Z)^2 + \Tr(WW) \Tr(Z)^3 \\
&\quad + 6 \Tr(WWZ) \Tr(ZZ) + 6 \Tr(W) \Tr(WZ) \Tr(ZZ) - 2 \Tr(W)^2 \Tr(Z) \Tr(ZZ) - 2 \Tr(WW) \Tr(Z) \Tr(ZZ) \\
&\quad - 2 \Tr(W)^2 \Tr(ZZZ) - 3 \Tr(WW) \Tr(ZZZ) \Bigr\} \\
&(R,\Lambda)= ([3,2], [3,2]), \\
&\quad 12 \, \Bigl\{ -3 \Tr(WWZZZ) + 3 \Tr(WZWZZ) - 2 \Tr(WWZZ) \Tr(Z) - \Tr(WZ)^2 \Tr(Z) + 2 \Tr(WZWZ) \Tr(Z) \\
&\quad - 2 \Tr(W) \Tr(WZZ) \Tr(Z) + \Tr(WWZ) \Tr(Z)^2 - 2 \Tr(W) \Tr(WZ) \Tr(Z)^2 + \Tr(WW) \Tr(Z)^3 \\
&\quad + \Tr(W)^2 \Tr(Z) \Tr(ZZ) + \Tr(WW) \Tr(Z) \Tr(ZZ) + \Tr(W)^2 \Tr(ZZZ) \Bigr\} \\
&(R,\Lambda)= ([3,1,1], [5]), \\
&\quad 6 \, \Bigl\{ 2 \Tr(WWZZZ) + 2 \Tr(WZWZZ) - 2 \Tr(WZ)^2 \Tr(Z) + \Tr(W)^2 \Tr(Z)^3 - 2 \Tr(W) \Tr(WZ) \Tr(ZZ) \\
&\quad - \Tr(WW) \Tr(Z) \Tr(ZZ) \Bigr\} \\
&(R,\Lambda)= ([3,1,1], [4,1]), \\
&\quad 12 \, \Bigl\{ \Tr(WZ) \Tr(WZZ) + 3 \Tr(W) \Tr(WZZZ) - 2 \Tr(WWZZ) \Tr(Z) - \Tr(WZWZ) \Tr(Z) + \Tr(W) \Tr(WZ) \Tr(Z)^2 \\
&\quad + \Tr(WW) \Tr(Z)^3 - 2 \Tr(WWZ) \Tr(ZZ) - 2 \Tr(W)^2 \Tr(Z) \Tr(ZZ) + \Tr(WW) \Tr(ZZZ) \Bigr\} \\
&(R,\Lambda)= ([3,1,1], [3,2]), \\
&\quad 3 \, \Bigl\{ 4 \Tr(WZ) \Tr(WZZ) + 4 \Tr(WWZZ) \Tr(Z) - 4 \Tr(WZWZ) \Tr(Z) - 2 \Tr(W) \Tr(WZ) \Tr(Z)^2 \\
&\quad + \Tr(WW) \Tr(Z)^3 - 2 \Tr(WWZ) \Tr(ZZ) + \Tr(W)^2 \Tr(Z) \Tr(ZZ) - 2 \Tr(WW) \Tr(ZZZ) \Bigr\} \\
&(R,\Lambda)= ([3,1,1], [3,2]), \\
&\quad 12 \, \Bigl\{ \Tr(WWZZZ) - \Tr(WZWZZ) + \Tr(WZ)^2 \Tr(Z) - 2 \Tr(W) \Tr(WZZ) \Tr(Z) + \Tr(WWZ) \Tr(Z)^2 \\
&\quad - \Tr(WW) \Tr(Z) \Tr(ZZ) + \Tr(W)^2 \Tr(ZZZ) \Bigr\} \\
&(R,\Lambda)= ([2,2,1], [5]), \\
&\quad -12 \Tr(WZ) \Tr(WZZ) + 12 \Tr(W) \Tr(WZZZ) + 12 \Tr(WWZZ) \Tr(Z) + 6 \Tr(WZ)^2 \Tr(Z) + 6 \Tr(WZWZ) \Tr(Z) \\
&\quad - 12 \Tr(W) \Tr(WZZ) \Tr(Z) - 6 \Tr(WWZ) \Tr(Z)^2 - 6 \Tr(W) \Tr(WZ) \Tr(Z)^2 + 5 \Tr(W)^2 \Tr(Z)^3 - \Tr(WW) \Tr(Z)^3 \\
&\quad - 6 \Tr(WWZ) \Tr(ZZ) + 6 \Tr(W) \Tr(WZ) \Tr(ZZ) - 3 \Tr(W)^2 \Tr(Z) \Tr(ZZ) + 3 \Tr(WW) \Tr(Z) \Tr(ZZ) \\
&\quad - 2 \Tr(W)^2 \Tr(ZZZ) - 2 \Tr(WW) \Tr(ZZZ) 
\end{alignedat}
$}

\resizebox{\linewidth}{!}{%
$
\begin{alignedat}{9}
&(R,\Lambda)= ([2,2,1], [4,1]), \\
&\quad 12 \, \Bigl\{ -3 \Tr(WZ) \Tr(WZZ) + 3 \Tr(W) \Tr(WZZZ) - 2 \Tr(WWZZ) \Tr(Z) + 4 \Tr(WZ)^2 \Tr(Z) - \Tr(WZWZ) \Tr(Z) \\
&\quad + 2 \Tr(W) \Tr(WZZ) \Tr(Z) - 4 \Tr(WWZ) \Tr(Z)^2 + \Tr(W) \Tr(WZ) \Tr(Z)^2 + \Tr(WW) \Tr(Z)^3 \\
&\quad + 6 \Tr(WWZ) \Tr(ZZ) - 6 \Tr(W) \Tr(WZ) \Tr(ZZ) - 2 \Tr(W)^2 \Tr(Z) \Tr(ZZ) + 2 \Tr(WW) \Tr(Z) \Tr(ZZ) \\
&\quad  + 2 \Tr(W)^2 \Tr(ZZZ) - 3 \Tr(WW) \Tr(ZZZ) \Bigr\} \\
&(R,\Lambda)= ([2,2,1], [3,2]), \\
&\quad 12 \, \Bigl\{ 3 \Tr(WWZZZ) - 3 \Tr(WZWZZ) - 2 \Tr(WWZZ) \Tr(Z) + \Tr(WZ)^2 \Tr(Z) + 2 \Tr(WZWZ) \Tr(Z) \\
&\quad + 2 \Tr(W) \Tr(WZZ) \Tr(Z) - \Tr(WWZ) \Tr(Z)^2 - 2 \Tr(W) \Tr(WZ) \Tr(Z)^2 + \Tr(WW) \Tr(Z)^3 \\
&\quad + \Tr(W)^2 \Tr(Z) \Tr(ZZ)  - \Tr(WW) \Tr(Z) \Tr(ZZ) - \Tr(W)^2 \Tr(ZZZ) \Bigr\} \\
&(R,\Lambda)= ([2,1,1,1], [5]), \\
&\quad 2 \, \Bigl\{ -6 \Tr(WWZZZ) - 6 \Tr(WZWZZ) + 6 \Tr(WZ) \Tr(WZZ) + 6 \Tr(W) \Tr(WZZ) \Tr(Z) + 3 \Tr(WWZ) \Tr(Z)^2 \\
&\quad - 6 \Tr(W) \Tr(WZ) \Tr(Z)^2 + 2 \Tr(W)^2 \Tr(Z)^3 - \Tr(WW) \Tr(Z)^3 + 3 \Tr(WWZ) \Tr(ZZ) - 3 \Tr(W)^2 \Tr(Z) \Tr(ZZ) \\
&\quad + \Tr(W)^2 \Tr(ZZZ) + \Tr(WW) \Tr(ZZZ) \Bigr\} \\
&(R,\Lambda)= ([2,1,1,1], [4,1]), \\
&\quad 6 \, \Bigl\{ -6 \Tr(W) \Tr(WZZZ) + 4 \Tr(WWZZ) \Tr(Z) - 2 \Tr(WZ)^2 \Tr(Z) + 2 \Tr(WZWZ) \Tr(Z) \\
&\quad + 2 \Tr(W) \Tr(WZZ) \Tr(Z) - 4 \Tr(WWZ) \Tr(Z)^2 + \Tr(W) \Tr(WZ) \Tr(Z)^2 + \Tr(WW) \Tr(Z)^3 \\
&\quad + 3 \Tr(W) \Tr(WZ) \Tr(ZZ) - 2 \Tr(W)^2 \Tr(Z) \Tr(ZZ) - \Tr(WW) \Tr(Z) \Tr(ZZ) + 2 \Tr(W)^2 \Tr(ZZZ) \Bigr\} \\
&(R,\Lambda)= ([2,1,1,1], [3,2]), \\
&\quad 6 \, \Bigl\{ -6 \Tr(WWZZZ) + 6 \Tr(WZWZZ) - 6 \Tr(WZ) \Tr(WZZ) + 4 \Tr(WWZZ) \Tr(Z) + 4 \Tr(WZ)^2 \Tr(Z) \\
&\quad - 4 \Tr(WZWZ) \Tr(Z) + 2 \Tr(W) \Tr(WZZ) \Tr(Z) - \Tr(WWZ) \Tr(Z)^2 - 2 \Tr(W) \Tr(WZ) \Tr(Z)^2 \\
&\quad + \Tr(WW) \Tr(Z)^3 + 3 \Tr(WWZ) \Tr(ZZ) + \Tr(W)^2 \Tr(Z) \Tr(ZZ) - 4 \Tr(WW) \Tr(Z) \Tr(ZZ) \\
&\quad - \Tr(W)^2 \Tr(ZZZ) + 3 \Tr(WW) \Tr(ZZZ) \Bigr\} \\
&(R,\Lambda)= ([1,1,1,1,1], [5]), \\
&\quad 12 \Tr(WWZZZ) + 12 \Tr(WZWZZ) - 12 \Tr(WZ) \Tr(WZZ) - 12 \Tr(W) \Tr(WZZZ) - 12 \Tr(WWZZ) \Tr(Z) \\
&\quad + 6 \Tr(WZ)^2 \Tr(Z) - 6 \Tr(WZWZ) \Tr(Z) + 12 \Tr(W) \Tr(WZZ) \Tr(Z) + 6 \Tr(WWZ) \Tr(Z)^2 \\
&\quad - 6 \Tr(W) \Tr(WZ) \Tr(Z)^2 + \Tr(W)^2 \Tr(Z)^3 - \Tr(WW) \Tr(Z)^3 - 6 \Tr(WWZ) \Tr(ZZ) + 6 \Tr(W) \Tr(WZ) \Tr(ZZ) \\
&\quad - 3 \Tr(W)^2 \Tr(Z) \Tr(ZZ) + 3 \Tr(WW) \Tr(Z) \Tr(ZZ) + 2 \Tr(W)^2 \Tr(ZZZ) - 2 \Tr(WW) \Tr(ZZZ)
\end{alignedat}
$}

\section{Data: Orthonormal bases for $\mathbb{C}[S_L]$}\label{sec:data}
In this appendix we give examples of solutions to the eigensystems in \ref{sec:EV CSL}, giving the Artin-Wedderburn and Kronecker decomposition of $\mathbb{C}[S_L]$, respectively.

\subsection{Matrix units of $S_L$}\label{app:MU basis SL}

The matrix unit of $S_L$ is defined by \eqref{def:AW basis},
\begin{equation}
Q^R_{IJ} = \frac{d_R}{|S_L|} \sum_{g \in S_L} D^R_{JI}(g) \, g^{-1} .
\label{app:AW basis}
\end{equation}
In the following we construct $Q^R_{IJ}$\,, and thus determine the matrix elements of the irreducible representation of $S_L$\,.\footnote{Our construction gives the Young seminormal representation, which is real and unitary. See the discussion at the end of Section \ref{sec:EV CSL}.} 

We argued in the main text that by solving the eigenvalue system \eqref{Eigen AW to MU}, the matrix unit can be determined uniquely up to the normalisation constant. 
If we follow the method of Section \ref{sec:integrality} using the Hermite normal form, we obtain the eigenvectors whose components are all integers. Let us call such an eigenvector the integer basis, denoted by
\begin{equation}
\sfQ^R_{IJ} \equiv \sfN^R_{IJ} \, Q^R_{IJ} 
\label{app:MU INT basis}
\end{equation}
where $\sfN^R_{IJ}$ is a normalisation constant.
The integer basis satisfies the condition
\begin{equation}
\sfQ^R_{IJ} = \sum_{g \in {\tt Basis} } c_g (\sfQ^R_{IJ}) \, g , \qquad  c_g (\sfQ^R_{IJ}) \in \bb{Z} \,.
\end{equation}
and the product relation
\begin{equation}
\sfQ^R_{IJ} \, \sfQ^S_{KL} = \frac{\sfN^R_{IJ} \, \sfN^R_{JL}}{\sfN^R_{IL}}  \, \delta^{RS} \, \delta_{JK} \, \sfQ^R_{IL} \,.
\label{app:matrix units unnnorm SL}
\end{equation}

Note that the explicit data shown below may not give a real and unitary $R$ in the sense that $D^R_{IJ} ( \sigma^{-1} ) \neq D^R_{JI} ( \sigma)$.\footnote{In other words, we did not impose the real unitarity condition on the integer basis $\sfQ^R_{IJ}$ with a generic $\sfN^R_{IJ}$\,.}
As discussed at the end of Section \ref{sec:decomp SL reg}, we need to rescale the eigenvectors to obtain the Young-Yamanouchi orthogonal representation, which is real and unitary.
However, the rescaling spoils the integer property.
Other $\bb{Z}$-valued representations, such as Young natural representation and Kazhdan-Lusztig
representations, also do not produce unitary matrices \cite{Bjorner2005c,Garsia198832}.

We wrote a {\tt Mathematica} code to compute the matrix units of $S_L$ explicitly up to $L=6$. 
Below we just present our results up to $L=4$.

\subsubsection{Case of $S_2$}

One finds
\begin{equation}
\sfQ^{[2]} = () + (12) \,, \qquad \sfQ^{[1,1]} = () - (12) , \qquad
\sfQ^R \, \sfQ^S = 2 \, \delta^{RS} Q^R 
\end{equation}
where we omit $I,J$ because all representations are one-dimensional.

\subsubsection{Case of $S_3$}

We express the matrix unit $Q^R_{IJ}$ as a pair of Young tableaux $\{ I,J \}$ of shape $R$.
We reorganise them into a vector
\ytableausetup{boxsize=1em,centertableaux}
\begin{align}
\scrE_\rho &\equiv \( 
\sfQ^{[3]}_{11} \,, 
\sfQ^{[2,1]}_{11} \,, 
\sfQ^{[2,1]}_{12}\,, 
\sfQ^{[2,1]}_{21} \,,
\sfQ^{[2,1]}_{22} \,,
\sfQ^{[1,1,1]}_{11} \)
\\[1mm]
&= \( {\scriptsize
\Bigl\{  \, \begin{ytableau} 1 & 2 & 3 \end{ytableau} \,,\,\begin{ytableau} 1 & 2 & 3 \end{ytableau} \, \Bigr\} ,
\Bigl\{  \, \begin{ytableau} 1 & 2 \\ 3 \end{ytableau} \,,\,\begin{ytableau} 1 & 2 \\ 3 \end{ytableau} \, \Bigr\} ,
\Bigl\{  \, \begin{ytableau} 1 & 2 \\ 3 \end{ytableau} \,,\,\begin{ytableau} 1 & 3 \\ 2 \end{ytableau}   \, \Bigr\} ,
\Bigl\{  \, \begin{ytableau} 1 & 3 \\ 2 \end{ytableau} \,,\,\begin{ytableau} 1 & 2 \\ 3 \end{ytableau} \, \Bigr\} ,
\Bigl\{  \, \begin{ytableau} 1 & 3 \\ 2 \end{ytableau} \,,\,\begin{ytableau} 1 & 3 \\ 2 \end{ytableau} \, \Bigr\} 
\Bigl\{  \, \begin{ytableau} 1 \\ 2 \\ 3 \end{ytableau} \,,\, \begin{ytableau} 1 \\ 2 \\ 3 \end{ytableau} \, \Bigr\}  
} \) .
\notag 
\end{align}
We choose a basis of $S_3$ as
\begin{equation}
{\tt Basis} = \{ (), (23),(12),(123),(132),(13) \}
\label{app:basis of S3}
\end{equation}
and expand $\scrE_\rho$ as
\begin{equation}
\scrE_\rho = \sum_{g \in {\tt Basis}} E_{\rho g} \, g .
\end{equation}
The coefficient matrix in the integer basis is given by
\begin{equation}
E_{\rho g} = \begin{pmatrix}
 1 & 1 & 1 & 1 & 1 & 1 \\
 2 & -1 & 2 & -1 & -1 & -1 \\
 0 & 1 & 0 & 1 & -1 & -1 \\
 0 & 1 & 0 & -1 & 1 & -1 \\
 2 & 1 & -2 & -1 & -1 & 1 \\
 1 & -1 & -1 & 1 & 1 & -1 \\
\end{pmatrix}
\end{equation}
We can write down the product relation \eqref{app:matrix units unnnorm SL} explicitly as
\begin{equation}
\scrE_\rho \, \scrE_{\rho'} = 
\begin{pmatrix}
 6 \, \scrE_1 & 0 & 0 & 0 & 0 & 0 \\
 0 & 6 \, \scrE_2 & 6 \, \scrE_3 & 0 & 0 & 0 \\
 0 & 0 & 0 & 2 \, \scrE_2 & 6 \, \scrE_3 & 0 \\
 0 & 6 \, \scrE_4 & 2 \, \scrE_5 & 0 & 0 & 0 \\
 0 & 0 & 0 & 6 \, \scrE_4 & 6 \, \scrE_5 & 0 \\
 0 & 0 & 0 & 0 & 0 & 6 \, \scrE_6 \\
\end{pmatrix} .
\label{S3 INT normalisation}
\end{equation}

In order to reproduce the standard matrix unit relation \eqref{def:matrix units SL}, we need to adjust the normalisation so that $\sfN^R_{IJ}=1$ for all eigenvectors. In this normalisation, we find
\begin{equation}
\widehat{\scrE}_\rho = \sum_{g \in {\tt Basis}} \hat E_{\rho g} \, g , \qquad
\hat E_{\rho g} = 
\begin{pmatrix}
 \frac{1}{6} & \frac{1}{6} & \frac{1}{6} & \frac{1}{6} & \frac{1}{6} & \frac{1}{6} \\[1mm]
 \frac{1}{3} & -\frac{1}{6} & \frac{1}{3} & -\frac{1}{6} & -\frac{1}{6} & -\frac{1}{6} \\[1mm]
 0 & \frac{1}{2 \sqrt{3}} & 0 & \frac{1}{2 \sqrt{3}} & -\frac{1}{2 \sqrt{3}} & -\frac{1}{2 \sqrt{3}} \\[1mm]
 0 & \frac{1}{2 \sqrt{3}} & 0 & -\frac{1}{2 \sqrt{3}} & \frac{1}{2 \sqrt{3}} & -\frac{1}{2 \sqrt{3}} \\[1mm]
 \frac{1}{3} & \frac{1}{6} & -\frac{1}{3} & -\frac{1}{6} & -\frac{1}{6} & \frac{1}{6} \\[1mm]
 \frac{1}{6} & -\frac{1}{6} & -\frac{1}{6} & \frac{1}{6} & \frac{1}{6} & -\frac{1}{6} \\
\end{pmatrix} .
\label{S3 MU Eag}
\end{equation}
For example,
\begin{equation}
\begin{alignedat}{9}
\widehat{\scrE}_1 &= \Bigl\{  \, \begin{ytableau} 1 & 2 & 3 \end{ytableau} \,,\,\begin{ytableau} 1 & 2 & 3 \end{ytableau} \, \Bigr\} 
& &= \frac{() + (23) + (12) + (123) + (132) + (13)}{6} 
\\
\widehat{\scrE}_6 &= \Bigl\{  \, \begin{ytableau} 1 \\ 2 \\ 3 \end{ytableau} \,,\, \begin{ytableau} 1 \\ 2 \\ 3 \end{ytableau} \, \Bigr\} 
& &= \frac{() - (23) - (12) + (123) + (132) - (13)}{6} \,.
\end{alignedat}
\end{equation}
The product of $\widehat{\scrE}$'s is given by
\begin{equation}
\widehat{\scrE}_\rho \, \widehat{\scrE}_{\rho'} = 
\begin{pmatrix}
 \widehat{\scrE}_1 & 0 & 0 & 0 & 0 & 0 \\
 0 & \widehat{\scrE}_2 & \widehat{\scrE}_3 & 0 & 0 & 0 \\
 0 & 0 & 0 & \widehat{\scrE}_2 & \widehat{\scrE}_3 & 0 \\
 0 & \widehat{\scrE}_4 & \widehat{\scrE}_5 & 0 & 0 & 0 \\
 0 & 0 & 0 & \widehat{\scrE}_4 & \widehat{\scrE}_5 & 0 \\
 0 & 0 & 0 & 0 & 0 & \widehat{\scrE}_6 \\
\end{pmatrix} 
\label{S3 MU normalisation}
\end{equation}
which agrees with \eqref{def:matrix units SL}.

\subsubsection{Case of $S_4$}

Define the vector of matrix units as
\ytableausetup{boxsize=1em,centertableaux}
\begin{align}
\hspace{-3mm}
\scrE_\rho \equiv &\Biggl( {\scriptsize
\Big\{ \, \begin{ytableau}1 & 2 & 3 & 4 \end{ytableau} \,,\, \begin{ytableau}1 & 2 & 3 & 4 \end{ytableau} \Big\} ,
\Big\{ \, \begin{ytableau}1 & 2 & 3 \\ 4 \end{ytableau} \,,\, \begin{ytableau}1 & 2 & 3 \\ 4 \end{ytableau} \Big\} ,
\Big\{ \, \begin{ytableau}1 & 2 & 3 \\ 4 \end{ytableau} \,,\, \begin{ytableau}1 & 2 & 4 \\ 3 \end{ytableau} \Big\} ,
\Big\{ \, \begin{ytableau}1 & 2 & 3 \\ 4 \end{ytableau} \,,\, \begin{ytableau}1 & 3 & 4 \\ 2 \end{ytableau} \Big\} ,
\Big\{ \, \begin{ytableau}1 & 2 & 4 \\ 3 \end{ytableau} \,,\, \begin{ytableau}1 & 2 & 3 \\ 4 \end{ytableau} \Big\} ,
}\notag\\[1mm]
&{\scriptsize
\Big\{ \, \begin{ytableau}1 & 2 & 4 \\ 3 \end{ytableau} \,,\, \begin{ytableau}1 & 2 & 4 \\ 3 \end{ytableau} \Big\} ,
\Big\{ \, \begin{ytableau}1 & 2 & 4 \\ 3 \end{ytableau} \,,\, \begin{ytableau}1 & 3 & 4 \\ 2 \end{ytableau} \Big\} ,
\Big\{ \, \begin{ytableau}1 & 3 & 4 \\ 2 \end{ytableau} \,,\, \begin{ytableau}1 & 2 & 3 \\ 4 \end{ytableau} \Big\} ,
\Big\{ \, \begin{ytableau}1 & 3 & 4 \\ 2 \end{ytableau} \,,\, \begin{ytableau}1 & 2 & 4 \\ 3 \end{ytableau} \Big\} ,
\Big\{ \, \begin{ytableau}1 & 3 & 4 \\ 2 \end{ytableau} \,,\, \begin{ytableau}1 & 3 & 4 \\ 2 \end{ytableau} \Big\} ,
}\notag\\[1mm]
&{\scriptsize
\Big\{ \, \begin{ytableau}1 & 2 \\ 3 & 4 \end{ytableau} \,,\, \begin{ytableau}1 & 2 \\ 3 & 4 \end{ytableau} \Big\} ,
\Big\{ \, \begin{ytableau}1 & 2 \\ 3 & 4 \end{ytableau} \,,\, \begin{ytableau}1 & 3 \\ 2 & 4 \end{ytableau} \Big\} ,
\Big\{ \, \begin{ytableau}1 & 3 \\ 2 & 4 \end{ytableau} \,,\, \begin{ytableau}1 & 2 \\ 3 & 4 \end{ytableau} \Big\} ,
\Big\{ \, \begin{ytableau}1 & 3 \\ 2 & 4 \end{ytableau} \,,\, \begin{ytableau}1 & 3 \\ 2 & 4 \end{ytableau} \Big\} ,
\Big\{ \, \begin{ytableau}1 & 2 \\ 3 \\ 4 \end{ytableau} \,,\, \begin{ytableau}1 & 2 \\ 3 \\ 4 \end{ytableau} \Big\} ,
}\notag\\[1mm]
&{\scriptsize
\Big\{ \, \begin{ytableau}1 & 2 \\ 3 \\ 4 \end{ytableau} \,,\, \begin{ytableau}1 & 3 \\ 2 \\ 4 \end{ytableau} \Big\} ,
\Big\{ \, \begin{ytableau}1 & 2 \\ 3 \\ 4 \end{ytableau} \,,\, \begin{ytableau}1 & 4 \\ 2 \\ 3 \end{ytableau} \Big\} ,
\Big\{ \, \begin{ytableau}1 & 3 \\ 2 \\ 4 \end{ytableau} \,,\, \begin{ytableau}1 & 2 \\ 3 \\ 4 \end{ytableau} \Big\} ,
\Big\{ \, \begin{ytableau}1 & 3 \\ 2 \\ 4 \end{ytableau} \,,\, \begin{ytableau}1 & 3 \\ 2 \\ 4 \end{ytableau} \Big\} ,
\Big\{ \, \begin{ytableau}1 & 3 \\ 2 \\ 4 \end{ytableau} \,,\, \begin{ytableau}1 & 4 \\ 2 \\ 3 \end{ytableau} \Big\} ,
}\notag\\[1mm]
&{\scriptsize
\Big\{ \, \begin{ytableau}1 & 4 \\ 2 \\ 3 \end{ytableau} \,,\, \begin{ytableau}1 & 2 \\ 3 \\ 4 \end{ytableau} \Big\} ,
\Big\{ \, \begin{ytableau}1 & 4 \\ 2 \\ 3 \end{ytableau} \,,\, \begin{ytableau}1 & 3 \\ 2 \\ 4 \end{ytableau} \Big\} ,
\Big\{ \, \begin{ytableau}1 & 4 \\ 2 \\ 3 \end{ytableau} \,,\, \begin{ytableau}1 & 4 \\ 2 \\ 3 \end{ytableau} \Big\} ,
\Big\{ \, \begin{ytableau}1 \\ 2 \\ 3 \\ 4 \end{ytableau} \,,\, \begin{ytableau}1 \\ 2 \\ 3 \\ 4 \end{ytableau} \Big\}
} \Biggr) 
\end{align}

and choose a basis of $S_4$ as
\begin{multline}
{\tt Basis} = \Bigl\{
(),  (34), (23), (234), (243), (24), 
\quad
(12), (12) (34), (123), (1234), (1243), (124),
\\
(132), (1342), (13), (134), (13) (24), (1324), 
\quad
(1432), (142), (143), (14), (1423), (14)(23) \Bigr\} .
\label{app:basis of S4}
\end{multline}
The coefficient matrix in the integer normalisation is given by
\begin{equation}
E_{\rho g} = {\scriptsize \left(
\begin{array}{@{\hskip 2pt}c@{\hskip 2pt}c@{\hskip 2pt}c@{\hskip 2pt}c@{\hskip 2pt}c@{\hskip 2pt}c|@{\hskip 2pt}c@{\hskip 2pt}c@{\hskip 2pt}c@{\hskip 2pt}c@{\hskip 2pt}c@{\hskip 2pt}c|@{\hskip 2pt}c@{\hskip 2pt}c@{\hskip 2pt}c@{\hskip 2pt}c@{\hskip 2pt}c@{\hskip 2pt}c|@{\hskip 2pt}c@{\hskip 2pt}c@{\hskip 2pt}c@{\hskip 2pt}c@{\hskip 2pt}c@{\hskip 2pt}c}
 1 & 1 & 1 & 1 & 1 & 1 & 1 & 1 & 1 & 1 & 1 & 1 & 1 & 1 & 1 & 1 & 1 & 1 & 1 & 1 & 1 & 1 & 1 & 1 \\
 3 & -1 & 3 & -1 & -1 & -1 & 3 & -1 & 3 & -1 & -1 & -1 & 3 & -1 & 3 & -1 & -1 & -1 & -1 & -1 & -1 & -1 & -1 & -1 \\
 0 & 2 & 0 & 2 & -1 & -1 & 0 & 2 & 0 & 2 & -1 & -1 & 0 & 2 & 0 & 2 & -1 & -1 & -1 & -1 & -1 & -1 & -1 & -1 \\
 0 & 0 & 0 & 0 & 1 & 1 & 0 & 0 & 0 & 0 & 1 & 1 & 0 & 0 & 0 & 0 & 1 & 1 & -1 & -1 & -1 & -1 & -1 & -1 \\
 0 & 2 & 0 & -1 & 2 & -1 & 0 & 2 & 0 & -1 & 2 & -1 & 0 & -1 & 0 & -1 & -1 & -1 & 2 & -1 & 2 & -1 & -1 & -1 \\
 6 & 2 & -3 & -1 & -1 & 5 & 6 & 2 & -3 & -1 & -1 & 5 & -3 & -1 & -3 & -1 & -4 & -4 & -1 & 5 & -1 & 5 & -4 & -4 \\\hline
 0 & 0 & 3 & 3 & 1 & 1 & 0 & 0 & 3 & 3 & 1 & 1 & -3 & -3 & -3 & -3 & -2 & -2 & -1 & -1 & -1 & -1 & 2 & 2 \\
 0 & 0 & 0 & 1 & 0 & 1 & 0 & 0 & 0 & -1 & 0 & -1 & 0 & 1 & 0 & -1 & 1 & -1 & 0 & 1 & 0 & -1 & 1 & -1 \\
 0 & 0 & 3 & 1 & 3 & 1 & 0 & 0 & -3 & -1 & -3 & -1 & 3 & 1 & -3 & -1 & -2 & 2 & 3 & 1 & -3 & -1 & -2 & 2 \\
 2 & 2 & 1 & 1 & 1 & 1 & -2 & -2 & -1 & -1 & -1 & -1 & -1 & -1 & 1 & 1 & 0 & 0 & -1 & -1 & 1 & 1 & 0 & 0 \\
 2 & 2 & -1 & -1 & -1 & -1 & 2 & 2 & -1 & -1 & -1 & -1 & -1 & -1 & -1 & -1 & 2 & 2 & -1 & -1 & -1 & -1 & 2 & 2 \\
 0 & 0 & 1 & -1 & 1 & -1 & 0 & 0 & 1 & -1 & 1 & -1 & -1 & 1 & -1 & 1 & 0 & 0 & -1 & 1 & -1 & 1 & 0 & 0 \\\hline
 0 & 0 & 1 & 1 & -1 & -1 & 0 & 0 & -1 & -1 & 1 & 1 & 1 & 1 & -1 & -1 & 0 & 0 & -1 & -1 & 1 & 1 & 0 & 0 \\
 2 & -2 & 1 & -1 & -1 & 1 & -2 & 2 & -1 & 1 & 1 & -1 & -1 & 1 & 1 & -1 & 2 & -2 & 1 & -1 & -1 & 1 & -2 & 2 \\
 2 & -2 & -1 & 1 & 1 & -1 & 2 & -2 & -1 & 1 & 1 & -1 & -1 & 1 & -1 & 1 & 0 & 0 & 1 & -1 & 1 & -1 & 0 & 0 \\
 0 & 0 & 3 & -1 & -3 & 1 & 0 & 0 & 3 & -1 & -3 & 1 & -3 & 1 & -3 & 1 & 2 & 2 & 3 & -1 & 3 & -1 & -2 & -2 \\
 0 & 0 & 0 & 1 & 0 & -1 & 0 & 0 & 0 & 1 & 0 & -1 & 0 & -1 & 0 & -1 & 1 & 1 & 0 & 1 & 0 & 1 & -1 & -1 \\
 0 & 0 & 3 & -3 & -1 & 1 & 0 & 0 & -3 & 3 & 1 & -1 & 3 & -3 & -3 & 3 & 2 & -2 & -1 & 1 & 1 & -1 & 2 & -2 \\\hline
 6 & -2 & 3 & -1 & -1 & -5 & -6 & 2 & -3 & 1 & 1 & 5 & -3 & 1 & 3 & -1 & -4 & 4 & 1 & 5 & -1 & -5 & 4 & -4 \\
 0 & 2 & 0 & 1 & -2 & -1 & 0 & -2 & 0 & -1 & 2 & 1 & 0 & -1 & 0 & 1 & 1 & -1 & 2 & 1 & -2 & -1 & -1 & 1 \\
 0 & 0 & 0 & 0 & 1 & -1 & 0 & 0 & 0 & 0 & -1 & 1 & 0 & 0 & 0 & 0 & 1 & -1 & 1 & -1 & -1 & 1 & 1 & -1 \\
 0 & 2 & 0 & -2 & 1 & -1 & 0 & -2 & 0 & 2 & -1 & 1 & 0 & 2 & 0 & -2 & 1 & -1 & -1 & 1 & 1 & -1 & -1 & 1 \\
 3 & 1 & -3 & -1 & -1 & 1 & -3 & -1 & 3 & 1 & 1 & -1 & 3 & 1 & -3 & -1 & -1 & 1 & 1 & -1 & -1 & 1 & 1 & -1 \\
 1 & -1 & -1 & 1 & 1 & -1 & -1 & 1 & 1 & -1 & -1 & 1 & 1 & -1 & -1 & 1 & 1 & -1 & -1 & 1 & 1 & -1 & -1 & 1 \\
\end{array}
\right). }
\end{equation}
The product $\scrE_\rho \, \scrE_{\rho'}$ is given by
\begin{equation*}
{\tiny \left(
\begin{array}{@{\hskip 1pt}c@{\hskip 1pt}c@{\hskip 1pt}c@{\hskip 1pt}c@{\hskip 1pt}c@{\hskip 1pt}c|@{\hskip 1pt}c@{\hskip 1pt}c@{\hskip 1pt}c@{\hskip 1pt}c@{\hskip 1pt}c@{\hskip 1pt}c|@{\hskip 1pt}c@{\hskip 1pt}c@{\hskip 1pt}c@{\hskip 1pt}c@{\hskip 1pt}c@{\hskip 1pt}c|@{\hskip 1pt}c@{\hskip 1pt}c@{\hskip 1pt}c@{\hskip 1pt}c@{\hskip 1pt}c@{\hskip 1pt}c}
 24 \scrE_1 & 0 & 0 & 0 & 0 & 0 & 0 & 0 & 0 & 0 & 0 & 0 & 0 & 0 & 0 & 0 & 0 & 0 & 0 & 0 & 0 & 0 & 0 & 0 \\
 0 & 24 \scrE_2 & 24 \scrE_3 & 24 \scrE_4 & 0 & 0 & 0 & 0 & 0 & 0 & 0 & 0 & 0 & 0 & 0 & 0 & 0 & 0 & 0 & 0 & 0 & 0 & 0 & 0 \\
 0 & 0 & 0 & 0 & 12 \scrE_2 & 48 \scrE_3 & 48 \scrE_4 & 0 & 0 & 0 & 0 & 0 & 0 & 0 & 0 & 0 & 0 & 0 & 0 & 0 & 0 & 0 & 0 & 0 \\
 0 & 0 & 0 & 0 & 0 & 0 & 0 & 4 \scrE_2 & 16 \scrE_3 & 16 \scrE_4 & 0 & 0 & 0 & 0 & 0 & 0 & 0 & 0 & 0 & 0 & 0 & 0 & 0 & 0 \\
 0 & 24 \scrE_5 & 6 \scrE_6 & 6 \scrE_7 & 0 & 0 & 0 & 0 & 0 & 0 & 0 & 0 & 0 & 0 & 0 & 0 & 0 & 0 & 0 & 0 & 0 & 0 & 0 & 0 \\
 0 & 0 & 0 & 0 & 48 \scrE_5 & 48 \scrE_6 & 48 \scrE_7 & 0 & 0 & 0 & 0 & 0 & 0 & 0 & 0 & 0 & 0 & 0 & 0 & 0 & 0 & 0 & 0 & 0 \\\hline
 0 & 0 & 0 & 0 & 0 & 0 & 0 & 16 \scrE_5 & 16 \scrE_6 & 16 \scrE_7 & 0 & 0 & 0 & 0 & 0 & 0 & 0 & 0 & 0 & 0 & 0 & 0 & 0 & 0 \\
 0 & 24 \scrE_8 & 6 \scrE_9 & 6 \scrE_{10} & 0 & 0 & 0 & 0 & 0 & 0 & 0 & 0 & 0 & 0 & 0 & 0 & 0 & 0 & 0 & 0 & 0 & 0 & 0 & 0 \\
 0 & 0 & 0 & 0 & 48 \scrE_8 & 48 \scrE_9 & 48 \scrE_{10} & 0 & 0 & 0 & 0 & 0 & 0 & 0 & 0 & 0 & 0 & 0 & 0 & 0 & 0 & 0 & 0 & 0 \\
 0 & 0 & 0 & 0 & 0 & 0 & 0 & 16 \scrE_8 & 16 \scrE_9 & 16 \scrE_{10} & 0 & 0 & 0 & 0 & 0 & 0 & 0 & 0 & 0 & 0 & 0 & 0 & 0 & 0 \\
 0 & 0 & 0 & 0 & 0 & 0 & 0 & 0 & 0 & 0 & 24 \scrE_{11} & 24 \scrE_{12} & 0 & 0 & 0 & 0 & 0 & 0 & 0 & 0 & 0 & 0 & 0 & 0 \\
 0 & 0 & 0 & 0 & 0 & 0 & 0 & 0 & 0 & 0 & 0 & 0 & 8 \scrE_{11} & 24 \scrE_{12} & 0 & 0 & 0 & 0 & 0 & 0 & 0 & 0 & 0 & 0 \\\hline
 0 & 0 & 0 & 0 & 0 & 0 & 0 & 0 & 0 & 0 & 24 \scrE_{13} & 8 \scrE_{14} & 0 & 0 & 0 & 0 & 0 & 0 & 0 & 0 & 0 & 0 & 0 & 0 \\
 0 & 0 & 0 & 0 & 0 & 0 & 0 & 0 & 0 & 0 & 0 & 0 & 24 \scrE_{13} & 24 \scrE_{14} & 0 & 0 & 0 & 0 & 0 & 0 & 0 & 0 & 0 & 0 \\
 0 & 0 & 0 & 0 & 0 & 0 & 0 & 0 & 0 & 0 & 0 & 0 & 0 & 0 & 16 \scrE_{15} & 16 \scrE_{16} & 16 \scrE_{17} & 0 & 0 & 0 & 0 & 0 & 0 & 0 \\
 0 & 0 & 0 & 0 & 0 & 0 & 0 & 0 & 0 & 0 & 0 & 0 & 0 & 0 & 0 & 0 & 0 & 48 \scrE_{15} & 48 \scrE_{16} & 48 \scrE_{17} & 0 & 0 & 0 & 0 \\
 0 & 0 & 0 & 0 & 0 & 0 & 0 & 0 & 0 & 0 & 0 & 0 & 0 & 0 & 0 & 0 & 0 & 0 & 0 & 0 & 6 \scrE_{15} & 6 \scrE_{16} & 24 \scrE_{17} & 0 \\
 0 & 0 & 0 & 0 & 0 & 0 & 0 & 0 & 0 & 0 & 0 & 0 & 0 & 0 & 16 \scrE_{18} & 16 \scrE_{19} & 16 \scrE_{20} & 0 & 0 & 0 & 0 & 0 & 0 & 0 \\\hline
 0 & 0 & 0 & 0 & 0 & 0 & 0 & 0 & 0 & 0 & 0 & 0 & 0 & 0 & 0 & 0 & 0 & 48 \scrE_{18} & 48 \scrE_{19} & 48 \scrE_{20} & 0 & 0 & 0 & 0 \\
 0 & 0 & 0 & 0 & 0 & 0 & 0 & 0 & 0 & 0 & 0 & 0 & 0 & 0 & 0 & 0 & 0 & 0 & 0 & 0 & 6 \scrE_{18} & 6 \scrE_{19} & 24 \scrE_{20} & 0 \\
 0 & 0 & 0 & 0 & 0 & 0 & 0 & 0 & 0 & 0 & 0 & 0 & 0 & 0 & 16 \scrE_{21} & 16 \scrE_{22} & 4 \scrE_{23} & 0 & 0 & 0 & 0 & 0 & 0 & 0 \\
 0 & 0 & 0 & 0 & 0 & 0 & 0 & 0 & 0 & 0 & 0 & 0 & 0 & 0 & 0 & 0 & 0 & 48 \scrE_{21} & 48 \scrE_{22} & 12 \scrE_{23} & 0 & 0 & 0 & 0 \\
 0 & 0 & 0 & 0 & 0 & 0 & 0 & 0 & 0 & 0 & 0 & 0 & 0 & 0 & 0 & 0 & 0 & 0 & 0 & 0 & 24 \scrE_{21} & 24 \scrE_{22} & 24 \scrE_{23} & 0 \\
 0 & 0 & 0 & 0 & 0 & 0 & 0 & 0 & 0 & 0 & 0 & 0 & 0 & 0 & 0 & 0 & 0 & 0 & 0 & 0 & 0 & 0 & 0 & 24 \scrE_{24} \\
\end{array}
\right)}
\end{equation*}

\subsection{Kronecker basis of $S_L$}\label{app:Kron basis SL}

Recall that the Kronecker basis of $S_L$ is defined by \eqref{def:kr basis SL},
\begin{equation}
\cQ^{R,\Lambda, \tau}_{\ \ K} = \frac{d_R}{|S_L|} \sum_{g \in S_L} \sum_{I,J=1}^{d_R} 
\CGs{R}{R}{\Lambda}{\tau}{I}{J}{K} \, D^R_{JI}(g) \, g^{-1}
= \frac{d_R}{|S_L|} \sum_{g \in S_L} \sum_{I,J=1}^{d_R} 
\CGs{R}{R}{\Lambda}{\tau}{I}{J}{K} \, D^R_{IJ}(g) \, g
\label{app:kr basis SL}
\end{equation}
We solve the eigenvalue system \eqref{Eigen Kron} using the Hermite normal form to compute the integer Kronecker basis, which is related to the original Kronecker basis by 
\begin{equation}
\sfQ^{R,\Lambda, \tau}_{\ \ K} \equiv \sfN^{R,\Lambda, \tau}_{\ \ K} \, \cQ^{R,\Lambda, \tau}_{\ \ K} 
\label{app:Kron INT basis}
\end{equation}
where $\sfN^{R,\Lambda, \tau}_{\ \ K}$ is a normalisation constant.

We will omit the multiplicity label $\tau$\,, because $C(R,R,\Lambda) \in \{ 0, 1 \}$ for $L \le 4$.

\subsubsection{Case of $S_2$}

One finds
\begin{equation}
\sfQ^{[2] ,[2]}  = () + (12) \,, \qquad \sfQ^{[1,1],[2]} = () - (12) , \qquad
\sfQ^{R, [2]} \sfQ^{S, [2]} = 2 \, \delta^{RS} \sfQ^{R, [2]} 
\end{equation}
where we omit $K$ because all representations are one-dimensional.
There is no basis element with $\Lambda=[1,1]$ because $C(R,R,[1,1])=0$ at $L=2$.

\subsubsection{Case of $S_3$}

We express the matrix unit $\cQ^{R,\Lambda, \tau}_{\ \ K}$ as a pair $\{ R, K \}$, where $R$ is a Young diagram and $K$ is a Young tableau $K$ of shape $\Lambda$.
We reorganise them into a vector
\ytableausetup{boxsize=1em,centertableaux}
\begin{equation}
\begin{aligned}
\scrF_\rho &\equiv \( \sfQ^{[1,1,1],[3]}_{\qquad \ 1} \,, 
\sfQ^{[3],[3]}_{\quad \, 1} \,, 
\sfQ^{[2,1],[1,1,1]}_{\quad \ \ \, 1} \,,
\sfQ^{[2,1],[3]}_{\quad \ \ \, 1} \,,
\sfQ^{[2,1],[2,1]}_{\quad \ \ \, 1} \,,
\sfQ^{[2,1],[2,1]}_{\quad \ \ \, 2}  \)
\\[1mm]
&= \( {\scriptsize
\Bigl\{  \, \ydiagram{3} \,,\,\begin{ytableau}1 & 2 & 3 \end{ytableau}\Bigr\} , 
\Bigl\{  \, \ydiagram{2, 1} \,,\,\begin{ytableau}1 & 2 & 3 \end{ytableau}\Bigr\} , 
\Bigl\{  \, \ydiagram{2, 1} \,,\,\begin{ytableau}1 & 2 \\ 3 \end{ytableau}\Bigr\} , 
\Bigl\{  \, \ydiagram{2, 1} \,,\,\begin{ytableau}1 & 3 \\ 2 \end{ytableau}\Bigr\} ,
\Bigl\{  \, \ydiagram{2, 1} \,,\,\begin{ytableau}1 \\ 2 \\ 3 \end{ytableau}\Bigr\} , 
\Bigl\{  \, \ydiagram{1, 1, 1} \,,\,\begin{ytableau}1 & 2 & 3 \end{ytableau}\Bigr\} 
} \) 
\end{aligned}
\end{equation}
and expand it in the basis of $S_3$ \eqref{app:basis of S3} as
\begin{equation}
\scrF_\rho = \sum_{g \in {\tt Basis}} F_{\rho g} \, g .
\end{equation}
The coefficient matrix in the integer basis is given by
\begin{equation}
F_{\rho g} = 
\begin{pmatrix}
 1 & 1 & 1 & 1 & 1 & 1 \\
 2 & 0 & 0 & -1 & -1 & 0 \\
 0 & 1 & -2 & 0 & 0 & 1 \\
 0 & 1 & 0 & 0 & 0 & -1 \\
 0 & 0 & 0 & 1 & -1 & 0 \\
 1 & -1 & -1 & 1 & 1 & -1 \\
\end{pmatrix} .
\label{S3 Kron Fag}
\end{equation}
For example,
\begin{equation}
\begin{alignedat}{9}
\scrF_1 &= \Bigl\{  \, \ydiagram{3} \,,\,\begin{ytableau}1 & 2 & 3 \end{ytableau}\Bigr\} 
& &= () + (23) + (12) + (123) + (132) + (13).
\\
\scrF_6 &= \Bigl\{  \, \ydiagram{1, 1, 1} \,,\,\begin{ytableau}1 & 2 & 3 \end{ytableau}\Bigr\} 
& &= () - (23) - (12) + (123) + (132) - (13) .
\end{alignedat}
\end{equation}
If we define $\scrF_\rho^{-1} = \sum_{g \in {\tt Basis}} F_{\rho g} \, g^{-1}$ and compute the $\delta$-function inner product, we find
\begin{equation}
\delta \Big( \scrF_\rho^{-1} \scrF_{\rho'} \Big) = \sfM_\rho \, \delta_{\rho \rho'} \,, \qquad
\sfM_\rho = \( 6,6,6,2,2,6 \).
\end{equation}

\subsubsection{Case of $S_4$}

Again, the multiplicity label $\tau$ is trivial at $L=4$.
Define the vector of Kronecker basis as
\ytableausetup{boxsize=1em,centertableaux}
\begin{align}
\hspace{-3mm}
\scrF_\rho \equiv &\Biggl( {\scriptsize
\Bigl\{ \, \ydiagram{4} \,, \, \begin{ytableau}1 & 2 & 3 & 4 \end{ytableau} \Bigr\} ,
\Bigl\{ \, \ydiagram{3, 1} \,, \, \begin{ytableau}1 & 2 & 3 & 4 \end{ytableau} \Bigr\} ,
\Bigl\{ \, \ydiagram{3, 1} \,, \, \begin{ytableau}1 & 2 & 3 \\ 4 \end{ytableau} \Bigr\} ,
\Bigl\{ \, \ydiagram{3, 1} \,, \, \begin{ytableau}1 & 2 & 4 \\ 3 \end{ytableau} \Bigr\} ,
}\notag\\[1mm]
&{\scriptsize
\Bigl\{ \, \ydiagram{3, 1} \,, \, \begin{ytableau}1 & 3 & 4 \\ 2 \end{ytableau} \Bigr\} ,
\Bigl\{ \, \ydiagram{3, 1} \,, \, \begin{ytableau}1 & 2 \\ 3 & 4 \end{ytableau} \Bigr\} ,
\Bigl\{ \, \ydiagram{3, 1} \,, \, \begin{ytableau}1 & 3 \\ 2 & 4 \end{ytableau} \Bigr\} ,
\Bigl\{ \, \ydiagram{3, 1} \,, \, \begin{ytableau}1 & 2 \\ 3 \\ 4 \end{ytableau} \Bigr\} ,
\Bigl\{ \, \ydiagram{3, 1} \,, \, \begin{ytableau}1 & 3 \\ 2 \\ 4 \end{ytableau} \Bigr\} ,
}\notag\\[1mm]
&{\scriptsize
\Bigl\{ \, \ydiagram{3, 1} \,, \, \begin{ytableau}1 & 4 \\ 2 \\ 3 \end{ytableau} \Bigr\} ,
\Bigl\{ \, \ydiagram{2, 2} \,, \, \begin{ytableau}1 & 2 & 3 & 4 \end{ytableau} \Bigr\} ,
\Bigl\{ \, \ydiagram{2, 2} \,, \, \begin{ytableau}1 & 2 \\ 3 & 4 \end{ytableau} \Bigr\} ,
\Bigl\{ \, \ydiagram{2, 2} \,, \, \begin{ytableau}1 & 3 \\ 2 & 4 \end{ytableau} \Bigr\} ,
\Bigl\{ \, \ydiagram{2, 2} \,, \, \begin{ytableau}1 \\ 2 \\ 3 \\ 4 \end{ytableau} \Bigr\} ,
}\notag\\[1mm]
&{\scriptsize
\Bigl\{ \, \ydiagram{2, 1, 1} \,, \, \begin{ytableau}1 & 2 & 3 & 4 \end{ytableau} \Bigr\} ,
\Bigl\{ \, \ydiagram{2, 1, 1} \,, \, \begin{ytableau}1 & 2 & 3 \\ 4 \end{ytableau} \Bigr\} ,
\Bigl\{ \, \ydiagram{2, 1, 1} \,, \, \begin{ytableau}1 & 2 & 4 \\ 3 \end{ytableau} \Bigr\} ,
\Bigl\{ \, \ydiagram{2, 1, 1} \,, \, \begin{ytableau}1 & 3 & 4 \\ 2 \end{ytableau} \Bigr\} ,
\Bigl\{ \, \ydiagram{2, 1, 1} \,, \, \begin{ytableau}1 & 2 \\ 3 & 4 \end{ytableau} \Bigr\} ,
}\notag\\[1mm]
&{\scriptsize
\Bigl\{ \, \ydiagram{2, 1, 1} \,, \, \begin{ytableau}1 & 3 \\ 2 & 4 \end{ytableau} \Bigr\} ,
\Bigl\{ \, \ydiagram{2, 1, 1} \,, \, \begin{ytableau}1 & 2 \\ 3 \\ 4 \end{ytableau} \Bigr\} ,
\Bigl\{ \, \ydiagram{2, 1, 1} \,, \, \begin{ytableau}1 & 3 \\ 2 \\ 4 \end{ytableau} \Bigr\} ,
\Bigl\{ \, \ydiagram{2, 1, 1} \,, \, \begin{ytableau}1 & 4 \\ 2 \\ 3 \end{ytableau} \Bigr\} ,
\Bigl\{ \, \ydiagram{1, 1, 1, 1} \,, \, \begin{ytableau}1 & 2 & 3 & 4 \end{ytableau} \Bigr\}
} \Biggr) 
\end{align}
and expand it in the basis of $S_4$ \eqref{app:basis of S4}.
The coefficient matrix in the integer basis is given by
\begin{equation}
F_{\rho g} = {\scriptsize \left(
\begin{array}{@{\hskip 2pt}c@{\hskip 2pt}c@{\hskip 2pt}c@{\hskip 2pt}c@{\hskip 2pt}c@{\hskip 2pt}c|@{\hskip 2pt}c@{\hskip 2pt}c@{\hskip 2pt}c@{\hskip 2pt}c@{\hskip 2pt}c@{\hskip 2pt}c|@{\hskip 2pt}c@{\hskip 2pt}c@{\hskip 2pt}c@{\hskip 2pt}c@{\hskip 2pt}c@{\hskip 2pt}c|@{\hskip 2pt}c@{\hskip 2pt}c@{\hskip 2pt}c@{\hskip 2pt}c@{\hskip 2pt}c@{\hskip 2pt}c}
 1 & 1 & 1 & 1 & 1 & 1 & 1 & 1 & 1 & 1 & 1 & 1 & 1 & 1 & 1 & 1 & 1 & 1 & 1 & 1 & 1 & 1 & 1 & 1 \\
 3 & 1 & 1 & 0 & 0 & 1 & 1 & -1 & 0 & -1 & -1 & 0 & 0 & -1 & 1 & 0 & -1 & -1 & -1 & 0 & 0 & 1 & -1 & -1 \\
 0 & 2 & -2 & 1 & 1 & 2 & -2 & 0 & -3 & 0 & 0 & 1 & -3 & 0 & -2 & 1 & 0 & 0 & 0 & 1 & 1 & 2 & 0 & 0 \\
 0 & 2 & 1 & 1 & 1 & -1 & -2 & 0 & 0 & 0 & 0 & -2 & 0 & 0 & 1 & 1 & 0 & 0 & 0 & -2 & 1 & -1 & 0 & 0 \\
 0 & 0 & 1 & 1 & 1 & 1 & 0 & 0 & 0 & 0 & 0 & 0 & 0 & 0 & -1 & -1 & 0 & 0 & 0 & 0 & -1 & -1 & 0 & 0 \\
 0 & 2 & -1 & 0 & 0 & -1 & 2 & 4 & 0 & 1 & 1 & 0 & 0 & 1 & -1 & 0 & -2 & -2 & 1 & 0 & 0 & -1 & -2 & -2 \\\hline
 0 & 0 & 1 & 0 & 0 & -1 & 0 & 0 & 0 & 1 & -1 & 0 & 0 & -1 & -1 & 0 & -2 & 0 & 1 & 0 & 0 & 1 & 0 & 2 \\
 0 & 0 & 0 & 1 & -1 & 0 & 0 & 0 & 0 & 1 & -1 & 0 & 0 & 1 & 0 & 1 & 0 & 0 & -1 & 0 & -1 & 0 & 0 & 0 \\
 0 & 0 & 0 & 1 & -1 & 0 & 0 & 0 & 0 & -1 & -1 & -2 & 0 & 1 & 0 & -1 & 0 & -2 & 1 & 2 & 1 & 0 & 2 & 0 \\
 0 & 0 & 0 & 1 & -1 & 0 & 0 & 0 & 3 & 2 & 2 & 1 & -3 & -2 & 0 & -1 & 0 & -2 & -2 & -1 & 1 & 0 & 2 & 0 \\
 2 & 0 & 0 & -1 & -1 & 0 & 0 & 2 & -1 & 0 & 0 & -1 & -1 & 0 & 0 & -1 & 2 & 0 & 0 & -1 & -1 & 0 & 0 & 2 \\
 0 & 2 & -1 & 0 & 0 & -1 & 2 & 0 & 0 & -1 & -1 & 0 & 0 & -1 & -1 & 0 & 0 & 2 & -1 & 0 & 0 & -1 & 2 & 0 \\\hline
 0 & 0 & 1 & 0 & 0 & -1 & 0 & 0 & 0 & -1 & 1 & 0 & 0 & 1 & -1 & 0 & 0 & 0 & -1 & 0 & 0 & 1 & 0 & 0 \\
 0 & 0 & 0 & 1 & -1 & 0 & 0 & 0 & -1 & 0 & 0 & 1 & 1 & 0 & 0 & -1 & 0 & 0 & 0 & -1 & 1 & 0 & 0 & 0 \\
 3 & -1 & -1 & 0 & 0 & -1 & -1 & -1 & 0 & 1 & 1 & 0 & 0 & 1 & -1 & 0 & -1 & 1 & 1 & 0 & 0 & -1 & 1 & -1 \\
 0 & 2 & -2 & -1 & -1 & 2 & -2 & 0 & 3 & 0 & 0 & -1 & 3 & 0 & -2 & -1 & 0 & 0 & 0 & -1 & -1 & 2 & 0 & 0 \\
 0 & 2 & 1 & -1 & -1 & -1 & -2 & 0 & 0 & 0 & 0 & 2 & 0 & 0 & 1 & -1 & 0 & 0 & 0 & 2 & -1 & -1 & 0 & 0 \\
 0 & 0 & 1 & -1 & -1 & 1 & 0 & 0 & 0 & 0 & 0 & 0 & 0 & 0 & -1 & 1 & 0 & 0 & 0 & 0 & 1 & -1 & 0 & 0 \\\hline
 0 & 2 & -1 & 0 & 0 & -1 & 2 & -4 & 0 & 1 & 1 & 0 & 0 & 1 & -1 & 0 & 2 & -2 & 1 & 0 & 0 & -1 & -2 & 2 \\
 0 & 0 & 1 & 0 & 0 & -1 & 0 & 0 & 0 & 1 & -1 & 0 & 0 & -1 & -1 & 0 & 2 & 0 & 1 & 0 & 0 & 1 & 0 & -2 \\
 0 & 0 & 0 & 1 & -1 & 0 & 0 & 0 & 0 & -1 & 1 & 0 & 0 & -1 & 0 & 1 & 0 & 0 & 1 & 0 & -1 & 0 & 0 & 0 \\
 0 & 0 & 0 & 1 & -1 & 0 & 0 & 0 & 0 & 1 & 1 & -2 & 0 & -1 & 0 & -1 & 0 & 2 & -1 & 2 & 1 & 0 & -2 & 0 \\
 0 & 0 & 0 & 1 & -1 & 0 & 0 & 0 & 3 & -2 & -2 & 1 & -3 & 2 & 0 & -1 & 0 & 2 & 2 & -1 & 1 & 0 & -2 & 0 \\
 1 & -1 & -1 & 1 & 1 & -1 & -1 & 1 & 1 & -1 & -1 & 1 & 1 & -1 & -1 & 1 & 1 & -1 & -1 & 1 & 1 & -1 & -1 & 1 \\
\end{array}
\right) }
\label{S4 Kron Fag}
\end{equation}
Their $\delta$-function inner product is given by  $\delta ( \scrF_\rho^{-1} \scrF_{\rho'} ) = \sfM_\rho \, \delta_{\rho \rho'}$\,, where
\begin{equation}
\sfM_\rho = \( 24, 24, 48, 24, 8, 48,\ \ 
16, 8, 24, 48, 24, 24, \ \ 
8, 8, 24, 48, 24, 8, \ \ 
48, 16, 8, 24, 48, 24 \).
\end{equation}

\newpage
\bibliographystyle{utphys}
\bibliography{bibgg}

\end{document}